\documentclass[
reprint,
superscriptaddress,
amsmath,amssymb,
aps, prl, 
]{revtex4-2}

\usepackage{graphicx}
\usepackage{physics}
\usepackage[
    colorlinks=true,    
    linkcolor=blue,     
    citecolor=blue,     
    urlcolor=blue,      
    hyperindex=true,
    breaklinks=true
]{hyperref}
\usepackage{adjustbox}

\newcommand{\lket}[1]{|#1\rangle\!\rangle}
\newcommand{\lbra}[1]{\langle\!\langle#1|}

\newcommand{\norder}[1]{{\mathopen{:}#1\mathclose{:}}}

\DeclareUnicodeCharacter{FFFD}{\"o}

\begin{document}

\title{Unveiling Hierarchical Invariants in Multiphoton Linear Optics}
\author{Baichuan Yang}
\thanks{These authors contributed equally to this work.}
\author{Hao Zhan}
\thanks{These authors contributed equally to this work.}
\author{Minghao Mi}
\affiliation{National Laboratory of Solid State Microstructures, Key Laboratory of Intelligent Optical Sensing and Manipulation, College of Engineering and Applied Sciences, Jiangsu Physical Science Research Center, and Collaborative Innovation Center of Advanced Microstructures, Nanjing University, Nanjing, Jiangsu 210093, China}
\author{Aonan Zhang}
\affiliation{Clarendon Laboratory, University of Oxford, Parks Road, Oxford OX1 3PU, United Kingdom}
\affiliation{Blackett Laboratory, Department of Physics, Imperial College London, Prince Consort Rd, London, SW7 2AZ, United Kingdom}
\author{Liang Xu}
\email{liangxu.ceas@nju.edu.cn}
\author{Lijian Zhang}
\email{lijian.zhang@nju.edu.cn}
\affiliation{National Laboratory of Solid State Microstructures, Key Laboratory of Intelligent Optical Sensing and Manipulation, College of Engineering and Applied Sciences, Jiangsu Physical Science Research Center, and Collaborative Innovation Center of Advanced Microstructures, Nanjing University, Nanjing, Jiangsu 210093, China}

\date{\today}

\begin{abstract}
Linear optical networks driven by quantum states of light are important building blocks of photonic quantum technologies. 
They access large bosonic Hilbert spaces through multiphoton interference.
At the same time, their dynamics are generated by single-particle mode transformations, thereby defining a highly structured subset of multiphoton unitaries and setting boundary on linear optics capability.
To elucidate this boundary, we reveal an underlying fine-grained symmetry structure that partitions the multiphoton operator space into invariant subspaces and generates a hierarchy of invariants.
We experimentally confirm the conservation of high-order invariants and 
demonstrate their operational utility in characterizing state reachability and  the metrological capability of multiphoton probes.
Our framework provides a symmetry-based perspective for understanding and harnessing structured multiphoton dynamics across photonic quantum technologies.
\end{abstract}

\maketitle

\textit{Introduction.---}Photonics has emerged as a versatile platform for quantum technologies, offering 
high-fidelity operations~\cite{senellart_highperformance_2017, wang_integrated_2020},
rich degrees of freedom~\cite{erhard_advances_2020}, and long-distance communication~\cite{wehner_quantum_2018, xu_secure_2020}.
Due to the weak interaction between photons, linear optical networks (LONs), constructed from beam splitters and phase shifters, form  the primary platform in photonics for manipulating multiphoton states~\cite{knill_scheme_2001, peruzzo_variational_2014, madsen_quantum_2022}.
By harnessing multiphoton interference, LONs generate  entanglement~\cite{killoran_extracting_2014, pan_multiphoton_2012, park_realization_2023} and access Hilbert spaces whose dimensions grow  exponentially with photon number~\cite{steinbrecher_quantum_2019}.
This capability is central to demonstrating quantum computational advantage such as boson sampling~\cite{aaronson_computational_2011, zhong_quantum_2020}, measurement-based quantum computation~\cite{nielsen_optical_2004, bartolucci_fusionbased_2023, hoch_quantum_2025}, and  Heisenberg-limited phase estimation using N00N states~\cite{boto_quantum_2000, nagata_beating_2007}.
However, despite the exponential growth of Hilbert space, the number of degrees of freedom in a LON scales only quadratically with mode number~\cite{reck_experimental_1994, clements_optimal_2016}. 
Consequently, LONs implement a family of highly structured multiphoton unitaries rather than arbitrary ones~\cite{oszmaniec_universal_2017, garcia-escartin_method_2019}, 
which limits deterministic state interconversion and defines the accessible preparation, processing, and measurement capabilities of multiphoton optical states~\cite{wangScalableAdvantageMultiphoton2025, ewert_$3_2014}.
\par
To characterize the capability landscape of multiphoton linear optics, previous works have sought to identify properties of multiphoton states that remain invariant under linear optics~\cite{migdal_multiphoton_2014}. 
Recently, an invariant subspace of the multiphoton operator space, termed the tangent space, was identified, with the associated tangent invariant providing a necessary condition for state interconversion~\cite{parellada_nogo_2023}. 
Subsequent theoretical efforts have extended the set of multiphoton invariants for general bosonic states based on observables~\cite{parellada_lie_2024}, and  further indicated the existence of a finite complete set of invariants~\cite{drauxInvariantsLinearOptics2025}.
The first-order Lie-algebra invariant introduced in~\cite{parellada_lie_2024} has been experimentally verified and related to the optical coherence~\cite{rodari_observation_2025}.
This naturally raises the question of whether the operator space admits a finer decomposition beyond the tangent space. 
At the same time, while existing invariants capture the intrinsic properties of bosonic states under linear optical evolution,
their operational role in characterizing the capabilities of multiphoton states in linear-optical tasks remains largely unexplored.
\par
In this work, guided by Noether's theorem~\cite{noether_invariant_1971}, we show that the structured dynamics of LONs gives rise to an underlying fine-grained symmetry structure.
This symmetry implies that coherence layers, defined as collections of coherence operators of specific order, are invariant under LON dynamics.
From this structure, we decompose multiphoton states onto the coherence layers, and identify a family of hierarchical invariants that generalizes the Lie-algebraic invariants in~\cite{parellada_nogo_2023}.
Furthermore, we highlight the utility of these invariants with two examples.
First, mismatches between invariants of different states quantify state reachability under LONs, which lower-bounds the minimum trace distance between a target state and the set of states reachable from a given input.
Second, invariants characterize the metrological capability of multiphoton probes, both for unknown parameter encodings and for LON probe optimization.
Experimentally, we generate a family of states with varying invariants via post-selection and confirm both the conservation of high-order invariants and their utility in characterizing state reachability and metrological capability.
Together, these results establish a hierarchical invariant framework for multiphoton linear optics, providing insights onto the capability of LONs in quantum information processing.

\par
\begin{figure}[htb]
    \centering
    \includegraphics[width=0.9\linewidth]{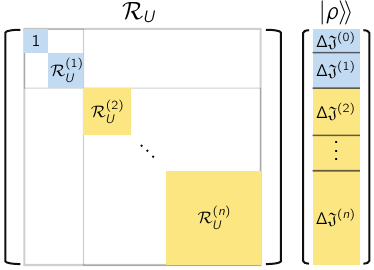}
    \caption{The Liouville representations of the LON unitary and the state. In the Liouville space, the state is vectorized as $\lket{\rho}$, and the LON unitary is represented by a block-diagonal superoperator $\mathcal{R}_U$, preserving each coherence layer.}
    \label{fig:schematic_new}
\end{figure}

\textit{Theory.---}
In the Heisenberg picture, a passive, lossless $m$-mode LON transforms the annihilation and creation operators as
$\hat{a}_i \mapsto \sum_{j} S_{ij}\hat{a}_j,$
where $S$ is the single-particle scattering matrix belonging to the unitary group $\mathrm{U}(m)$.
Any element of $\mathrm{U}(m)$ can be physically implemented via universal interferometer architectures~\cite{reck_experimental_1994, carolan_universal_2015, clements_optimal_2016}.
For clarity, we first focus on the fixed $n$-photon sector $\mathcal{H}_{n,m}$ of dimension $M = \binom{m+n-1}{n}$, spanned by Fock states $\ket{n_1, \dots, n_m}$ with $\sum_i n_i = n$.
The scattering matrix $S$ induces an $n$-photon representation $\varphi_n: \mathrm{U}(m) \to \mathrm{U}(M)$ on $\mathcal{H}_{n,m}$, whose image defines the LON-accessible subgroup $\mathrm{U_L} \subset \mathrm{U}(M)$ of dimension $m^2 \ll M^2$~\cite{aaronson_computational_2011, garcia-escartin_method_2019}.
This dimensional gap is elucidated from the Lie-algebraic perspective.
LON Hamiltonians are defined via the Jordan--Schwinger (JS) map as $\mathcal J^{(1)}(h) = \sum_{s,t=1}^m h_{st} \,\hat{a}_s^\dagger \hat{a}_t$~\cite{jordan_zusammenhang_1935,schwinger_angular_1952,garcia-escartin_multiple_2019},
where $h$ is the $m\times m$ Hermitian matrix generating the scattering matrix $S=\exp(ih)$.
This Hamiltonian structure is restricted to operators containing exactly one creation–annihilation pair, precluding the nonlinear higher-order interactions required to generate arbitrary multiphoton unitaries.
\par
While general unitary dynamics naturally preserves the state purity and spectrum, the Hamiltonian structure of LONs implies a fine-grained symmetry structure.
Specifically, under the Heisenberg transformation induced by an LON, the operator subspace $\mathfrak{J}^{(1)}$ spanned by $\mathcal{J}^{(1)}(h)$ is invariant under LON unitaries~\cite{parellada_nogo_2023}:
$U\mathcal{J}^{(1)}(h)U^{\dagger}=\mathcal{J}^{(1)}(S^\dagger h S)\in \mathfrak{J}^{(1)}, \ \forall U\in \mathrm{U_L}$.
To unveil the invariant structure of the remaining operator space, we generalize the JS map as follows:
\begin{equation}
\mathcal{J}^{(j)}(h_1,\dots,h_j) = \norder{ \prod_{i=1}^j \mathcal{J}^{(1)}(h_i) },
\quad  j \le n,
\label{eq:Jj-main}
\end{equation}
where $h_1,\dots,h_j$ are $m\times m$ Hermitian matrices, and the surrounding colons denote the normal ordering of creation and annihilation operators.
We define the order-\(j\) Jordan space as 
\(\mathfrak J^{(j)}:=\mathrm{span}_{\mathbb R}\{\mathcal J^{(j)}(h_1,\ldots,h_j): h_1, \cdots, h_j \in \mathrm{Herm}(m)\}\).
Physically, $\mathfrak{J}^{(j)}$ collects all Hermitian operators built from $j$ creation--annihilation pairs, i.e.\ all $j$th-order coherence operators~\cite{glauber_quantum_1963}.
In particular, the highest-order space $\mathfrak{J}^{(n)}$ coincides with the full Hermitian operator space $\mathrm{Herm}(\mathcal{H}_{n,m})$ (see Supplementary Material (SM) Sec.~A~\cite{Supplementary}).

In the $n$-photon sector, Jordan spaces form a hierarchy:
$\mathfrak J^{(1)}\subset \mathfrak J^{(2)}\subset \cdots \subset \mathfrak J^{(n)}.$
This follows directly from the definition of $\mathcal{J}^{(j)}$. By setting $h_{j+1}=\tfrac{1}{n-j}I_m$ with $I_m$ as the $m\times m$ identity matrix, Eq.~\eqref{eq:Jj-main} yields that, for all $\ket{\psi}\in\mathcal H_{n,m}$,
\(
\mathcal J^{(j+1)}\big(h_1,\dots,h_j,\tfrac{1}{n-j}I_m\big)\ket{\psi}
=\mathcal J^{(j)}(h_1,\dots,h_j)\ket{\psi}.
\)
This implies that all lower-order coherences are effectively contained within higher-order ones.
To isolate the ``genuine'' order-$j$ coherence, we define the coherence layer $\Delta\mathfrak J^{(j)}$ as the Hilbert--Schmidt orthogonal complement of $\mathfrak J^{(j-1)}$ within $\mathfrak J^{(j)}$ for $j \ge 1$, and set $\Delta\mathfrak J^{(0)}:=\mathbb R\cdot I_M$.
This yields the hierarchical decomposition of the full Hermitian operator space:
\begin{equation}
\label{eq:orth-dec}
\mathrm{Herm}(\mathcal{H}_{n, m})=\bigoplus_{j=0}^{n}\Delta\mathfrak J^{(j)}.
\end{equation}
Let $\{H_\mu^{(j)}\}$ be an orthonormal basis adapt to this decomposition, with  $\mathrm{tr}(H_\mu^{(j)} H_\nu^{(k)}) = \delta_{jk}\delta_{\mu,\nu}$ and $\Delta \mathfrak{J}^{(j)} = \mathrm{span}_{\mathbb{R}} \{H_\mu^{(j)}\}$.
Then any density matrix can be decomposed into its coherence layer components:
$\rho^{(j)} := \sum_{\mu} \mathrm{tr}(H_\mu^{(j)} \rho) H_\mu^{(j)}$.

Remarkably, LON dynamics preserve this hierarchical structure.
Under the Heisenberg transformation, the order-$j$ JS map is covariant 
under any $U\in \mathrm{U_{L}}$:
$U \mathcal{J}^{(j)}(h_1,\dots,h_j) U^\dagger = \mathcal{J}^{(j)}(S^\dagger h_1 S, \dots, S^\dagger h_j S).$
It can be further shown (see SM Sec.~C~\cite{Supplementary}) that each coherence layer is individually invariant under LONs~\cite{wilkens_benchmarking_2024, arienzo_bosonic_2025}\footnote{This mathematical decomposition is also derived in these works for the purpose of filtered randomized benchmarking; here, we establish its physical interpretation as the conservation of hierarchical coherence.}
\begin{equation}
\label{eq:DeltaJ-invariance-main}
U\,\Delta\mathfrak J^{(j)}\,U^\dagger = \Delta\mathfrak J^{(j)}, \quad \forall U\in \mathrm{U_L}.
\end{equation}

This layer invariance can be made fully explicit in a Liouville representation~\cite{havel_procedures_2003, greenbaum_introduction_2015,gyamfi2020Fundamentals}.
The state is vectorized as $\lket{\rho}:= \sum_{\mu j} \mathrm{tr}( H_\mu^{(j)} \rho) \lket{H_\mu^{(j)}}$.
The LON unitary $U\in \mathrm{U_L}$ is then represented by the superoperator $\mathcal{R}_U:=\mathrm{tr}(H_\mu^{(j)} U H_\nu^{(k)} U^\dagger) \lket{ H_\nu^{(k)}}\! \lbra{H_\mu^{(j)}}$.
Crucially, Eq.~\eqref{eq:DeltaJ-invariance-main} ensures $\mathrm{tr}(H_\mu^{(j)} U H_\nu^{(k)} U^\dagger) = 0$ for $j\ne k$, so $\mathcal{R}_U$ is block diagonal with respect to coherence layers (see Fig.~\ref{fig:schematic_new}).
By defining the superoperator $\mathcal{P}_j := \sum_\mu \lket{H_\mu^{(j)}}\! \lbra{H_\mu^{(j)}}$ as the projection onto $\Delta\mathfrak J^{(j)}$, we obtain:
\begin{equation}
\label{eq:commutation}
    [\mathcal{R}_U, \mathcal{P}_j] = 0, \quad \forall U\in \mathrm{U_L}.
\end{equation}
Physically, this commutation relation unveils a fine-grained continuous symmetry of the LON dynamics.
In particular, it strictly forbids coherence transfer between different hierarchical orders.
Because LON dynamics conserves total photon number, a general bosonic state can be decomposed into invariant photon-number sectors~\cite{bartlett_reference_2007}. Within each fixed $n$-photon sector, our construction yields the hierarchical decomposition into coherence layers. Thus, the coherence layer structure provides a further refinement of the photon-number decomposition, and our framework extends naturally from fixed $n$-photon states to general bosonic states.

This fine-grained symmetry enforces the conservation of hierarchical properties of states.
The commutation relation in Eq.~\eqref{eq:commutation} implies $(U\rho U^\dagger)^{(j)} = U\rho^{(j)}U^\dagger$, meaning that the $j$-th coherence-layer component evolves by unitary conjugation,
thereby unveiling the layer purity and layer spectrum as hierarchical invariants
\begin{equation}
\label{eq:inv}
I_j:=\tr\!\big[(\rho^{(j)})^2\big],
\quad
\boldsymbol{\lambda}_j := \big(\lambda_1^{(j)}, \dots, \lambda_M^{(j)}\big)^\top,
\end{equation}
where $\lambda_{1}^{(j)}, \dots, \lambda_{M}^{(j)}$ are the eigenvalues of $\rho^{(j)}$ sorted in ascending order.
These hierarchical invariants recover the tangent invariant~\cite{parellada_nogo_2023} as $I_t = I_0 + I_1$ and the observable-based invariant~\cite{parellada_lie_2024} as $I_o = \binom{m+n}{m+1} I_1 + n^2/m$ (derived in SM Sec.~F~\cite{Supplementary}).

These conservation laws provide quantitative criteria for state reachability in LONs.
For instance, the Fock states $\ket{20}$ and $\ket{11}$ possess different first-layer purities,  precluding exact deterministic interconversion via LONs.
Moreover, the layer spectrum provides a finer characterization: the Fock state $\ket{22}$ and the N00N state $(\ket{40} + e^{i\phi}\ket{04})/\sqrt{2}$ share identical layer purities, yet differ in their layer spectra (see SM Sec.~E~\cite{Supplementary}).
To quantify approximate reachability, we define the spectral distance $D_{\mathrm{spec}}$, which lower-bounds the minimum trace distance between a target state $\sigma$ and the reachable set $\{U\rho U^\dagger: U\in U_L\}$ from an input state $\rho$, with the proof~\footnote{The derivation follows the strategy in Ref.~\cite{parellada_lie_2024}, adapted here to the coherence layer structure.} given in SM Sec.~H~\cite{Supplementary}:
\begin{subequations}
\label{eq:distance_bound}
\begin{align}
D_{\mathrm{spec}}^2(\rho, \sigma) &:= \frac{1}{4} \sum_{j=0}^n \big\| \boldsymbol{\lambda}_j(\rho) - \boldsymbol{\lambda}_j(\sigma) \big\|^2, \label{eq:def_Dspec} \\
D_{\mathrm{spec}}(\rho, \sigma) &\le \min_{U\in \mathrm{U_{L}}} D_{\mathrm{tr}}(U\rho U^\dagger, \sigma). \label{eq:bound_Dspec}
\end{align}
\end{subequations}
Here, $\|\cdot\|$ denotes the Euclidean norm. 


Beyond characterizing the state reachability, these invariants we unveil also quantify the metrological capability of multiphoton probes.
Consider a multiphoton probe $\ket{\psi}$ undergoing an LON unitary encoding $\exp(i\theta H)$ with $H\in \mathfrak{J}^{(1)}$.
In two complementary scenarios, we show that these hierarchical invariants determine the probe's metrological power characterized by quantum Fisher information (QFI)~\cite{1976quantum, PhysRevLett.72.3439}:
\begin{subequations}
\label{eq:metrology_limits}
\begin{alignat}{2}
\overline{F}_Q &= \alpha_{n, m} - \beta_{n, m} I_1(\psi)  &\quad& (\text{general } n, m), \label{eq:FQ_avg} \\
F_Q^{\text{max}} &= 2\Big(1 + \sqrt{1-2I_1(\psi)}\Big) && (\text{for } n=m=2). \label{eq:FQ_opt}
\end{alignat}
\end{subequations}
See SM Sec.~I~\cite{Supplementary} for proof.
For an unknown encoding Hamiltonian, Eq.~\eqref{eq:FQ_avg} proves that the average QFI over all $H$ for a fixed probe is determined by the first-layer purity $I_1$ ($\alpha_{n, m}$ and $\beta_{n, m}$ are positive constants).
This implies that increasing higher-order coherence, equivalently decreasing $I_1$, improves the average sensitivity of the probe. In particular, the N00N state, for which $I_1=0$, attains the highest average QFI.
Turning to the known Hamiltonian scenario, Eq.~\eqref{eq:FQ_opt} yields the maximal QFI achievable via optimal LON preprocessing for two-photon two-mode states.
Indeed, multiphoton states with different invariants exhibit different degrees of identical‑particle entanglement~\cite{morris_entanglement_2020}, reflecting differences in their metrological capabilities.

\begin{figure}[t]
    \centering
    \includegraphics[width=1\linewidth]{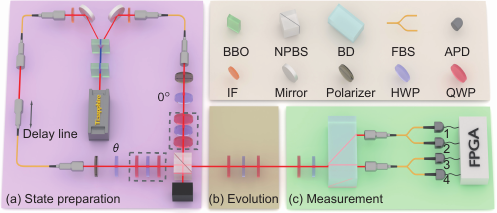}
    \caption{Experimental setup for observing hierarchical invariants. 
    (a) State Preparation: Photon pairs are generated via type-II spontaneous parametric down-conversion (SPDC) and then pass through two interference filters (IFs). Two-photon states with varying invariant values depending on $\theta$, are prepared using HOM interference followed by post-selection.
    The green dashed boxes indicate two QWP–HWP–QWP (QHQ) wave plate groups, which correct for non-ideal phase and polarization changes induced by the non-polarizing beam splitter (NPBS) (see SM Sec.~J~\cite{Supplementary}).
    (b) Evolution: Arbitrary two-mode unitary evolutions can be realized using a QHQ wave plate group.  
    (c) Measurement: Projective measurements on two-photon two-mode basis states are performed using a QWP and an HWP to set measurement basis, followed by a beam displacer (BD) and two pseudo photon-number-resolving detectors (PPNRDs). Each PPNRD comprises a fiber beam splitter (FBS) and two avalanche photodiodes (APDs), with detection counts processed by a field-programmable gate array (FPGA). BBO: $\beta$-barium borate crystal. QWP: quarter-wave plate. HWP: half-wave plate.
    }
\label{fig:exp}
\end{figure}
\begin{figure*}[t]
    \centering
\includegraphics[width=1\linewidth]{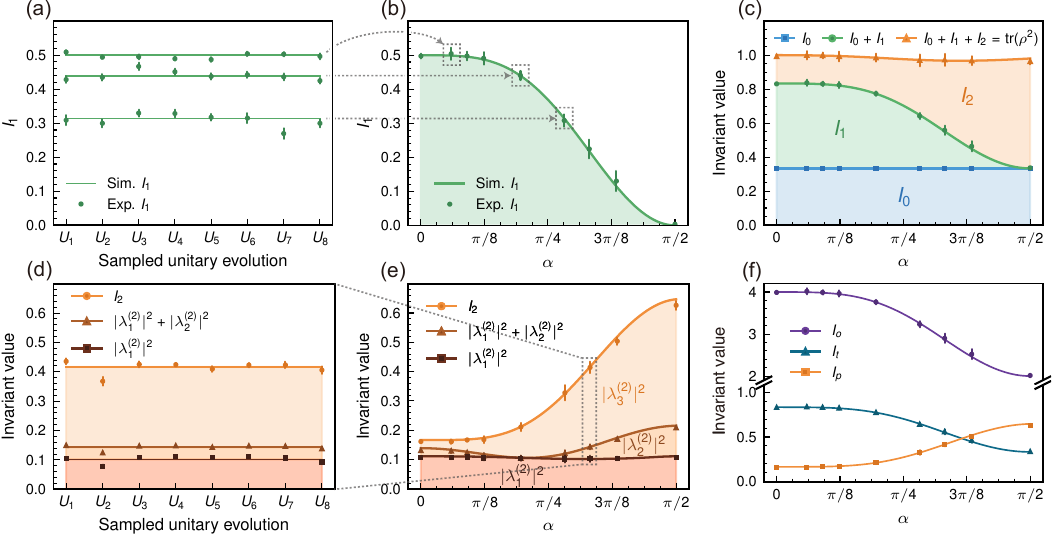}
    \caption{
    Experimental verification of the conservation of hierarchical invariants.
(a) Directly measured $I_1$ (markers) for three representative inputs [marked in (b)] across eight sampled unitaries $\{U_k\}$. Solid lines represent simulations considering the imperfect HOM visibility ($96.8\%$) (see SM Sec.~J~\cite{Supplementary}), applicable to all panels.
(b) Unitary-averaged invariant $\bar I_1(\alpha)$ (markers) versus state parameter $\alpha$.
(c) Decomposition of purity $\tr(\rho^2)=I_0+I_1+I_2$. Markers show the unitary-averaged cumulative sums, while shaded areas indicate the simulated contribution of each coherence layer.
(d) Measured second-layer spectrum across $\{U_k\}$ for the specific input marked in (e).
(e) Decomposition of the second-layer spectrum. Markers represent unitary-averaged cumulative sums.
(f) Observation of Lie-algebraic invariants ($I_o, I_t, I_p$)~\cite{parellada_nogo_2023, parellada_lie_2024}.
$I_t$ and $I_o$ are directly measured, while $I_p$ is derived from tomography.
In (b), (c), (e), and (f), markers denote averages over $\{U_k\}$ and error bars indicate the standard deviation; in (a) and (d), error bars are simulated via Poissonian counting statistics.
    }
 \label{fig:result}
\end{figure*}
\textit{Experiments.}---We experimentally verify the conservation of the proposed invariants and demonstrate their operational utility in state interconversion and metrology.
To observe these invariants, the standard approach is to reconstruct the full state via quantum state tomography~\cite{banchi_multiphoton_2018}, followed by direct computation.
The first-layer purity $I_1$ can also be efficiently determined from $m^2{-}1$ observable values, thereby circumventing the overhead of full tomography.
We construct observables $\hat{O}_\alpha = \mathcal{J}^{(1)}(\Gamma_\alpha)$ from the $m^2-1$ generalized Gell-Mann matrices $\{\Gamma_\alpha\}$.
Crucially, these operators form an orthogonal basis for the first coherence layer $\Delta\mathfrak{J}^{(1)}$, so that the invariant can be directly obtained from the squared expectation values up to a normalization factor, $I_1 \propto \sum_\alpha \langle \hat{O}_\alpha \rangle^2$ (see SM Sec.~F~\cite{Supplementary}).
Experimentally, the diagonal observables are measured by direct photon counting, while the off-diagonal terms are accessed by interfering two modes on a beam splitter followed by photon counting~\footnote{This measurement strategy aligns with Ref.~\cite{parellada_lie_2024}. The generalized Gell-Mann basis links our hierarchical definition to their observable-based invariants, ensuring that both share the same experimental accessibility.}.

The experimental setup for our two-photon two-mode demonstration is illustrated in Fig.~\ref{fig:exp}. It comprises three modules: state preparation (a), evolution (b), and measurement (c).
In the state preparation module, photon pairs are generated via type-II beam-like SPDC, pumped by a frequency-doubled Ti:sapphire laser. 
One photon of each pair is prepared as $\cos 2\theta\ket{H}+\sin2\theta\ket{V}$ via a HWP, while the other is initialized in $\ket{H}$, where $\ket{H}$ and $\ket{V}$ denote horizontal and vertical polarizations, respectively.
To generate two-photon states with distinct invariants, we perform Hong-Ou-Mandel (HOM) interference~\cite{hong_measurement_1987, branczyk_hongoumandel_2017} at an NPBS and postselect events in which both photons exit the same output port.
This prepares a family of states $\ket{\psi_\alpha}=\cos\alpha\,\ket{2_H,0_V}+\sin\alpha\,\ket{1_H,1_V}$ with $0\le\alpha\le\pi/2$, where $\alpha$ is set by $\theta$ (see SM Sec.~J~\cite{Supplementary}).
In the evolution module, arbitrary linear optical unitary evolution on the two polarization modes can be realized using a QHQ wave plate group, where a HWP is sandwiched between two QWPs.
For each input state $\rho_\alpha = \ket{\psi_\alpha}\!\!\bra{\psi_\alpha}$, we apply eight representative LON unitaries $U_k$ (see SM Sec.~K~\cite{Supplementary}), generating the output states $\rho_{\alpha, k}:= U_k \rho_{\alpha}  U^\dagger_k$.
In the measurement module, a beam displacer (BD) followed by two pseudo photon-number-resolving detectors (PPNRDs) enables projections onto the two-photon two-mode basis states. 
Coincidence events among the four APDs correspond to projections onto specific Fock states: \{1, 2\} for \(\ket{2_H, 0_V}\), \{3, 4\} for \(\ket{0_H, 2_V}\), and all others for \(\ket{1_H, 1_V}\).
A QWP and an HWP placed before the BD rotate the measurement basis, and six QWP--HWP settings are used for quantum state tomography~\cite{adamson_multiparticle_2007, banchi_multiphoton_2018} (see SM Sec.~L~\cite{Supplementary}).
\par
We first experimentally confirm the conservation of hierarchical invariants by tracking the first-layer purity $I_1$ and the second-layer spectrum $\boldsymbol{\lambda}_2$ across sampled unitaries [Figs.~\hyperref[fig:result]{3(a)} and \hyperref[fig:result]{3(d)}], where the observed stability within experimental uncertainty confirms the conservation law.
To characterize the invariant statistics for a given input state, we employ a unitary-averaging strategy to compute the mean $\bar{I}(\alpha) :=\frac{1}{K}\sum_{k} I(\rho_{\alpha, k})$ and the standard deviation $\sigma(\alpha) := \sqrt{\frac{1}{K}\sum_{k} [I(\rho_{\alpha, k}) - \bar{I}(\alpha)]^2}$ with $K=8$.
The mean shows the experimental agreement with theoretical predictions, while the standard deviation quantifies the robustness of the conservation under different sampled unitaries.
Visualizing these statistics (mean as markers, standard deviation as error bars), Fig.~\hyperref[fig:result]{3(c)} validates the decomposition of global purity into layer contributions, $\mathrm{tr}(\rho^2)=\sum_{j=0}^2 I_j$, while Fig.~\hyperref[fig:result]{3(e)} further resolves the second-layer purity $I_2$ into its constituent spectral invariants.
Fig.~\hyperref[fig:result]{3(f)} presents the experimental observation of the Lie-algebraic invariants $(I_t, I_p, I_o)$.


\begin{figure}[t]
    \centering
    \includegraphics[width=1\linewidth]{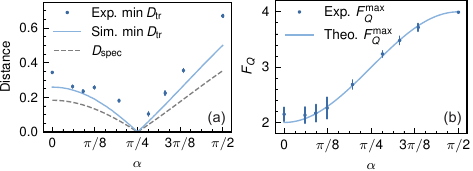}
    \caption{\label{fig:fig4}
Operational utility of hierarchical invariants.
(a) State reachability. Markers represent the minimal trace distance to the target state $\rho_{\pi/4}$ for an input state $\rho_\alpha$ observed across sampled unitaries $\{U_k\}$. The solid curve shows the numerical minimum estimated from $10^5$ Haar-random sampled unitaries, while the dashed curve is the lower bound set by the spectral distance $D_{\mathrm{spec}}$ [Eq.~\eqref{eq:def_Dspec}].
(b) Metrological capability for a representative known parameter encoding Hamiltonian $\hat J_y=(\hat a_H^\dagger \hat a_V - \hat a_V^\dagger \hat a_H)/(2i)$.
Markers denote the maximum QFI for the $\hat J_y$ encoding found across $\{U_k\}$, derived from the experimentally reconstructed density matrices.
The solid curve plots the theoretical maximal QFI bound for two-photon two-mode states [Eq.~\eqref{eq:FQ_opt}]. 
Error bars are simulated via Poissonian counting statistics.}
\end{figure}

Having verified the conservation laws, we further illustrate the operational utility of these invariants.
Fig.~\hyperref[fig:fig4]{4(a)} characterizes state reachability in LONs. The smallest experimentally observed trace distances to the target state are bounded from below by the spectral distance $D_{\mathrm{spec}}$, in agreement with the theoretical prediction in Eq.~\eqref{eq:bound_Dspec}.
Fig.~\hyperref[fig:fig4]{4(b)} illustrates the metrological capability attainable under LON preprocessing.
The maximal QFI for a known encoding, evaluated from the reconstructed states, follows the theoretical bound in Eq.~\eqref{eq:FQ_opt}.


\textit{Conclusion.---}In summary, we have unveiled a fine-grained symmetry structure inherent to linear optical dynamics, which enforces the conservation of hierarchical invariants.
This structure is compatible with photon-number conservation and provides a symmetry-based framework for general bosonic states.
These hierarchical invariants have direct operational utility. 
In the examples we demonstrate, they provide quantitative tools for characterizing the state reachability in LONs and the attainable precision in quantum metrology.
Looking ahead, these invariants may provide a promising toolkit for analyzing the expressivity of quantum neural networks~\cite{ono_demonstration_2023, gan_fock_2022}, characterizing multiphoton entanglement~\cite{somma_quantum_2005, bastin_operational_2009}, and extending resource theories of optical coherence~\cite{baumgratz_quantifying_2014, tan_quantifying_2017}.

\begin{acknowledgments}
\textit{Acknowledgments.---}This work was supported by National Natural Science Foundation   of   China (Grants No. U24A2017, No. 12347104 and No. 12461160276), the National Key Research and Development Program of China (Grants No. 2023YFC2205802), Natural Science Foundation of Jiangsu Province (Grants No. BK20243060 and No. BK20233001).
\end{acknowledgments}

\bibliographystyle{apsrev4-2} 
\bibliography{ref} 

\begin{thebibliography}{59}%
\makeatletter
\providecommand \@ifxundefined [1]{%
 \@ifx{#1\undefined}
}%
\providecommand \@ifnum [1]{%
 \ifnum #1\expandafter \@firstoftwo
 \else \expandafter \@secondoftwo
 \fi
}%
\providecommand \@ifx [1]{%
 \ifx #1\expandafter \@firstoftwo
 \else \expandafter \@secondoftwo
 \fi
}%
\providecommand \natexlab [1]{#1}%
\providecommand \enquote  [1]{``#1''}%
\providecommand \bibnamefont  [1]{#1}%
\providecommand \bibfnamefont [1]{#1}%
\providecommand \citenamefont [1]{#1}%
\providecommand \href@noop [0]{\@secondoftwo}%
\providecommand \href [0]{\begingroup \@sanitize@url \@href}%
\providecommand \@href[1]{\@@startlink{#1}\@@href}%
\providecommand \@@href[1]{\endgroup#1\@@endlink}%
\providecommand \@sanitize@url [0]{\catcode `\\12\catcode `\$12\catcode
  `\&12\catcode `\#12\catcode `\^12\catcode `\_12\catcode `\%12\relax}%
\providecommand \@@startlink[1]{}%
\providecommand \@@endlink[0]{}%
\providecommand \url  [0]{\begingroup\@sanitize@url \@url }%
\providecommand \@url [1]{\endgroup\@href {#1}{\urlprefix }}%
\providecommand \urlprefix  [0]{URL }%
\providecommand \Eprint [0]{\href }%
\providecommand \doibase [0]{https://doi.org/}%
\providecommand \selectlanguage [0]{\@gobble}%
\providecommand \bibinfo  [0]{\@secondoftwo}%
\providecommand \bibfield  [0]{\@secondoftwo}%
\providecommand \translation [1]{[#1]}%
\providecommand \BibitemOpen [0]{}%
\providecommand \bibitemStop [0]{}%
\providecommand \bibitemNoStop [0]{.\EOS\space}%
\providecommand \EOS [0]{\spacefactor3000\relax}%
\providecommand \BibitemShut  [1]{\csname bibitem#1\endcsname}%
\let\auto@bib@innerbib\@empty
\bibitem [{\citenamefont {Senellart}\ \emph {et~al.}(2017)\citenamefont
  {Senellart}, \citenamefont {Solomon},\ and\ \citenamefont
  {White}}]{senellart_highperformance_2017}%
  \BibitemOpen
  \bibfield  {author} {\bibinfo {author} {\bibfnamefont {P.}~\bibnamefont
  {Senellart}}, \bibinfo {author} {\bibfnamefont {G.}~\bibnamefont {Solomon}},\
  and\ \bibinfo {author} {\bibfnamefont {A.}~\bibnamefont {White}},\ }\href
  {https://doi.org/10.1038/nnano.2017.218} {\bibfield  {journal} {\bibinfo
  {journal} {Nature Nanotech}\ }\textbf {\bibinfo {volume} {12}},\ \bibinfo
  {pages} {1026} (\bibinfo {year} {2017})}\BibitemShut {NoStop}%
\bibitem [{\citenamefont {Wang}\ \emph {et~al.}(2020)\citenamefont {Wang},
  \citenamefont {Sciarrino}, \citenamefont {Laing},\ and\ \citenamefont
  {Thompson}}]{wang_integrated_2020}%
  \BibitemOpen
  \bibfield  {author} {\bibinfo {author} {\bibfnamefont {J.}~\bibnamefont
  {Wang}}, \bibinfo {author} {\bibfnamefont {F.}~\bibnamefont {Sciarrino}},
  \bibinfo {author} {\bibfnamefont {A.}~\bibnamefont {Laing}},\ and\ \bibinfo
  {author} {\bibfnamefont {M.~G.}\ \bibnamefont {Thompson}},\ }\href
  {https://doi.org/10.1038/s41566-019-0532-1} {\bibfield  {journal} {\bibinfo
  {journal} {Nat. Photonics}\ }\textbf {\bibinfo {volume} {14}},\ \bibinfo
  {pages} {273} (\bibinfo {year} {2020})}\BibitemShut {NoStop}%
\bibitem [{\citenamefont {Erhard}\ \emph {et~al.}(2020)\citenamefont {Erhard},
  \citenamefont {Krenn},\ and\ \citenamefont
  {Zeilinger}}]{erhard_advances_2020}%
  \BibitemOpen
  \bibfield  {author} {\bibinfo {author} {\bibfnamefont {M.}~\bibnamefont
  {Erhard}}, \bibinfo {author} {\bibfnamefont {M.}~\bibnamefont {Krenn}},\ and\
  \bibinfo {author} {\bibfnamefont {A.}~\bibnamefont {Zeilinger}},\ }\href
  {https://doi.org/10.1038/s42254-020-0193-5} {\bibfield  {journal} {\bibinfo
  {journal} {Nat Rev Phys}\ }\textbf {\bibinfo {volume} {2}},\ \bibinfo {pages}
  {365} (\bibinfo {year} {2020})}\BibitemShut {NoStop}%
\bibitem [{\citenamefont {Wehner}\ \emph {et~al.}(2018)\citenamefont {Wehner},
  \citenamefont {Elkouss},\ and\ \citenamefont {Hanson}}]{wehner_quantum_2018}%
  \BibitemOpen
  \bibfield  {author} {\bibinfo {author} {\bibfnamefont {S.}~\bibnamefont
  {Wehner}}, \bibinfo {author} {\bibfnamefont {D.}~\bibnamefont {Elkouss}},\
  and\ \bibinfo {author} {\bibfnamefont {R.}~\bibnamefont {Hanson}},\ }\href
  {https://doi.org/10.1126/science.aam9288} {\bibfield  {journal} {\bibinfo
  {journal} {Science}\ }\textbf {\bibinfo {volume} {362}},\ \bibinfo {pages}
  {eaam9288} (\bibinfo {year} {2018})}\BibitemShut {NoStop}%
\bibitem [{\citenamefont {Xu}\ \emph {et~al.}(2020)\citenamefont {Xu},
  \citenamefont {Ma}, \citenamefont {Zhang}, \citenamefont {Lo},\ and\
  \citenamefont {Pan}}]{xu_secure_2020}%
  \BibitemOpen
  \bibfield  {author} {\bibinfo {author} {\bibfnamefont {F.}~\bibnamefont
  {Xu}}, \bibinfo {author} {\bibfnamefont {X.}~\bibnamefont {Ma}}, \bibinfo
  {author} {\bibfnamefont {Q.}~\bibnamefont {Zhang}}, \bibinfo {author}
  {\bibfnamefont {H.-K.}\ \bibnamefont {Lo}},\ and\ \bibinfo {author}
  {\bibfnamefont {J.-W.}\ \bibnamefont {Pan}},\ }\href
  {https://doi.org/10.1103/RevModPhys.92.025002} {\bibfield  {journal}
  {\bibinfo  {journal} {Rev. Mod. Phys.}\ }\textbf {\bibinfo {volume} {92}},\
  \bibinfo {pages} {025002} (\bibinfo {year} {2020})}\BibitemShut {NoStop}%
\bibitem [{\citenamefont {Knill}\ \emph {et~al.}(2001)\citenamefont {Knill},
  \citenamefont {Laflamme},\ and\ \citenamefont {Milburn}}]{knill_scheme_2001}%
  \BibitemOpen
  \bibfield  {author} {\bibinfo {author} {\bibfnamefont {E.}~\bibnamefont
  {Knill}}, \bibinfo {author} {\bibfnamefont {R.}~\bibnamefont {Laflamme}},\
  and\ \bibinfo {author} {\bibfnamefont {G.~J.}\ \bibnamefont {Milburn}},\
  }\href {https://doi.org/10.1038/35051009} {\bibfield  {journal} {\bibinfo
  {journal} {Nature}\ }\textbf {\bibinfo {volume} {409}},\ \bibinfo {pages}
  {46} (\bibinfo {year} {2001})}\BibitemShut {NoStop}%
\bibitem [{\citenamefont {Peruzzo}\ \emph {et~al.}(2014)\citenamefont
  {Peruzzo}, \citenamefont {McClean}, \citenamefont {Shadbolt}, \citenamefont
  {Yung}, \citenamefont {Zhou}, \citenamefont {Love}, \citenamefont
  {{Aspuru-Guzik}},\ and\ \citenamefont {O'Brien}}]{peruzzo_variational_2014}%
  \BibitemOpen
  \bibfield  {author} {\bibinfo {author} {\bibfnamefont {A.}~\bibnamefont
  {Peruzzo}}, \bibinfo {author} {\bibfnamefont {J.}~\bibnamefont {McClean}},
  \bibinfo {author} {\bibfnamefont {P.}~\bibnamefont {Shadbolt}}, \bibinfo
  {author} {\bibfnamefont {M.-H.}\ \bibnamefont {Yung}}, \bibinfo {author}
  {\bibfnamefont {X.-Q.}\ \bibnamefont {Zhou}}, \bibinfo {author}
  {\bibfnamefont {P.~J.}\ \bibnamefont {Love}}, \bibinfo {author}
  {\bibfnamefont {A.}~\bibnamefont {{Aspuru-Guzik}}},\ and\ \bibinfo {author}
  {\bibfnamefont {J.~L.}\ \bibnamefont {O'Brien}},\ }\href
  {https://doi.org/10.1038/ncomms5213} {\bibfield  {journal} {\bibinfo
  {journal} {Nat Commun}\ }\textbf {\bibinfo {volume} {5}},\ \bibinfo {pages}
  {4213} (\bibinfo {year} {2014})}\BibitemShut {NoStop}%
\bibitem [{\citenamefont {Madsen}\ \emph {et~al.}(2022)\citenamefont {Madsen},
  \citenamefont {Laudenbach}, \citenamefont {Askarani}, \citenamefont
  {Rortais}, \citenamefont {Vincent}, \citenamefont {Bulmer}, \citenamefont
  {Miatto}, \citenamefont {Neuhaus}, \citenamefont {Helt}, \citenamefont
  {Collins}, \citenamefont {Lita}, \citenamefont {Gerrits}, \citenamefont
  {Nam}, \citenamefont {Vaidya}, \citenamefont {Menotti}, \citenamefont
  {Dhand}, \citenamefont {Vernon}, \citenamefont {Quesada},\ and\ \citenamefont
  {Lavoie}}]{madsen_quantum_2022}%
  \BibitemOpen
  \bibfield  {author} {\bibinfo {author} {\bibfnamefont {L.~S.}\ \bibnamefont
  {Madsen}}, \bibinfo {author} {\bibfnamefont {F.}~\bibnamefont {Laudenbach}},
  \bibinfo {author} {\bibfnamefont {M.~F.}\ \bibnamefont {Askarani}}, \bibinfo
  {author} {\bibfnamefont {F.}~\bibnamefont {Rortais}}, \bibinfo {author}
  {\bibfnamefont {T.}~\bibnamefont {Vincent}}, \bibinfo {author} {\bibfnamefont
  {J.~F.~F.}\ \bibnamefont {Bulmer}}, \bibinfo {author} {\bibfnamefont {F.~M.}\
  \bibnamefont {Miatto}}, \bibinfo {author} {\bibfnamefont {L.}~\bibnamefont
  {Neuhaus}}, \bibinfo {author} {\bibfnamefont {L.~G.}\ \bibnamefont {Helt}},
  \bibinfo {author} {\bibfnamefont {M.~J.}\ \bibnamefont {Collins}}, \bibinfo
  {author} {\bibfnamefont {A.~E.}\ \bibnamefont {Lita}}, \bibinfo {author}
  {\bibfnamefont {T.}~\bibnamefont {Gerrits}}, \bibinfo {author} {\bibfnamefont
  {S.~W.}\ \bibnamefont {Nam}}, \bibinfo {author} {\bibfnamefont {V.~D.}\
  \bibnamefont {Vaidya}}, \bibinfo {author} {\bibfnamefont {M.}~\bibnamefont
  {Menotti}}, \bibinfo {author} {\bibfnamefont {I.}~\bibnamefont {Dhand}},
  \bibinfo {author} {\bibfnamefont {Z.}~\bibnamefont {Vernon}}, \bibinfo
  {author} {\bibfnamefont {N.}~\bibnamefont {Quesada}},\ and\ \bibinfo {author}
  {\bibfnamefont {J.}~\bibnamefont {Lavoie}},\ }\href
  {https://doi.org/10.1038/s41586-022-04725-x} {\bibfield  {journal} {\bibinfo
  {journal} {Nature}\ }\textbf {\bibinfo {volume} {606}},\ \bibinfo {pages}
  {75} (\bibinfo {year} {2022})}\BibitemShut {NoStop}%
\bibitem [{\citenamefont {Killoran}\ \emph {et~al.}(2014)\citenamefont
  {Killoran}, \citenamefont {Cramer},\ and\ \citenamefont
  {Plenio}}]{killoran_extracting_2014}%
  \BibitemOpen
  \bibfield  {author} {\bibinfo {author} {\bibfnamefont {N.}~\bibnamefont
  {Killoran}}, \bibinfo {author} {\bibfnamefont {M.}~\bibnamefont {Cramer}},\
  and\ \bibinfo {author} {\bibfnamefont {M.~B.}\ \bibnamefont {Plenio}},\
  }\href {https://doi.org/10.1103/PhysRevLett.112.150501} {\bibfield  {journal}
  {\bibinfo  {journal} {Phys. Rev. Lett.}\ }\textbf {\bibinfo {volume} {112}},\
  \bibinfo {pages} {150501} (\bibinfo {year} {2014})}\BibitemShut {NoStop}%
\bibitem [{\citenamefont {Pan}\ \emph {et~al.}(2012)\citenamefont {Pan},
  \citenamefont {Chen}, \citenamefont {Lu}, \citenamefont {Weinfurter},
  \citenamefont {Zeilinger},\ and\ \citenamefont
  {{\.Z}ukowski}}]{pan_multiphoton_2012}%
  \BibitemOpen
  \bibfield  {author} {\bibinfo {author} {\bibfnamefont {J.-W.}\ \bibnamefont
  {Pan}}, \bibinfo {author} {\bibfnamefont {Z.-B.}\ \bibnamefont {Chen}},
  \bibinfo {author} {\bibfnamefont {C.-Y.}\ \bibnamefont {Lu}}, \bibinfo
  {author} {\bibfnamefont {H.}~\bibnamefont {Weinfurter}}, \bibinfo {author}
  {\bibfnamefont {A.}~\bibnamefont {Zeilinger}},\ and\ \bibinfo {author}
  {\bibfnamefont {M.}~\bibnamefont {{\.Z}ukowski}},\ }\href
  {https://doi.org/10.1103/RevModPhys.84.777} {\bibfield  {journal} {\bibinfo
  {journal} {Rev. Mod. Phys.}\ }\textbf {\bibinfo {volume} {84}},\ \bibinfo
  {pages} {777} (\bibinfo {year} {2012})}\BibitemShut {NoStop}%
\bibitem [{\citenamefont {Park}\ \emph {et~al.}(2023)\citenamefont {Park},
  \citenamefont {Matsumoto}, \citenamefont {Kiyohara}, \citenamefont {Hofmann},
  \citenamefont {Okamoto},\ and\ \citenamefont
  {Takeuchi}}]{park_realization_2023}%
  \BibitemOpen
  \bibfield  {author} {\bibinfo {author} {\bibfnamefont {G.}~\bibnamefont
  {Park}}, \bibinfo {author} {\bibfnamefont {I.}~\bibnamefont {Matsumoto}},
  \bibinfo {author} {\bibfnamefont {T.}~\bibnamefont {Kiyohara}}, \bibinfo
  {author} {\bibfnamefont {H.~F.}\ \bibnamefont {Hofmann}}, \bibinfo {author}
  {\bibfnamefont {R.}~\bibnamefont {Okamoto}},\ and\ \bibinfo {author}
  {\bibfnamefont {S.}~\bibnamefont {Takeuchi}},\ }\href
  {https://doi.org/10.1126/sciadv.adj8146} {\bibfield  {journal} {\bibinfo
  {journal} {Sci. Adv.}\ }\textbf {\bibinfo {volume} {9}},\ \bibinfo {pages}
  {eadj8146} (\bibinfo {year} {2023})}\BibitemShut {NoStop}%
\bibitem [{\citenamefont {Steinbrecher}\ \emph {et~al.}(2019)\citenamefont
  {Steinbrecher}, \citenamefont {Olson}, \citenamefont {Englund},\ and\
  \citenamefont {Carolan}}]{steinbrecher_quantum_2019}%
  \BibitemOpen
  \bibfield  {author} {\bibinfo {author} {\bibfnamefont {G.~R.}\ \bibnamefont
  {Steinbrecher}}, \bibinfo {author} {\bibfnamefont {J.~P.}\ \bibnamefont
  {Olson}}, \bibinfo {author} {\bibfnamefont {D.}~\bibnamefont {Englund}},\
  and\ \bibinfo {author} {\bibfnamefont {J.}~\bibnamefont {Carolan}},\ }\href
  {https://doi.org/10.1038/s41534-019-0174-7} {\bibfield  {journal} {\bibinfo
  {journal} {npj Quantum Inf}\ }\textbf {\bibinfo {volume} {5}},\ \bibinfo
  {pages} {60} (\bibinfo {year} {2019})}\BibitemShut {NoStop}%
\bibitem [{\citenamefont {Aaronson}\ and\ \citenamefont
  {Arkhipov}(2011)}]{aaronson_computational_2011}%
  \BibitemOpen
  \bibfield  {author} {\bibinfo {author} {\bibfnamefont {S.}~\bibnamefont
  {Aaronson}}\ and\ \bibinfo {author} {\bibfnamefont {A.}~\bibnamefont
  {Arkhipov}},\ }in\ \href {https://doi.org/10.1145/1993636.1993682} {\emph
  {\bibinfo {booktitle} {Proc. {{Forty-Third Annu}}. {{ACM Symp}}. {{Theory
  Comput}}.}}},\ \bibinfo {series and number} {{{STOC}} '11}\ (\bibinfo
  {publisher} {Association for Computing Machinery},\ \bibinfo {address} {New
  York, NY, USA},\ \bibinfo {year} {2011})\ pp.\ \bibinfo {pages}
  {333--342}\BibitemShut {NoStop}%
\bibitem [{\citenamefont {Zhong}\ \emph {et~al.}(2020)\citenamefont {Zhong},
  \citenamefont {Wang}, \citenamefont {Deng}, \citenamefont {Chen},
  \citenamefont {Peng}, \citenamefont {Luo}, \citenamefont {Qin}, \citenamefont
  {Wu}, \citenamefont {Ding}, \citenamefont {Hu}, \citenamefont {Hu},
  \citenamefont {Yang}, \citenamefont {Zhang}, \citenamefont {Li},
  \citenamefont {Li}, \citenamefont {Jiang}, \citenamefont {Gan}, \citenamefont
  {Yang}, \citenamefont {You}, \citenamefont {Wang}, \citenamefont {Li},
  \citenamefont {Liu}, \citenamefont {Lu},\ and\ \citenamefont
  {Pan}}]{zhong_quantum_2020}%
  \BibitemOpen
  \bibfield  {author} {\bibinfo {author} {\bibfnamefont {H.-S.}\ \bibnamefont
  {Zhong}}, \bibinfo {author} {\bibfnamefont {H.}~\bibnamefont {Wang}},
  \bibinfo {author} {\bibfnamefont {Y.-H.}\ \bibnamefont {Deng}}, \bibinfo
  {author} {\bibfnamefont {M.-C.}\ \bibnamefont {Chen}}, \bibinfo {author}
  {\bibfnamefont {L.-C.}\ \bibnamefont {Peng}}, \bibinfo {author}
  {\bibfnamefont {Y.-H.}\ \bibnamefont {Luo}}, \bibinfo {author} {\bibfnamefont
  {J.}~\bibnamefont {Qin}}, \bibinfo {author} {\bibfnamefont {D.}~\bibnamefont
  {Wu}}, \bibinfo {author} {\bibfnamefont {X.}~\bibnamefont {Ding}}, \bibinfo
  {author} {\bibfnamefont {Y.}~\bibnamefont {Hu}}, \bibinfo {author}
  {\bibfnamefont {P.}~\bibnamefont {Hu}}, \bibinfo {author} {\bibfnamefont
  {X.-Y.}\ \bibnamefont {Yang}}, \bibinfo {author} {\bibfnamefont {W.-J.}\
  \bibnamefont {Zhang}}, \bibinfo {author} {\bibfnamefont {H.}~\bibnamefont
  {Li}}, \bibinfo {author} {\bibfnamefont {Y.}~\bibnamefont {Li}}, \bibinfo
  {author} {\bibfnamefont {X.}~\bibnamefont {Jiang}}, \bibinfo {author}
  {\bibfnamefont {L.}~\bibnamefont {Gan}}, \bibinfo {author} {\bibfnamefont
  {G.}~\bibnamefont {Yang}}, \bibinfo {author} {\bibfnamefont {L.}~\bibnamefont
  {You}}, \bibinfo {author} {\bibfnamefont {Z.}~\bibnamefont {Wang}}, \bibinfo
  {author} {\bibfnamefont {L.}~\bibnamefont {Li}}, \bibinfo {author}
  {\bibfnamefont {N.-L.}\ \bibnamefont {Liu}}, \bibinfo {author} {\bibfnamefont
  {C.-Y.}\ \bibnamefont {Lu}},\ and\ \bibinfo {author} {\bibfnamefont {J.-W.}\
  \bibnamefont {Pan}},\ }\href {https://doi.org/10.1126/science.abe8770}
  {\bibfield  {journal} {\bibinfo  {journal} {Science}\ }\textbf {\bibinfo
  {volume} {370}},\ \bibinfo {pages} {1460} (\bibinfo {year}
  {2020})}\BibitemShut {NoStop}%
\bibitem [{\citenamefont {Nielsen}(2004)}]{nielsen_optical_2004}%
  \BibitemOpen
  \bibfield  {author} {\bibinfo {author} {\bibfnamefont {M.~A.}\ \bibnamefont
  {Nielsen}},\ }\bibfield  {journal} {\bibinfo  {journal} {Phys. Rev. Lett.}\
  }\textbf {\bibinfo {volume} {93}},\ \href
  {https://doi.org/10.1103/PhysRevLett.93.040503}
  {10.1103/PhysRevLett.93.040503} (\bibinfo {year} {2004})\BibitemShut
  {NoStop}%
\bibitem [{\citenamefont {Bartolucci}\ \emph {et~al.}(2023)\citenamefont
  {Bartolucci}, \citenamefont {Birchall}, \citenamefont {Bomb{\'i}n},
  \citenamefont {Cable}, \citenamefont {Dawson}, \citenamefont
  {{Gimeno-Segovia}}, \citenamefont {Johnston}, \citenamefont {Kieling},
  \citenamefont {Nickerson}, \citenamefont {Pant}, \citenamefont {Pastawski},
  \citenamefont {Rudolph},\ and\ \citenamefont
  {Sparrow}}]{bartolucci_fusionbased_2023}%
  \BibitemOpen
  \bibfield  {author} {\bibinfo {author} {\bibfnamefont {S.}~\bibnamefont
  {Bartolucci}}, \bibinfo {author} {\bibfnamefont {P.}~\bibnamefont
  {Birchall}}, \bibinfo {author} {\bibfnamefont {H.}~\bibnamefont
  {Bomb{\'i}n}}, \bibinfo {author} {\bibfnamefont {H.}~\bibnamefont {Cable}},
  \bibinfo {author} {\bibfnamefont {C.}~\bibnamefont {Dawson}}, \bibinfo
  {author} {\bibfnamefont {M.}~\bibnamefont {{Gimeno-Segovia}}}, \bibinfo
  {author} {\bibfnamefont {E.}~\bibnamefont {Johnston}}, \bibinfo {author}
  {\bibfnamefont {K.}~\bibnamefont {Kieling}}, \bibinfo {author} {\bibfnamefont
  {N.}~\bibnamefont {Nickerson}}, \bibinfo {author} {\bibfnamefont
  {M.}~\bibnamefont {Pant}}, \bibinfo {author} {\bibfnamefont {F.}~\bibnamefont
  {Pastawski}}, \bibinfo {author} {\bibfnamefont {T.}~\bibnamefont {Rudolph}},\
  and\ \bibinfo {author} {\bibfnamefont {C.}~\bibnamefont {Sparrow}},\ }\href
  {https://doi.org/10.1038/s41467-023-36493-1} {\bibfield  {journal} {\bibinfo
  {journal} {Nat Commun}\ }\textbf {\bibinfo {volume} {14}},\ \bibinfo {pages}
  {912} (\bibinfo {year} {2023})}\BibitemShut {NoStop}%
\bibitem [{\citenamefont {Hoch}\ \emph {et~al.}(2025)\citenamefont {Hoch},
  \citenamefont {Caruccio}, \citenamefont {Rodari}, \citenamefont
  {Francalanci}, \citenamefont {Suprano}, \citenamefont {Giordani},
  \citenamefont {Carvacho}, \citenamefont {Spagnolo}, \citenamefont {Koudia},
  \citenamefont {Proietti}, \citenamefont {Liorni}, \citenamefont {Cerocchi},
  \citenamefont {Albiero}, \citenamefont {Di~Giano}, \citenamefont {Gardina},
  \citenamefont {Ceccarelli}, \citenamefont {Corrielli}, \citenamefont
  {Chabaud}, \citenamefont {Osellame}, \citenamefont {Dispenza},\ and\
  \citenamefont {Sciarrino}}]{hoch_quantum_2025}%
  \BibitemOpen
  \bibfield  {author} {\bibinfo {author} {\bibfnamefont {F.}~\bibnamefont
  {Hoch}}, \bibinfo {author} {\bibfnamefont {E.}~\bibnamefont {Caruccio}},
  \bibinfo {author} {\bibfnamefont {G.}~\bibnamefont {Rodari}}, \bibinfo
  {author} {\bibfnamefont {T.}~\bibnamefont {Francalanci}}, \bibinfo {author}
  {\bibfnamefont {A.}~\bibnamefont {Suprano}}, \bibinfo {author} {\bibfnamefont
  {T.}~\bibnamefont {Giordani}}, \bibinfo {author} {\bibfnamefont
  {G.}~\bibnamefont {Carvacho}}, \bibinfo {author} {\bibfnamefont
  {N.}~\bibnamefont {Spagnolo}}, \bibinfo {author} {\bibfnamefont
  {S.}~\bibnamefont {Koudia}}, \bibinfo {author} {\bibfnamefont
  {M.}~\bibnamefont {Proietti}}, \bibinfo {author} {\bibfnamefont
  {C.}~\bibnamefont {Liorni}}, \bibinfo {author} {\bibfnamefont
  {F.}~\bibnamefont {Cerocchi}}, \bibinfo {author} {\bibfnamefont
  {R.}~\bibnamefont {Albiero}}, \bibinfo {author} {\bibfnamefont
  {N.}~\bibnamefont {Di~Giano}}, \bibinfo {author} {\bibfnamefont
  {M.}~\bibnamefont {Gardina}}, \bibinfo {author} {\bibfnamefont
  {F.}~\bibnamefont {Ceccarelli}}, \bibinfo {author} {\bibfnamefont
  {G.}~\bibnamefont {Corrielli}}, \bibinfo {author} {\bibfnamefont
  {U.}~\bibnamefont {Chabaud}}, \bibinfo {author} {\bibfnamefont
  {R.}~\bibnamefont {Osellame}}, \bibinfo {author} {\bibfnamefont
  {M.}~\bibnamefont {Dispenza}},\ and\ \bibinfo {author} {\bibfnamefont
  {F.}~\bibnamefont {Sciarrino}},\ }\href
  {https://doi.org/10.1038/s41467-025-55877-z} {\bibfield  {journal} {\bibinfo
  {journal} {Nature Communications}\ }\textbf {\bibinfo {volume} {16}},\
  \bibinfo {pages} {902} (\bibinfo {year} {2025})}\BibitemShut {NoStop}%
\bibitem [{\citenamefont {Boto}\ \emph {et~al.}(2000)\citenamefont {Boto},
  \citenamefont {Kok}, \citenamefont {Abrams}, \citenamefont {Braunstein},
  \citenamefont {Williams},\ and\ \citenamefont {Dowling}}]{boto_quantum_2000}%
  \BibitemOpen
  \bibfield  {author} {\bibinfo {author} {\bibfnamefont {A.~N.}\ \bibnamefont
  {Boto}}, \bibinfo {author} {\bibfnamefont {P.}~\bibnamefont {Kok}}, \bibinfo
  {author} {\bibfnamefont {D.~S.}\ \bibnamefont {Abrams}}, \bibinfo {author}
  {\bibfnamefont {S.~L.}\ \bibnamefont {Braunstein}}, \bibinfo {author}
  {\bibfnamefont {C.~P.}\ \bibnamefont {Williams}},\ and\ \bibinfo {author}
  {\bibfnamefont {J.~P.}\ \bibnamefont {Dowling}},\ }\href
  {https://doi.org/10.1103/PhysRevLett.85.2733} {\bibfield  {journal} {\bibinfo
   {journal} {Phys. Rev. Lett.}\ }\textbf {\bibinfo {volume} {85}},\ \bibinfo
  {pages} {2733} (\bibinfo {year} {2000})}\BibitemShut {NoStop}%
\bibitem [{\citenamefont {Nagata}\ \emph {et~al.}(2007)\citenamefont {Nagata},
  \citenamefont {Okamoto}, \citenamefont {O'Brien}, \citenamefont {Sasaki},\
  and\ \citenamefont {Takeuchi}}]{nagata_beating_2007}%
  \BibitemOpen
  \bibfield  {author} {\bibinfo {author} {\bibfnamefont {T.}~\bibnamefont
  {Nagata}}, \bibinfo {author} {\bibfnamefont {R.}~\bibnamefont {Okamoto}},
  \bibinfo {author} {\bibfnamefont {J.~L.}\ \bibnamefont {O'Brien}}, \bibinfo
  {author} {\bibfnamefont {K.}~\bibnamefont {Sasaki}},\ and\ \bibinfo {author}
  {\bibfnamefont {S.}~\bibnamefont {Takeuchi}},\ }\href
  {https://doi.org/10.1126/science.1138007} {\bibfield  {journal} {\bibinfo
  {journal} {Science}\ }\textbf {\bibinfo {volume} {316}},\ \bibinfo {pages}
  {726} (\bibinfo {year} {2007})}\BibitemShut {NoStop}%
\bibitem [{\citenamefont {Reck}\ \emph {et~al.}(1994)\citenamefont {Reck},
  \citenamefont {Zeilinger}, \citenamefont {Bernstein},\ and\ \citenamefont
  {Bertani}}]{reck_experimental_1994}%
  \BibitemOpen
  \bibfield  {author} {\bibinfo {author} {\bibfnamefont {M.}~\bibnamefont
  {Reck}}, \bibinfo {author} {\bibfnamefont {A.}~\bibnamefont {Zeilinger}},
  \bibinfo {author} {\bibfnamefont {H.~J.}\ \bibnamefont {Bernstein}},\ and\
  \bibinfo {author} {\bibfnamefont {P.}~\bibnamefont {Bertani}},\ }\href
  {https://doi.org/10.1103/PhysRevLett.73.58} {\bibfield  {journal} {\bibinfo
  {journal} {Phys. Rev. Lett.}\ }\textbf {\bibinfo {volume} {73}},\ \bibinfo
  {pages} {58} (\bibinfo {year} {1994})}\BibitemShut {NoStop}%
\bibitem [{\citenamefont {Clements}\ \emph {et~al.}(2016)\citenamefont
  {Clements}, \citenamefont {Humphreys}, \citenamefont {Metcalf}, \citenamefont
  {Kolthammer},\ and\ \citenamefont {Walsmley}}]{clements_optimal_2016}%
  \BibitemOpen
  \bibfield  {author} {\bibinfo {author} {\bibfnamefont {W.~R.}\ \bibnamefont
  {Clements}}, \bibinfo {author} {\bibfnamefont {P.~C.}\ \bibnamefont
  {Humphreys}}, \bibinfo {author} {\bibfnamefont {B.~J.}\ \bibnamefont
  {Metcalf}}, \bibinfo {author} {\bibfnamefont {W.~S.}\ \bibnamefont
  {Kolthammer}},\ and\ \bibinfo {author} {\bibfnamefont {I.~A.}\ \bibnamefont
  {Walsmley}},\ }\href {https://doi.org/10.1364/OPTICA.3.001460} {\bibfield
  {journal} {\bibinfo  {journal} {Optica}\ }\textbf {\bibinfo {volume} {3}},\
  \bibinfo {pages} {1460} (\bibinfo {year} {2016})}\BibitemShut {NoStop}%
\bibitem [{\citenamefont {Oszmaniec}\ and\ \citenamefont
  {Zimbor{\'a}s}(2017)}]{oszmaniec_universal_2017}%
  \BibitemOpen
  \bibfield  {author} {\bibinfo {author} {\bibfnamefont {M.}~\bibnamefont
  {Oszmaniec}}\ and\ \bibinfo {author} {\bibfnamefont {Z.}~\bibnamefont
  {Zimbor{\'a}s}},\ }\href {https://doi.org/10.1103/PhysRevLett.119.220502}
  {\bibfield  {journal} {\bibinfo  {journal} {Phys. Rev. Lett.}\ }\textbf
  {\bibinfo {volume} {119}},\ \bibinfo {pages} {220502} (\bibinfo {year}
  {2017})}\BibitemShut {NoStop}%
\bibitem [{\citenamefont {{Garcia-Escartin}}\ \emph
  {et~al.}(2019{\natexlab{a}})\citenamefont {{Garcia-Escartin}}, \citenamefont
  {Gimeno},\ and\ \citenamefont
  {{Moyano-Fern{\'a}ndez}}}]{garcia-escartin_method_2019}%
  \BibitemOpen
  \bibfield  {author} {\bibinfo {author} {\bibfnamefont {J.~C.}\ \bibnamefont
  {{Garcia-Escartin}}}, \bibinfo {author} {\bibfnamefont {V.}~\bibnamefont
  {Gimeno}},\ and\ \bibinfo {author} {\bibfnamefont {J.~J.}\ \bibnamefont
  {{Moyano-Fern{\'a}ndez}}},\ }\href
  {https://doi.org/10.1103/PhysRevA.100.022301} {\bibfield  {journal} {\bibinfo
   {journal} {Phys. Rev. A}\ }\textbf {\bibinfo {volume} {100}},\ \bibinfo
  {pages} {022301} (\bibinfo {year} {2019}{\natexlab{a}})}\BibitemShut
  {NoStop}%
\bibitem [{\citenamefont {Wang}\ \emph {et~al.}(2025)\citenamefont {Wang},
  \citenamefont {Yin}, \citenamefont {Haug}, \citenamefont {Pentangelo},
  \citenamefont {Piacentini}, \citenamefont {Crespi}, \citenamefont
  {Ceccarelli}, \citenamefont {Osellame},\ and\ \citenamefont
  {Walther}}]{wangScalableAdvantageMultiphoton2025}%
  \BibitemOpen
  \bibfield  {author} {\bibinfo {author} {\bibfnamefont {Y.}~\bibnamefont
  {Wang}}, \bibinfo {author} {\bibfnamefont {Z.}~\bibnamefont {Yin}}, \bibinfo
  {author} {\bibfnamefont {T.}~\bibnamefont {Haug}}, \bibinfo {author}
  {\bibfnamefont {C.}~\bibnamefont {Pentangelo}}, \bibinfo {author}
  {\bibfnamefont {S.}~\bibnamefont {Piacentini}}, \bibinfo {author}
  {\bibfnamefont {A.}~\bibnamefont {Crespi}}, \bibinfo {author} {\bibfnamefont
  {F.}~\bibnamefont {Ceccarelli}}, \bibinfo {author} {\bibfnamefont
  {R.}~\bibnamefont {Osellame}},\ and\ \bibinfo {author} {\bibfnamefont
  {P.}~\bibnamefont {Walther}},\ }\href
  {https://doi.org/10.48550/arXiv.2511.21951} {\bibinfo {title} {A scalable
  advantage in multi-photon quantum machine learning}} (\bibinfo {year}
  {2025}),\ \Eprint {https://arxiv.org/abs/2511.21951} {arXiv:2511.21951
  [quant-ph]} \BibitemShut {NoStop}%
\bibitem [{\citenamefont {Ewert}\ and\ \citenamefont {{van
  Loock}}(2014)}]{ewert_$3_2014}%
  \BibitemOpen
  \bibfield  {author} {\bibinfo {author} {\bibfnamefont {F.}~\bibnamefont
  {Ewert}}\ and\ \bibinfo {author} {\bibfnamefont {P.}~\bibnamefont {{van
  Loock}}},\ }\href {https://doi.org/10.1103/PhysRevLett.113.140403} {\bibfield
   {journal} {\bibinfo  {journal} {Phys. Rev. Lett.}\ }\textbf {\bibinfo
  {volume} {113}},\ \bibinfo {pages} {140403} (\bibinfo {year}
  {2014})}\BibitemShut {NoStop}%
\bibitem [{\citenamefont {Migda{\l}}\ \emph {et~al.}(2014)\citenamefont
  {Migda{\l}}, \citenamefont {{Rodr{\'i}guez-Laguna}}, \citenamefont
  {Oszmaniec},\ and\ \citenamefont {Lewenstein}}]{migdal_multiphoton_2014}%
  \BibitemOpen
  \bibfield  {author} {\bibinfo {author} {\bibfnamefont {P.}~\bibnamefont
  {Migda{\l}}}, \bibinfo {author} {\bibfnamefont {J.}~\bibnamefont
  {{Rodr{\'i}guez-Laguna}}}, \bibinfo {author} {\bibfnamefont {M.}~\bibnamefont
  {Oszmaniec}},\ and\ \bibinfo {author} {\bibfnamefont {M.}~\bibnamefont
  {Lewenstein}},\ }\href {https://doi.org/10.1103/PhysRevA.89.062329}
  {\bibfield  {journal} {\bibinfo  {journal} {Phys. Rev. A}\ }\textbf {\bibinfo
  {volume} {89}},\ \bibinfo {pages} {062329} (\bibinfo {year}
  {2014})}\BibitemShut {NoStop}%
\bibitem [{\citenamefont {Parellada}\ \emph {et~al.}(2023)\citenamefont
  {Parellada}, \citenamefont {{Gimeno i Garcia}}, \citenamefont
  {{Moyano-Fern{\'a}ndez}},\ and\ \citenamefont
  {{Garcia-Escartin}}}]{parellada_nogo_2023}%
  \BibitemOpen
  \bibfield  {author} {\bibinfo {author} {\bibfnamefont {P.~V.}\ \bibnamefont
  {Parellada}}, \bibinfo {author} {\bibfnamefont {V.}~\bibnamefont {{Gimeno i
  Garcia}}}, \bibinfo {author} {\bibfnamefont {J.~J.}\ \bibnamefont
  {{Moyano-Fern{\'a}ndez}}},\ and\ \bibinfo {author} {\bibfnamefont {J.~C.}\
  \bibnamefont {{Garcia-Escartin}}},\ }\href
  {https://doi.org/10.1016/j.rinp.2023.107108} {\bibfield  {journal} {\bibinfo
  {journal} {Results in Physics}\ }\textbf {\bibinfo {volume} {54}},\ \bibinfo
  {pages} {107108} (\bibinfo {year} {2023})}\BibitemShut {NoStop}%
\bibitem [{\citenamefont {Parellada}\ \emph {et~al.}(2024)\citenamefont
  {Parellada}, \citenamefont {i~Garcia}, \citenamefont
  {{Moyano-Fern{\'a}ndez}},\ and\ \citenamefont
  {{Garcia-Escartin}}}]{parellada_lie_2024}%
  \BibitemOpen
  \bibfield  {author} {\bibinfo {author} {\bibfnamefont {P.~V.}\ \bibnamefont
  {Parellada}}, \bibinfo {author} {\bibfnamefont {V.~G.}\ \bibnamefont
  {i~Garcia}}, \bibinfo {author} {\bibfnamefont {J.~J.}\ \bibnamefont
  {{Moyano-Fern{\'a}ndez}}},\ and\ \bibinfo {author} {\bibfnamefont {J.~C.}\
  \bibnamefont {{Garcia-Escartin}}},\ }\href@noop {} {\bibinfo {title} {Lie
  algebraic invariants in quantum linear optics}} (\bibinfo {year} {2024}),\
  \Eprint {https://arxiv.org/abs/2409.12223} {arXiv:2409.12223} \BibitemShut
  {NoStop}%
\bibitem [{\citenamefont {Draux}\ \emph {et~al.}(2025)\citenamefont {Draux},
  \citenamefont {Perdrix}, \citenamefont {Jeandel},\ and\ \citenamefont
  {Mansfield}}]{drauxInvariantsLinearOptics2025}%
  \BibitemOpen
  \bibfield  {author} {\bibinfo {author} {\bibfnamefont {S.}~\bibnamefont
  {Draux}}, \bibinfo {author} {\bibfnamefont {S.}~\bibnamefont {Perdrix}},
  \bibinfo {author} {\bibfnamefont {E.}~\bibnamefont {Jeandel}},\ and\ \bibinfo
  {author} {\bibfnamefont {S.}~\bibnamefont {Mansfield}},\ }\href
  {https://doi.org/10.48550/arXiv.2509.02211} {\bibinfo {title} {Invariants in
  {{Linear Optics}}}} (\bibinfo {year} {2025}),\ \Eprint
  {https://arxiv.org/abs/2509.02211} {arXiv:2509.02211 [quant-ph]} \BibitemShut
  {NoStop}%
\bibitem [{\citenamefont {Rodari}\ \emph {et~al.}(2025)\citenamefont {Rodari},
  \citenamefont {Francalanci}, \citenamefont {Caruccio}, \citenamefont {Hoch},
  \citenamefont {Carvacho}, \citenamefont {Giordani}, \citenamefont {Spagnolo},
  \citenamefont {Albiero}, \citenamefont {Di~Giano}, \citenamefont
  {Ceccarelli}, \citenamefont {Corrielli}, \citenamefont {Crespi},
  \citenamefont {Osellame}, \citenamefont {Chabaud},\ and\ \citenamefont
  {Sciarrino}}]{rodari_observation_2025}%
  \BibitemOpen
  \bibfield  {author} {\bibinfo {author} {\bibfnamefont {G.}~\bibnamefont
  {Rodari}}, \bibinfo {author} {\bibfnamefont {T.}~\bibnamefont {Francalanci}},
  \bibinfo {author} {\bibfnamefont {E.}~\bibnamefont {Caruccio}}, \bibinfo
  {author} {\bibfnamefont {F.}~\bibnamefont {Hoch}}, \bibinfo {author}
  {\bibfnamefont {G.}~\bibnamefont {Carvacho}}, \bibinfo {author}
  {\bibfnamefont {T.}~\bibnamefont {Giordani}}, \bibinfo {author}
  {\bibfnamefont {N.}~\bibnamefont {Spagnolo}}, \bibinfo {author}
  {\bibfnamefont {R.}~\bibnamefont {Albiero}}, \bibinfo {author} {\bibfnamefont
  {N.}~\bibnamefont {Di~Giano}}, \bibinfo {author} {\bibfnamefont
  {F.}~\bibnamefont {Ceccarelli}}, \bibinfo {author} {\bibfnamefont
  {G.}~\bibnamefont {Corrielli}}, \bibinfo {author} {\bibfnamefont
  {A.}~\bibnamefont {Crespi}}, \bibinfo {author} {\bibfnamefont
  {R.}~\bibnamefont {Osellame}}, \bibinfo {author} {\bibfnamefont
  {U.}~\bibnamefont {Chabaud}},\ and\ \bibinfo {author} {\bibfnamefont
  {F.}~\bibnamefont {Sciarrino}},\ }\href {https://doi.org/10.1103/7961-hg2q}
  {\bibfield  {journal} {\bibinfo  {journal} {Phys. Rev. Research}\ }\textbf
  {\bibinfo {volume} {7}},\ \bibinfo {pages} {043325} (\bibinfo {year}
  {2025})}\BibitemShut {NoStop}%
\bibitem [{\citenamefont {Noether}(1971)}]{noether_invariant_1971}%
  \BibitemOpen
  \bibfield  {author} {\bibinfo {author} {\bibfnamefont {E.}~\bibnamefont
  {Noether}},\ }\href {https://doi.org/10.1080/00411457108231446} {\bibfield
  {journal} {\bibinfo  {journal} {Transp. Theory Stat. Phys.}\ }\textbf
  {\bibinfo {volume} {1}},\ \bibinfo {pages} {186} (\bibinfo {year}
  {1971})}\BibitemShut {NoStop}%
\bibitem [{\citenamefont {Carolan}\ \emph {et~al.}(2015)\citenamefont
  {Carolan}, \citenamefont {Harrold}, \citenamefont {Sparrow}, \citenamefont
  {{Mart{\'i}n-L{\'o}pez}}, \citenamefont {Russell}, \citenamefont
  {Silverstone}, \citenamefont {Shadbolt}, \citenamefont {Matsuda},
  \citenamefont {Oguma}, \citenamefont {Itoh}, \citenamefont {Marshall},
  \citenamefont {Thompson}, \citenamefont {Matthews}, \citenamefont
  {Hashimoto}, \citenamefont {O'Brien},\ and\ \citenamefont
  {Laing}}]{carolan_universal_2015}%
  \BibitemOpen
  \bibfield  {author} {\bibinfo {author} {\bibfnamefont {J.}~\bibnamefont
  {Carolan}}, \bibinfo {author} {\bibfnamefont {C.}~\bibnamefont {Harrold}},
  \bibinfo {author} {\bibfnamefont {C.}~\bibnamefont {Sparrow}}, \bibinfo
  {author} {\bibfnamefont {E.}~\bibnamefont {{Mart{\'i}n-L{\'o}pez}}}, \bibinfo
  {author} {\bibfnamefont {N.~J.}\ \bibnamefont {Russell}}, \bibinfo {author}
  {\bibfnamefont {J.~W.}\ \bibnamefont {Silverstone}}, \bibinfo {author}
  {\bibfnamefont {P.~J.}\ \bibnamefont {Shadbolt}}, \bibinfo {author}
  {\bibfnamefont {N.}~\bibnamefont {Matsuda}}, \bibinfo {author} {\bibfnamefont
  {M.}~\bibnamefont {Oguma}}, \bibinfo {author} {\bibfnamefont
  {M.}~\bibnamefont {Itoh}}, \bibinfo {author} {\bibfnamefont {G.~D.}\
  \bibnamefont {Marshall}}, \bibinfo {author} {\bibfnamefont {M.~G.}\
  \bibnamefont {Thompson}}, \bibinfo {author} {\bibfnamefont {J.~C.~F.}\
  \bibnamefont {Matthews}}, \bibinfo {author} {\bibfnamefont {T.}~\bibnamefont
  {Hashimoto}}, \bibinfo {author} {\bibfnamefont {J.~L.}\ \bibnamefont
  {O'Brien}},\ and\ \bibinfo {author} {\bibfnamefont {A.}~\bibnamefont
  {Laing}},\ }\href {https://doi.org/10.1126/science.aab3642} {\bibfield
  {journal} {\bibinfo  {journal} {Science}\ }\textbf {\bibinfo {volume}
  {349}},\ \bibinfo {pages} {711} (\bibinfo {year} {2015})}\BibitemShut
  {NoStop}%
\bibitem [{\citenamefont {Jordan}(1935)}]{jordan_zusammenhang_1935}%
  \BibitemOpen
  \bibfield  {author} {\bibinfo {author} {\bibfnamefont {P.}~\bibnamefont
  {Jordan}},\ }\href {https://doi.org/10.1007/BF01330618} {\bibfield  {journal}
  {\bibinfo  {journal} {Z. Physik}\ }\textbf {\bibinfo {volume} {94}},\
  \bibinfo {pages} {531} (\bibinfo {year} {1935})}\BibitemShut {NoStop}%
\bibitem [{\citenamefont {Schwinger}(1952)}]{schwinger_angular_1952}%
  \BibitemOpen
  \bibfield  {author} {\bibinfo {author} {\bibfnamefont {J.}~\bibnamefont
  {Schwinger}},\ }\href@noop {} {\emph {\bibinfo {title} {{{ON ANGULAR
  MOMENTUM}}}}},\ \bibinfo {type} {Tech. Rep.}\ \bibinfo {number} {NYO-3071}\
  (\bibinfo  {institution} {Harvard Univ., Cambridge, MA (United States);
  Nuclear Development Associates, Inc. (US)},\ \bibinfo {year}
  {1952})\BibitemShut {NoStop}%
\bibitem [{\citenamefont {{Garcia-Escartin}}\ \emph
  {et~al.}(2019{\natexlab{b}})\citenamefont {{Garcia-Escartin}}, \citenamefont
  {Gimeno},\ and\ \citenamefont
  {{Moyano-Fern{\'a}ndez}}}]{garcia-escartin_multiple_2019}%
  \BibitemOpen
  \bibfield  {author} {\bibinfo {author} {\bibfnamefont {J.~C.}\ \bibnamefont
  {{Garcia-Escartin}}}, \bibinfo {author} {\bibfnamefont {V.}~\bibnamefont
  {Gimeno}},\ and\ \bibinfo {author} {\bibfnamefont {J.~J.}\ \bibnamefont
  {{Moyano-Fern{\'a}ndez}}},\ }\href
  {https://doi.org/10.1016/j.optcom.2018.08.082} {\bibfield  {journal}
  {\bibinfo  {journal} {Optics Communications}\ }\textbf {\bibinfo {volume}
  {430}},\ \bibinfo {pages} {434} (\bibinfo {year}
  {2019}{\natexlab{b}})}\BibitemShut {NoStop}%
\bibitem [{\citenamefont {Glauber}(1963)}]{glauber_quantum_1963}%
  \BibitemOpen
  \bibfield  {author} {\bibinfo {author} {\bibfnamefont {R.~J.}\ \bibnamefont
  {Glauber}},\ }\href {https://doi.org/10.1103/PhysRev.130.2529} {\bibfield
  {journal} {\bibinfo  {journal} {Phys. Rev.}\ }\textbf {\bibinfo {volume}
  {130}},\ \bibinfo {pages} {2529} (\bibinfo {year} {1963})}\BibitemShut
  {NoStop}%
\bibitem [{Sup()}]{Supplementary}%
  \BibitemOpen
  \href@noop {} {}\bibinfo {note} {Supplementary Information for ``Unveiling
  Hierarchical Invariants in Multiphoton Linear Optics''}\BibitemShut {NoStop}%
\bibitem [{\citenamefont {Wilkens}\ \emph {et~al.}(2024)\citenamefont
  {Wilkens}, \citenamefont {Ioannou}, \citenamefont {Derbyshire}, \citenamefont
  {Eisert}, \citenamefont {Hangleiter}, \citenamefont {Roth},\ and\
  \citenamefont {Haferkamp}}]{wilkens_benchmarking_2024}%
  \BibitemOpen
  \bibfield  {author} {\bibinfo {author} {\bibfnamefont {J.}~\bibnamefont
  {Wilkens}}, \bibinfo {author} {\bibfnamefont {M.}~\bibnamefont {Ioannou}},
  \bibinfo {author} {\bibfnamefont {E.}~\bibnamefont {Derbyshire}}, \bibinfo
  {author} {\bibfnamefont {J.}~\bibnamefont {Eisert}}, \bibinfo {author}
  {\bibfnamefont {D.}~\bibnamefont {Hangleiter}}, \bibinfo {author}
  {\bibfnamefont {I.}~\bibnamefont {Roth}},\ and\ \bibinfo {author}
  {\bibfnamefont {J.}~\bibnamefont {Haferkamp}},\ }\href
  {https://doi.org/10.48550/arXiv.2408.11105} {\bibinfo {title} {Benchmarking
  bosonic and fermionic dynamics}} (\bibinfo {year} {2024}),\ \Eprint
  {https://arxiv.org/abs/2408.11105} {arXiv:2408.11105 [quant-ph]} \BibitemShut
  {NoStop}%
\bibitem [{\citenamefont {Arienzo}\ \emph {et~al.}(2025)\citenamefont
  {Arienzo}, \citenamefont {Grinko}, \citenamefont {Kliesch},\ and\
  \citenamefont {Heinrich}}]{arienzo_bosonic_2025}%
  \BibitemOpen
  \bibfield  {author} {\bibinfo {author} {\bibfnamefont {M.}~\bibnamefont
  {Arienzo}}, \bibinfo {author} {\bibfnamefont {D.}~\bibnamefont {Grinko}},
  \bibinfo {author} {\bibfnamefont {M.}~\bibnamefont {Kliesch}},\ and\ \bibinfo
  {author} {\bibfnamefont {M.}~\bibnamefont {Heinrich}},\ }\href
  {https://doi.org/10.1103/PRXQuantum.6.020305} {\bibfield  {journal} {\bibinfo
   {journal} {PRX Quantum}\ }\textbf {\bibinfo {volume} {6}},\ \bibinfo {pages}
  {020305} (\bibinfo {year} {2025})}\BibitemShut {NoStop}%
\bibitem [{Note1()}]{Note1}%
  \BibitemOpen
  \bibinfo {note} {This mathematical decomposition is also derived in these
  works for the purpose of filtered randomized benchmarking; here, we establish
  its physical interpretation as the conservation of hierarchical
  coherence.}\BibitemShut {Stop}%
\bibitem [{\citenamefont {Havel}(2003)}]{havel_procedures_2003}%
  \BibitemOpen
  \bibfield  {author} {\bibinfo {author} {\bibfnamefont {T.~F.}\ \bibnamefont
  {Havel}},\ }\href {https://doi.org/10.1063/1.1518555} {\bibfield  {journal}
  {\bibinfo  {journal} {J. Math. Phys.}\ }\textbf {\bibinfo {volume} {44}},\
  \bibinfo {pages} {534} (\bibinfo {year} {2003})},\ \Eprint
  {https://arxiv.org/abs/quant-ph/0201127} {arXiv:quant-ph/0201127}
  \BibitemShut {NoStop}%
\bibitem [{\citenamefont {Greenbaum}(2015)}]{greenbaum_introduction_2015}%
  \BibitemOpen
  \bibfield  {author} {\bibinfo {author} {\bibfnamefont {D.}~\bibnamefont
  {Greenbaum}},\ }\href {https://doi.org/10.48550/arXiv.1509.02921} {\bibinfo
  {title} {Introduction to {{Quantum Gate Set Tomography}}}} (\bibinfo {year}
  {2015}),\ \Eprint {https://arxiv.org/abs/1509.02921} {arXiv:1509.02921}
  \BibitemShut {NoStop}%
\bibitem [{\citenamefont {Gyamfi}(2020)}]{gyamfi2020Fundamentals}%
  \BibitemOpen
  \bibfield  {author} {\bibinfo {author} {\bibfnamefont {J.~A.}\ \bibnamefont
  {Gyamfi}},\ }\href {https://doi.org/10.1088/1361-6404/ab9fdd} {\bibfield
  {journal} {\bibinfo  {journal} {Eur. J. Phys.}\ }\textbf {\bibinfo {volume}
  {41}},\ \bibinfo {pages} {063002} (\bibinfo {year} {2020})}\BibitemShut
  {NoStop}%
\bibitem [{\citenamefont {Bartlett}\ \emph {et~al.}(2007)\citenamefont
  {Bartlett}, \citenamefont {Rudolph},\ and\ \citenamefont
  {Spekkens}}]{bartlett_reference_2007}%
  \BibitemOpen
  \bibfield  {author} {\bibinfo {author} {\bibfnamefont {S.~D.}\ \bibnamefont
  {Bartlett}}, \bibinfo {author} {\bibfnamefont {T.}~\bibnamefont {Rudolph}},\
  and\ \bibinfo {author} {\bibfnamefont {R.~W.}\ \bibnamefont {Spekkens}},\
  }\href {https://doi.org/10.1103/RevModPhys.79.555} {\bibfield  {journal}
  {\bibinfo  {journal} {Rev. Mod. Phys.}\ }\textbf {\bibinfo {volume} {79}},\
  \bibinfo {pages} {555} (\bibinfo {year} {2007})}\BibitemShut {NoStop}%
\bibitem [{Note2()}]{Note2}%
  \BibitemOpen
  \bibinfo {note} {The derivation follows the strategy in Ref.~\cite
  {parellada_lie_2024}, adapted here to the coherence layer
  structure.}\BibitemShut {Stop}%
\bibitem [{197(1976)}]{1976quantum}%
  \BibitemOpen
  \href {https://books.google.com.hk/books?id=Ne3iT_QLcsMC} {\emph {\bibinfo
  {title} {Quantum Detection and Estimation Theory}}},\ Mathematics in Science
  and Engineering\ (\bibinfo  {publisher} {Academic Press},\ \bibinfo {year}
  {1976})\BibitemShut {NoStop}%
\bibitem [{\citenamefont {Braunstein}\ and\ \citenamefont
  {Caves}(1994)}]{PhysRevLett.72.3439}%
  \BibitemOpen
  \bibfield  {author} {\bibinfo {author} {\bibfnamefont {S.~L.}\ \bibnamefont
  {Braunstein}}\ and\ \bibinfo {author} {\bibfnamefont {C.~M.}\ \bibnamefont
  {Caves}},\ }\href {https://doi.org/10.1103/PhysRevLett.72.3439} {\bibfield
  {journal} {\bibinfo  {journal} {Phys. Rev. Lett.}\ }\textbf {\bibinfo
  {volume} {72}},\ \bibinfo {pages} {3439} (\bibinfo {year}
  {1994})}\BibitemShut {NoStop}%
\bibitem [{\citenamefont {Morris}\ \emph {et~al.}(2020)\citenamefont {Morris},
  \citenamefont {Yadin}, \citenamefont {Fadel}, \citenamefont {Zibold},
  \citenamefont {Treutlein},\ and\ \citenamefont
  {Adesso}}]{morris_entanglement_2020}%
  \BibitemOpen
  \bibfield  {author} {\bibinfo {author} {\bibfnamefont {B.}~\bibnamefont
  {Morris}}, \bibinfo {author} {\bibfnamefont {B.}~\bibnamefont {Yadin}},
  \bibinfo {author} {\bibfnamefont {M.}~\bibnamefont {Fadel}}, \bibinfo
  {author} {\bibfnamefont {T.}~\bibnamefont {Zibold}}, \bibinfo {author}
  {\bibfnamefont {P.}~\bibnamefont {Treutlein}},\ and\ \bibinfo {author}
  {\bibfnamefont {G.}~\bibnamefont {Adesso}},\ }\href
  {https://doi.org/10.1103/PhysRevX.10.041012} {\bibfield  {journal} {\bibinfo
  {journal} {Phys. Rev. X}\ }\textbf {\bibinfo {volume} {10}},\ \bibinfo
  {pages} {041012} (\bibinfo {year} {2020})}\BibitemShut {NoStop}%
\bibitem [{\citenamefont {Banchi}\ \emph {et~al.}(2018)\citenamefont {Banchi},
  \citenamefont {Kolthammer},\ and\ \citenamefont
  {Kim}}]{banchi_multiphoton_2018}%
  \BibitemOpen
  \bibfield  {author} {\bibinfo {author} {\bibfnamefont {L.}~\bibnamefont
  {Banchi}}, \bibinfo {author} {\bibfnamefont {W.~S.}\ \bibnamefont
  {Kolthammer}},\ and\ \bibinfo {author} {\bibfnamefont {M.~S.}\ \bibnamefont
  {Kim}},\ }\href {https://doi.org/10.1103/PhysRevLett.121.250402} {\bibfield
  {journal} {\bibinfo  {journal} {Phys. Rev. Lett.}\ }\textbf {\bibinfo
  {volume} {121}},\ \bibinfo {pages} {250402} (\bibinfo {year}
  {2018})}\BibitemShut {NoStop}%
\bibitem [{Note3()}]{Note3}%
  \BibitemOpen
  \bibinfo {note} {This measurement strategy aligns with Ref.~\cite
  {parellada_lie_2024}. The generalized Gell-Mann basis links our hierarchical
  definition to their observable-based invariants, ensuring that both share the
  same experimental accessibility.}\BibitemShut {Stop}%
\bibitem [{\citenamefont {Hong}\ \emph {et~al.}(1987)\citenamefont {Hong},
  \citenamefont {Ou},\ and\ \citenamefont {Mandel}}]{hong_measurement_1987}%
  \BibitemOpen
  \bibfield  {author} {\bibinfo {author} {\bibfnamefont {C.~K.}\ \bibnamefont
  {Hong}}, \bibinfo {author} {\bibfnamefont {Z.~Y.}\ \bibnamefont {Ou}},\ and\
  \bibinfo {author} {\bibfnamefont {L.}~\bibnamefont {Mandel}},\ }\href
  {https://doi.org/10.1103/PhysRevLett.59.2044} {\bibfield  {journal} {\bibinfo
   {journal} {Phys. Rev. Lett.}\ }\textbf {\bibinfo {volume} {59}},\ \bibinfo
  {pages} {2044} (\bibinfo {year} {1987})}\BibitemShut {NoStop}%
\bibitem [{\citenamefont {Bra{\'n}czyk}(2017)}]{branczyk_hongoumandel_2017}%
  \BibitemOpen
  \bibfield  {author} {\bibinfo {author} {\bibfnamefont {A.~M.}\ \bibnamefont
  {Bra{\'n}czyk}},\ }\href@noop {} {\bibinfo {title} {Hong-{{Ou-Mandel
  Interference}}}} (\bibinfo {year} {2017}),\ \Eprint
  {https://arxiv.org/abs/1711.00080} {arXiv:1711.00080} \BibitemShut {NoStop}%
\bibitem [{\citenamefont {Adamson}\ \emph {et~al.}(2007)\citenamefont
  {Adamson}, \citenamefont {Shalm}, \citenamefont {Mitchell},\ and\
  \citenamefont {Steinberg}}]{adamson_multiparticle_2007}%
  \BibitemOpen
  \bibfield  {author} {\bibinfo {author} {\bibfnamefont {R.~B.~A.}\
  \bibnamefont {Adamson}}, \bibinfo {author} {\bibfnamefont {L.~K.}\
  \bibnamefont {Shalm}}, \bibinfo {author} {\bibfnamefont {M.~W.}\ \bibnamefont
  {Mitchell}},\ and\ \bibinfo {author} {\bibfnamefont {A.~M.}\ \bibnamefont
  {Steinberg}},\ }\href {https://doi.org/10.1103/PhysRevLett.98.043601}
  {\bibfield  {journal} {\bibinfo  {journal} {Phys. Rev. Lett.}\ }\textbf
  {\bibinfo {volume} {98}},\ \bibinfo {pages} {043601} (\bibinfo {year}
  {2007})}\BibitemShut {NoStop}%
\bibitem [{\citenamefont {Ono}\ \emph {et~al.}(2023)\citenamefont {Ono},
  \citenamefont {Roga}, \citenamefont {Wakui}, \citenamefont {Fujiwara},
  \citenamefont {Miki}, \citenamefont {Terai},\ and\ \citenamefont
  {Takeoka}}]{ono_demonstration_2023}%
  \BibitemOpen
  \bibfield  {author} {\bibinfo {author} {\bibfnamefont {T.}~\bibnamefont
  {Ono}}, \bibinfo {author} {\bibfnamefont {W.}~\bibnamefont {Roga}}, \bibinfo
  {author} {\bibfnamefont {K.}~\bibnamefont {Wakui}}, \bibinfo {author}
  {\bibfnamefont {M.}~\bibnamefont {Fujiwara}}, \bibinfo {author}
  {\bibfnamefont {S.}~\bibnamefont {Miki}}, \bibinfo {author} {\bibfnamefont
  {H.}~\bibnamefont {Terai}},\ and\ \bibinfo {author} {\bibfnamefont
  {M.}~\bibnamefont {Takeoka}},\ }\href
  {https://doi.org/10.1103/PhysRevLett.131.013601} {\bibfield  {journal}
  {\bibinfo  {journal} {Phys. Rev. Lett.}\ }\textbf {\bibinfo {volume} {131}},\
  \bibinfo {pages} {013601} (\bibinfo {year} {2023})}\BibitemShut {NoStop}%
\bibitem [{\citenamefont {Gan}\ \emph {et~al.}(2022)\citenamefont {Gan},
  \citenamefont {Leykam},\ and\ \citenamefont {Angelakis}}]{gan_fock_2022}%
  \BibitemOpen
  \bibfield  {author} {\bibinfo {author} {\bibfnamefont {B.~Y.}\ \bibnamefont
  {Gan}}, \bibinfo {author} {\bibfnamefont {D.}~\bibnamefont {Leykam}},\ and\
  \bibinfo {author} {\bibfnamefont {D.~G.}\ \bibnamefont {Angelakis}},\ }\href
  {https://doi.org/10.1140/epjqt/s40507-022-00135-0} {\bibfield  {journal}
  {\bibinfo  {journal} {EPJ Quantum Technol.}\ }\textbf {\bibinfo {volume}
  {9}},\ \bibinfo {pages} {1} (\bibinfo {year} {2022})}\BibitemShut {NoStop}%
\bibitem [{\citenamefont {Somma}(2005)}]{somma_quantum_2005}%
  \BibitemOpen
  \bibfield  {author} {\bibinfo {author} {\bibfnamefont {R.~D.}\ \bibnamefont
  {Somma}},\ }\href {https://doi.org/10.48550/arXiv.quant-ph/0512209} {\bibinfo
  {title} {Quantum {{Computation}}, {{Complexity}}, and {{Many-Body Physics}}}}
  (\bibinfo {year} {2005}),\ \Eprint {https://arxiv.org/abs/quant-ph/0512209}
  {arXiv:quant-ph/0512209} \BibitemShut {NoStop}%
\bibitem [{\citenamefont {Bastin}\ \emph {et~al.}(2009)\citenamefont {Bastin},
  \citenamefont {Krins}, \citenamefont {Mathonet}, \citenamefont {Godefroid},
  \citenamefont {Lamata},\ and\ \citenamefont
  {Solano}}]{bastin_operational_2009}%
  \BibitemOpen
  \bibfield  {author} {\bibinfo {author} {\bibfnamefont {T.}~\bibnamefont
  {Bastin}}, \bibinfo {author} {\bibfnamefont {S.}~\bibnamefont {Krins}},
  \bibinfo {author} {\bibfnamefont {P.}~\bibnamefont {Mathonet}}, \bibinfo
  {author} {\bibfnamefont {M.}~\bibnamefont {Godefroid}}, \bibinfo {author}
  {\bibfnamefont {L.}~\bibnamefont {Lamata}},\ and\ \bibinfo {author}
  {\bibfnamefont {E.}~\bibnamefont {Solano}},\ }\href
  {https://doi.org/10.1103/PhysRevLett.103.070503} {\bibfield  {journal}
  {\bibinfo  {journal} {Phys. Rev. Lett.}\ }\textbf {\bibinfo {volume} {103}},\
  \bibinfo {pages} {070503} (\bibinfo {year} {2009})}\BibitemShut {NoStop}%
\bibitem [{\citenamefont {Baumgratz}\ \emph {et~al.}(2014)\citenamefont
  {Baumgratz}, \citenamefont {Cramer},\ and\ \citenamefont
  {Plenio}}]{baumgratz_quantifying_2014}%
  \BibitemOpen
  \bibfield  {author} {\bibinfo {author} {\bibfnamefont {T.}~\bibnamefont
  {Baumgratz}}, \bibinfo {author} {\bibfnamefont {M.}~\bibnamefont {Cramer}},\
  and\ \bibinfo {author} {\bibfnamefont {M.~B.}\ \bibnamefont {Plenio}},\
  }\href {https://doi.org/10.1103/PhysRevLett.113.140401} {\bibfield  {journal}
  {\bibinfo  {journal} {Phys. Rev. Lett.}\ }\textbf {\bibinfo {volume} {113}},\
  \bibinfo {pages} {140401} (\bibinfo {year} {2014})}\BibitemShut {NoStop}%
\bibitem [{\citenamefont {Tan}\ \emph {et~al.}(2017)\citenamefont {Tan},
  \citenamefont {Volkoff}, \citenamefont {Kwon},\ and\ \citenamefont
  {Jeong}}]{tan_quantifying_2017}%
  \BibitemOpen
  \bibfield  {author} {\bibinfo {author} {\bibfnamefont {K.~C.}\ \bibnamefont
  {Tan}}, \bibinfo {author} {\bibfnamefont {T.}~\bibnamefont {Volkoff}},
  \bibinfo {author} {\bibfnamefont {H.}~\bibnamefont {Kwon}},\ and\ \bibinfo
  {author} {\bibfnamefont {H.}~\bibnamefont {Jeong}},\ }\href
  {https://doi.org/10.1103/PhysRevLett.119.190405} {\bibfield  {journal}
  {\bibinfo  {journal} {Phys. Rev. Lett.}\ }\textbf {\bibinfo {volume} {119}},\
  \bibinfo {pages} {190405} (\bibinfo {year} {2017})}\BibitemShut {NoStop}%
\end{thebibliography}%


\begin{thebibliography}{24}%
\makeatletter
\providecommand \@ifxundefined [1]{%
 \@ifx{#1\undefined}
}%
\providecommand \@ifnum [1]{%
 \ifnum #1\expandafter \@firstoftwo
 \else \expandafter \@secondoftwo
 \fi
}%
\providecommand \@ifx [1]{%
 \ifx #1\expandafter \@firstoftwo
 \else \expandafter \@secondoftwo
 \fi
}%
\providecommand \natexlab [1]{#1}%
\providecommand \enquote  [1]{``#1''}%
\providecommand \bibnamefont  [1]{#1}%
\providecommand \bibfnamefont [1]{#1}%
\providecommand \citenamefont [1]{#1}%
\providecommand \href@noop [0]{\@secondoftwo}%
\providecommand \href [0]{\begingroup \@sanitize@url \@href}%
\providecommand \@href[1]{\@@startlink{#1}\@@href}%
\providecommand \@@href[1]{\endgroup#1\@@endlink}%
\providecommand \@sanitize@url [0]{\catcode `\\12\catcode `\$12\catcode
  `\&12\catcode `\#12\catcode `\^12\catcode `\_12\catcode `\%12\relax}%
\providecommand \@@startlink[1]{}%
\providecommand \@@endlink[0]{}%
\providecommand \url  [0]{\begingroup\@sanitize@url \@url }%
\providecommand \@url [1]{\endgroup\@href {#1}{\urlprefix }}%
\providecommand \urlprefix  [0]{URL }%
\providecommand \Eprint [0]{\href }%
\providecommand \doibase [0]{https://doi.org/}%
\providecommand \selectlanguage [0]{\@gobble}%
\providecommand \bibinfo  [0]{\@secondoftwo}%
\providecommand \bibfield  [0]{\@secondoftwo}%
\providecommand \translation [1]{[#1]}%
\providecommand \BibitemOpen [0]{}%
\providecommand \bibitemStop [0]{}%
\providecommand \bibitemNoStop [0]{.\EOS\space}%
\providecommand \EOS [0]{\spacefactor3000\relax}%
\providecommand \BibitemShut  [1]{\csname bibitem#1\endcsname}%
\let\auto@bib@innerbib\@empty
\bibitem [{\citenamefont {Reck}\ \emph {et~al.}(1994)\citenamefont {Reck},
  \citenamefont {Zeilinger}, \citenamefont {Bernstein},\ and\ \citenamefont
  {Bertani}}]{reck_experimental_1994}%
  \BibitemOpen
  \bibfield  {author} {\bibinfo {author} {\bibfnamefont {M.}~\bibnamefont
  {Reck}}, \bibinfo {author} {\bibfnamefont {A.}~\bibnamefont {Zeilinger}},
  \bibinfo {author} {\bibfnamefont {H.~J.}\ \bibnamefont {Bernstein}},\ and\
  \bibinfo {author} {\bibfnamefont {P.}~\bibnamefont {Bertani}},\ }\href
  {https://doi.org/10.1103/PhysRevLett.73.58} {\bibfield  {journal} {\bibinfo
  {journal} {Phys. Rev. Lett.}\ }\textbf {\bibinfo {volume} {73}},\ \bibinfo
  {pages} {58} (\bibinfo {year} {1994})}\BibitemShut {NoStop}%
\bibitem [{\citenamefont {Bouland}\ and\ \citenamefont
  {Aaronson}(2014)}]{bouland_generation_2014}%
  \BibitemOpen
  \bibfield  {author} {\bibinfo {author} {\bibfnamefont {A.}~\bibnamefont
  {Bouland}}\ and\ \bibinfo {author} {\bibfnamefont {S.}~\bibnamefont
  {Aaronson}},\ }\href {https://doi.org/10.1103/PhysRevA.89.062316} {\bibfield
  {journal} {\bibinfo  {journal} {Phys. Rev. A}\ }\textbf {\bibinfo {volume}
  {89}},\ \bibinfo {pages} {062316} (\bibinfo {year} {2014})}\BibitemShut
  {NoStop}%
\bibitem [{\citenamefont {Carolan}\ \emph {et~al.}(2015)\citenamefont
  {Carolan}, \citenamefont {Harrold}, \citenamefont {Sparrow}, \citenamefont
  {Martín-López}, \citenamefont {Russell}, \citenamefont {Silverstone},
  \citenamefont {Shadbolt}, \citenamefont {Matsuda}, \citenamefont {Oguma},
  \citenamefont {Itoh}, \citenamefont {Marshall}, \citenamefont {Thompson},
  \citenamefont {Matthews}, \citenamefont {Hashimoto}, \citenamefont
  {O’Brien},\ and\ \citenamefont {Laing}}]{carolan_universal_2015}%
  \BibitemOpen
  \bibfield  {author} {\bibinfo {author} {\bibfnamefont {J.}~\bibnamefont
  {Carolan}}, \bibinfo {author} {\bibfnamefont {C.}~\bibnamefont {Harrold}},
  \bibinfo {author} {\bibfnamefont {C.}~\bibnamefont {Sparrow}}, \bibinfo
  {author} {\bibfnamefont {E.}~\bibnamefont {Martín-López}}, \bibinfo
  {author} {\bibfnamefont {N.~J.}\ \bibnamefont {Russell}}, \bibinfo {author}
  {\bibfnamefont {J.~W.}\ \bibnamefont {Silverstone}}, \bibinfo {author}
  {\bibfnamefont {P.~J.}\ \bibnamefont {Shadbolt}}, \bibinfo {author}
  {\bibfnamefont {N.}~\bibnamefont {Matsuda}}, \bibinfo {author} {\bibfnamefont
  {M.}~\bibnamefont {Oguma}}, \bibinfo {author} {\bibfnamefont
  {M.}~\bibnamefont {Itoh}}, \bibinfo {author} {\bibfnamefont {G.~D.}\
  \bibnamefont {Marshall}}, \bibinfo {author} {\bibfnamefont {M.~G.}\
  \bibnamefont {Thompson}}, \bibinfo {author} {\bibfnamefont {J.~C.~F.}\
  \bibnamefont {Matthews}}, \bibinfo {author} {\bibfnamefont {T.}~\bibnamefont
  {Hashimoto}}, \bibinfo {author} {\bibfnamefont {J.~L.}\ \bibnamefont
  {O’Brien}},\ and\ \bibinfo {author} {\bibfnamefont {A.}~\bibnamefont
  {Laing}},\ }\href {https://doi.org/10.1126/science.aab3642} {\bibfield
  {journal} {\bibinfo  {journal} {Science}\ }\textbf {\bibinfo {volume}
  {349}},\ \bibinfo {pages} {711} (\bibinfo {year} {2015})}\BibitemShut
  {NoStop}%
\bibitem [{\citenamefont {Clements}\ \emph {et~al.}(2016)\citenamefont
  {Clements}, \citenamefont {Humphreys}, \citenamefont {Metcalf}, \citenamefont
  {Kolthammer},\ and\ \citenamefont {Walmsley}}]{clements_optimal_2016}%
  \BibitemOpen
  \bibfield  {author} {\bibinfo {author} {\bibfnamefont {W.~R.}\ \bibnamefont
  {Clements}}, \bibinfo {author} {\bibfnamefont {P.~C.}\ \bibnamefont
  {Humphreys}}, \bibinfo {author} {\bibfnamefont {B.~J.}\ \bibnamefont
  {Metcalf}}, \bibinfo {author} {\bibfnamefont {W.~S.}\ \bibnamefont
  {Kolthammer}},\ and\ \bibinfo {author} {\bibfnamefont {I.~A.}\ \bibnamefont
  {Walmsley}},\ }\href {https://doi.org/10.1364/OPTICA.3.001460} {\bibfield
  {journal} {\bibinfo  {journal} {Optica}\ }\textbf {\bibinfo {volume} {3}},\
  \bibinfo {pages} {1460} (\bibinfo {year} {2016})}\BibitemShut {NoStop}%
\bibitem [{\citenamefont {Aaronson}\ and\ \citenamefont
  {Arkhipov}(2011)}]{aaronson_computational_2011}%
  \BibitemOpen
  \bibfield  {author} {\bibinfo {author} {\bibfnamefont {S.}~\bibnamefont
  {Aaronson}}\ and\ \bibinfo {author} {\bibfnamefont {A.}~\bibnamefont
  {Arkhipov}},\ }in\ \href {https://doi.org/10.1145/1993636.1993682} {\emph
  {\bibinfo {booktitle} {Proceedings of the Forty-Third Annual ACM Symposium on
  Theory of Computing}}},\ \bibinfo {series and number} {STOC '11}\ (\bibinfo
  {publisher} {Association for Computing Machinery},\ \bibinfo {address} {New
  York, NY, USA},\ \bibinfo {year} {2011})\ p.\ \bibinfo {pages}
  {333–342}\BibitemShut {NoStop}%
\bibitem [{\citenamefont {Garcia-Escartin}\ \emph
  {et~al.}(2019{\natexlab{a}})\citenamefont {Garcia-Escartin}, \citenamefont
  {Gimeno},\ and\ \citenamefont
  {Moyano-Fern\'andez}}]{garcia-escartin_method_2019}%
  \BibitemOpen
  \bibfield  {author} {\bibinfo {author} {\bibfnamefont {J.~C.}\ \bibnamefont
  {Garcia-Escartin}}, \bibinfo {author} {\bibfnamefont {V.}~\bibnamefont
  {Gimeno}},\ and\ \bibinfo {author} {\bibfnamefont {J.~J.}\ \bibnamefont
  {Moyano-Fern\'andez}},\ }\href {https://doi.org/10.1103/PhysRevA.100.022301}
  {\bibfield  {journal} {\bibinfo  {journal} {Phys. Rev. A}\ }\textbf {\bibinfo
  {volume} {100}},\ \bibinfo {pages} {022301} (\bibinfo {year}
  {2019}{\natexlab{a}})}\BibitemShut {NoStop}%
\bibitem [{\citenamefont {Jordan}(1935)}]{jordan_zusammenhang_1935}%
  \BibitemOpen
  \bibfield  {author} {\bibinfo {author} {\bibfnamefont {P.}~\bibnamefont
  {Jordan}},\ }\href {https://doi.org/10.1007/BF01330618} {\bibfield  {journal}
  {\bibinfo  {journal} {Zeitschrift f{\"u}r Physik}\ }\textbf {\bibinfo
  {volume} {94}},\ \bibinfo {pages} {531} (\bibinfo {year} {1935})}\BibitemShut
  {NoStop}%
\bibitem [{\citenamefont {Schwinger}(1952)}]{schwinger_angular_1952}%
  \BibitemOpen
  \bibfield  {author} {\bibinfo {author} {\bibfnamefont {J.}~\bibnamefont
  {Schwinger}},\ }\href {https://doi.org/10.2172/4389568} {\emph {\bibinfo
  {title} {ON ANGULAR MOMENTUM}}},\ \bibinfo {type} {Tech. Rep.}\ (\bibinfo
  {institution} {Harvard Univ., Cambridge, MA (United States); Nuclear
  Development Associates, Inc. (US)},\ \bibinfo {year} {1952})\BibitemShut
  {NoStop}%
\bibitem [{\citenamefont {Garcia-Escartin}\ \emph
  {et~al.}(2019{\natexlab{b}})\citenamefont {Garcia-Escartin}, \citenamefont
  {Gimeno},\ and\ \citenamefont
  {Moyano-Fernández}}]{garcia-escartin_multiple_2019}%
  \BibitemOpen
  \bibfield  {author} {\bibinfo {author} {\bibfnamefont {J.~C.}\ \bibnamefont
  {Garcia-Escartin}}, \bibinfo {author} {\bibfnamefont {V.}~\bibnamefont
  {Gimeno}},\ and\ \bibinfo {author} {\bibfnamefont {J.~J.}\ \bibnamefont
  {Moyano-Fernández}},\ }\href
  {https://doi.org/https://doi.org/10.1016/j.optcom.2018.08.082} {\bibfield
  {journal} {\bibinfo  {journal} {Optics Communications}\ }\textbf {\bibinfo
  {volume} {430}},\ \bibinfo {pages} {434} (\bibinfo {year}
  {2019}{\natexlab{b}})}\BibitemShut {NoStop}%
\bibitem [{\citenamefont {Parellada}\ \emph {et~al.}(2024)\citenamefont
  {Parellada}, \citenamefont {i~Garcia}, \citenamefont {Moyano-Fernández},\
  and\ \citenamefont {Garcia-Escartin}}]{parellada_lie_2024}%
  \BibitemOpen
  \bibfield  {author} {\bibinfo {author} {\bibfnamefont {P.~V.}\ \bibnamefont
  {Parellada}}, \bibinfo {author} {\bibfnamefont {V.~G.}\ \bibnamefont
  {i~Garcia}}, \bibinfo {author} {\bibfnamefont {J.~J.}\ \bibnamefont
  {Moyano-Fernández}},\ and\ \bibinfo {author} {\bibfnamefont {J.~C.}\
  \bibnamefont {Garcia-Escartin}},\ }\href {https://arxiv.org/abs/2409.12223}
  {\bibinfo {title} {Lie algebraic invariants in quantum linear optics}}
  (\bibinfo {year} {2024}),\ \Eprint {https://arxiv.org/abs/2409.12223}
  {arXiv:2409.12223 [quant-ph]} \BibitemShut {NoStop}%
\bibitem [{\citenamefont {Bertlmann}\ and\ \citenamefont
  {Krammer}(2008)}]{bertlmann_bloch_2008}%
  \BibitemOpen
  \bibfield  {author} {\bibinfo {author} {\bibfnamefont {R.~A.}\ \bibnamefont
  {Bertlmann}}\ and\ \bibinfo {author} {\bibfnamefont {P.}~\bibnamefont
  {Krammer}},\ }\href {https://doi.org/10.1088/1751-8113/41/23/235303}
  {\bibfield  {journal} {\bibinfo  {journal} {Journal of Physics A:
  Mathematical and Theoretical}\ }\textbf {\bibinfo {volume} {41}},\ \bibinfo
  {pages} {235303} (\bibinfo {year} {2008})}\BibitemShut {NoStop}%
\bibitem [{\citenamefont {Parellada}\ \emph {et~al.}(2023)\citenamefont
  {Parellada}, \citenamefont {{Gimeno i Garcia}}, \citenamefont
  {Moyano-FernÃ¡ndez},\ and\ \citenamefont
  {Garcia-Escartin}}]{parellada_no-go_2023}%
  \BibitemOpen
  \bibfield  {author} {\bibinfo {author} {\bibfnamefont {P.~V.}\ \bibnamefont
  {Parellada}}, \bibinfo {author} {\bibfnamefont {V.}~\bibnamefont {{Gimeno i
  Garcia}}}, \bibinfo {author} {\bibfnamefont {J.~J.}\ \bibnamefont
  {Moyano-FernÃ¡ndez}},\ and\ \bibinfo {author} {\bibfnamefont {J.~C.}\
  \bibnamefont {Garcia-Escartin}},\ }\href
  {https://doi.org/https://doi.org/10.1016/j.rinp.2023.107108} {\bibfield
  {journal} {\bibinfo  {journal} {Results in Physics}\ }\textbf {\bibinfo
  {volume} {54}},\ \bibinfo {pages} {107108} (\bibinfo {year}
  {2023})}\BibitemShut {NoStop}%
\bibitem [{\citenamefont {Hoffman}\ and\ \citenamefont
  {Wielandt}(1953)}]{hoffman_variation_1953}%
  \BibitemOpen
  \bibfield  {author} {\bibinfo {author} {\bibfnamefont {A.~J.}\ \bibnamefont
  {Hoffman}}\ and\ \bibinfo {author} {\bibfnamefont {H.~W.}\ \bibnamefont
  {Wielandt}},\ }\href {https://doi.org/10.1215/S0012-7094-53-02004-3}
  {\bibfield  {journal} {\bibinfo  {journal} {Duke Mathematical Journal}\
  }\textbf {\bibinfo {volume} {20}},\ \bibinfo {pages} {37 } (\bibinfo {year}
  {1953})}\BibitemShut {NoStop}%
\bibitem [{\citenamefont {Braunstein}\ and\ \citenamefont
  {Caves}(1994)}]{braunstein_statistical_1994}%
  \BibitemOpen
  \bibfield  {author} {\bibinfo {author} {\bibfnamefont {S.~L.}\ \bibnamefont
  {Braunstein}}\ and\ \bibinfo {author} {\bibfnamefont {C.~M.}\ \bibnamefont
  {Caves}},\ }\href {https://doi.org/10.1103/PhysRevLett.72.3439} {\bibfield
  {journal} {\bibinfo  {journal} {Physical Review Letters}\ }\textbf {\bibinfo
  {volume} {72}},\ \bibinfo {pages} {3439} (\bibinfo {year} {1994})},\ \bibinfo
  {note} {publisher: American Physical Society}\BibitemShut {NoStop}%
\bibitem [{\citenamefont {Giovannetti}\ \emph {et~al.}(2011)\citenamefont
  {Giovannetti}, \citenamefont {Lloyd},\ and\ \citenamefont
  {Maccone}}]{giovannetti_advances_2011}%
  \BibitemOpen
  \bibfield  {author} {\bibinfo {author} {\bibfnamefont {V.}~\bibnamefont
  {Giovannetti}}, \bibinfo {author} {\bibfnamefont {S.}~\bibnamefont {Lloyd}},\
  and\ \bibinfo {author} {\bibfnamefont {L.}~\bibnamefont {Maccone}},\ }\href
  {https://doi.org/10.1038/nphoton.2011.35} {\bibfield  {journal} {\bibinfo
  {journal} {Nature Photonics}\ }\textbf {\bibinfo {volume} {5}},\ \bibinfo
  {pages} {222} (\bibinfo {year} {2011})},\ \bibinfo {note} {publisher: Nature
  Publishing Group}\BibitemShut {NoStop}%
\bibitem [{\citenamefont {Sun}\ \emph {et~al.}(2010)\citenamefont {Sun},
  \citenamefont {Ma}, \citenamefont {Lu},\ and\ \citenamefont
  {Wang}}]{sun_fisher_2010}%
  \BibitemOpen
  \bibfield  {author} {\bibinfo {author} {\bibfnamefont {Z.}~\bibnamefont
  {Sun}}, \bibinfo {author} {\bibfnamefont {J.}~\bibnamefont {Ma}}, \bibinfo
  {author} {\bibfnamefont {X.-M.}\ \bibnamefont {Lu}},\ and\ \bibinfo {author}
  {\bibfnamefont {X.}~\bibnamefont {Wang}},\ }\href
  {https://doi.org/10.1103/PhysRevA.82.022306} {\bibfield  {journal} {\bibinfo
  {journal} {Physical Review A}\ }\textbf {\bibinfo {volume} {82}},\ \bibinfo
  {pages} {022306} (\bibinfo {year} {2010})}\BibitemShut {NoStop}%
\bibitem [{\citenamefont {Bartlett}\ \emph {et~al.}(2007)\citenamefont
  {Bartlett}, \citenamefont {Rudolph},\ and\ \citenamefont
  {Spekkens}}]{bartlett_reference_2007}%
  \BibitemOpen
  \bibfield  {author} {\bibinfo {author} {\bibfnamefont {S.~D.}\ \bibnamefont
  {Bartlett}}, \bibinfo {author} {\bibfnamefont {T.}~\bibnamefont {Rudolph}},\
  and\ \bibinfo {author} {\bibfnamefont {R.~W.}\ \bibnamefont {Spekkens}},\
  }\href {https://doi.org/10.1103/RevModPhys.79.555} {\bibfield  {journal}
  {\bibinfo  {journal} {Reviews of Modern Physics}\ }\textbf {\bibinfo {volume}
  {79}},\ \bibinfo {pages} {555} (\bibinfo {year} {2007})}\BibitemShut
  {NoStop}%
\bibitem [{\citenamefont {Wick}\ \emph {et~al.}(1952)\citenamefont {Wick},
  \citenamefont {Wightman},\ and\ \citenamefont
  {Wigner}}]{wick_intrinsic_1952}%
  \BibitemOpen
  \bibfield  {author} {\bibinfo {author} {\bibfnamefont {G.~C.}\ \bibnamefont
  {Wick}}, \bibinfo {author} {\bibfnamefont {A.~S.}\ \bibnamefont {Wightman}},\
  and\ \bibinfo {author} {\bibfnamefont {E.~P.}\ \bibnamefont {Wigner}},\
  }\href {https://doi.org/10.1103/PhysRev.88.101} {\bibfield  {journal}
  {\bibinfo  {journal} {Physical Review}\ }\textbf {\bibinfo {volume} {88}},\
  \bibinfo {pages} {101} (\bibinfo {year} {1952})}\BibitemShut {NoStop}%
\bibitem [{\citenamefont {Hong}\ \emph {et~al.}(1987)\citenamefont {Hong},
  \citenamefont {Ou},\ and\ \citenamefont {Mandel}}]{hong_measurement_1987}%
  \BibitemOpen
  \bibfield  {author} {\bibinfo {author} {\bibfnamefont {C.~K.}\ \bibnamefont
  {Hong}}, \bibinfo {author} {\bibfnamefont {Z.~Y.}\ \bibnamefont {Ou}},\ and\
  \bibinfo {author} {\bibfnamefont {L.}~\bibnamefont {Mandel}},\ }\href
  {https://doi.org/10.1103/PhysRevLett.59.2044} {\bibfield  {journal} {\bibinfo
   {journal} {Phys. Rev. Lett.}\ }\textbf {\bibinfo {volume} {59}},\ \bibinfo
  {pages} {2044} (\bibinfo {year} {1987})}\BibitemShut {NoStop}%
\bibitem [{\citenamefont {Brańczyk}(2017)}]{branczyk_hong-ou-mandel_2017}%
  \BibitemOpen
  \bibfield  {author} {\bibinfo {author} {\bibfnamefont {A.~M.}\ \bibnamefont
  {Brańczyk}},\ }\href {https://arxiv.org/abs/1711.00080} {\bibinfo {title}
  {Hong-ou-mandel interference}} (\bibinfo {year} {2017}),\ \Eprint
  {https://arxiv.org/abs/1711.00080} {arXiv:1711.00080 [quant-ph]} \BibitemShut
  {NoStop}%
\bibitem [{\citenamefont {Zyczkowski}\ and\ \citenamefont
  {Kus}(1994)}]{zyczkowski_random_1994}%
  \BibitemOpen
  \bibfield  {author} {\bibinfo {author} {\bibfnamefont {K.}~\bibnamefont
  {Zyczkowski}}\ and\ \bibinfo {author} {\bibfnamefont {M.}~\bibnamefont
  {Kus}},\ }\href {https://doi.org/10.1088/0305-4470/27/12/028} {\bibfield
  {journal} {\bibinfo  {journal} {Journal of Physics A: Mathematical and
  General}\ }\textbf {\bibinfo {volume} {27}},\ \bibinfo {pages} {4235}
  (\bibinfo {year} {1994})}\BibitemShut {NoStop}%
\bibitem [{\citenamefont {Simon}\ and\ \citenamefont
  {Mukunda}(1989)}]{simon_universal_1989}%
  \BibitemOpen
  \bibfield  {author} {\bibinfo {author} {\bibfnamefont {R.}~\bibnamefont
  {Simon}}\ and\ \bibinfo {author} {\bibfnamefont {N.}~\bibnamefont
  {Mukunda}},\ }\href
  {https://doi.org/https://doi.org/10.1016/0375-9601(89)90748-2} {\bibfield
  {journal} {\bibinfo  {journal} {Physics Letters A}\ }\textbf {\bibinfo
  {volume} {138}},\ \bibinfo {pages} {474} (\bibinfo {year}
  {1989})}\BibitemShut {NoStop}%
\bibitem [{\citenamefont {Banchi}\ \emph {et~al.}(2018)\citenamefont {Banchi},
  \citenamefont {Kolthammer},\ and\ \citenamefont
  {Kim}}]{banchi_multiphoton_2018}%
  \BibitemOpen
  \bibfield  {author} {\bibinfo {author} {\bibfnamefont {L.}~\bibnamefont
  {Banchi}}, \bibinfo {author} {\bibfnamefont {W.~S.}\ \bibnamefont
  {Kolthammer}},\ and\ \bibinfo {author} {\bibfnamefont {M.~S.}\ \bibnamefont
  {Kim}},\ }\href {https://doi.org/10.1103/PhysRevLett.121.250402} {\bibfield
  {journal} {\bibinfo  {journal} {Phys. Rev. Lett.}\ }\textbf {\bibinfo
  {volume} {121}},\ \bibinfo {pages} {250402} (\bibinfo {year}
  {2018})}\BibitemShut {NoStop}%
\bibitem [{\citenamefont {Adamson}\ \emph {et~al.}(2007)\citenamefont
  {Adamson}, \citenamefont {Shalm}, \citenamefont {Mitchell},\ and\
  \citenamefont {Steinberg}}]{adamson_multiparticle_2007}%
  \BibitemOpen
  \bibfield  {author} {\bibinfo {author} {\bibfnamefont {R.~B.~A.}\
  \bibnamefont {Adamson}}, \bibinfo {author} {\bibfnamefont {L.~K.}\
  \bibnamefont {Shalm}}, \bibinfo {author} {\bibfnamefont {M.~W.}\ \bibnamefont
  {Mitchell}},\ and\ \bibinfo {author} {\bibfnamefont {A.~M.}\ \bibnamefont
  {Steinberg}},\ }\href {https://doi.org/10.1103/PhysRevLett.98.043601}
  {\bibfield  {journal} {\bibinfo  {journal} {Phys. Rev. Lett.}\ }\textbf
  {\bibinfo {volume} {98}},\ \bibinfo {pages} {043601} (\bibinfo {year}
  {2007})}\BibitemShut {NoStop}%
\end{thebibliography}%

\end{document}


\title{Supplemental Material for \\
``Unveiling Hierarchical Invariants in Multiphoton Linear Optics''}

\author{Baichuan Yang}
\thanks{These authors contributed equally to this work.}
\author{Hao Zhan}
\thanks{These authors contributed equally to this work.}
\author{Minghao Mi}
\author{Aonan Zhang}
\author{Liang Xu}
\email{liangxu.ceas@nju.edu.cn}
\author{Lijian Zhang}
\email{lijian.zhang@nju.edu.cn}

\maketitle

\onecolumngrid

\tableofcontents
\clearpage

\section*{Preliminaries}

In this Supplemental Material we collect the definitions, technical lemmas, and proofs underlying the main text.
We begin by fixing notation for the fixed-\(n\)-photon sector, the Hilbert--Schmidt structure on operators, and the action of passive, lossless linear optical networks.

We consider \(m\) bosonic modes with creation and annihilation operators \(\hat a_j^\dagger,\hat a_j\) obeying the canonical commutation relations
\begin{equation}
[\hat a_j,\hat a_k^\dagger]=\delta_{jk},\qquad
[\hat a_j,\hat a_k]=[\hat a_j^\dagger,\hat a_k^\dagger]=0.
\label{eq:CCR}
\end{equation}
The total photon-number operator is
\[
\hat N=\sum_{j=1}^m \hat a_j^\dagger \hat a_j.
\]
The fixed \(n\)-photon sector is
\begin{equation}
\mathcal H_{n,m}
:=
\Span_{\mathbb C}\left\{
\ket{\mathbf n}
=
\prod_{j=1}^{m}\frac{(\hat a_j^{\dagger})^{n_j}}{\sqrt{n_j!}}\ket{0}
\;:\;
\mathbf n=(n_1,\dots,n_m)\in\mathbb N^m,\ 
\sum_{j=1}^m n_j=n
\right\},
\label{eq:Hnm-def}
\end{equation}
with dimension
\[
M=\binom{m+n-1}{n}.
\]
We equip operators on \(\mathcal H_{n,m}\) with the Hilbert--Schmidt (HS) inner product
\begin{equation}
\langle A,B\rangle_{\mathrm{HS}}:=\tr(A^\dagger B).
\label{eq:HS-inner}
\end{equation}

We consider passive, lossless linear optical networks (LONs) on these \(m\) modes.
Such a network is specified by a single-particle scattering matrix \(S\in\mathrm U(m)\), which acts in the Heisenberg picture as
\begin{equation}
\hat a_{j}^{\dagger}\ \longmapsto\ \sum_{k=1}^m S_{kj}\,\hat a_k^{\dagger}.
\label{eq:Heis-S}
\end{equation}
Any \(S\in\mathrm U(m)\) can be implemented using universal interferometer architectures built from beam splitters and phase shifters~\cite{reck_experimental_1994,bouland_generation_2014,carolan_universal_2015,clements_optimal_2016}.

The scattering matrix \(S\) induces an \(n\)-photon unitary
\[
U=\varphi_n(S)\in\mathrm U(M)
\]
acting on \(\mathcal H_{n,m}\)~\cite{aaronson_computational_2011}.
Accordingly, the set of all LON-realizable unitaries forms the Lie subgroup
\[
\mathrm{U_L} :=\varphi_n(\mathrm U(m))\subset \mathrm U(M),
\]
whose dimension is \(m^2\), whereas \(\dim\mathrm U(M)=M^2\)~\cite{garcia-escartin_method_2019}.
Thus passive linear optics cannot deterministically realize arbitrary unitaries on \(\mathcal H_{n,m}\).

To prepare for the hierarchy developed in the next section, we recall the standard Jordan--Schwinger map,
\begin{equation}
\mathcal J^{(1)}(h)=\sum_{s,t=1}^m h_{st}\,\hat a_s^\dagger \hat a_t,
\qquad h\in\Herm(m),
\label{eq:J1-def}
\end{equation}
which generates passive linear-optical dynamics on the fixed-\(n\) sector.
In particular, if \(S=\exp(ih)\), then the induced \(n\)-photon unitary is
\[
U=\varphi_n(S)=\exp\!\big(i\,\mathcal J^{(1)}(h)\big)
\]
on \(\mathcal H_{n,m}\)~\cite{jordan_zusammenhang_1935,schwinger_angular_1952,garcia-escartin_multiple_2019}.

\section{Higher-order Jordan--Schwinger Maps and Jordan Spaces}

\SecIntro{
Building on the first-order Jordan--Schwinger map recalled above, we now introduce its higher-order generalization \(\mathcal J^{(j)}\) and the associated order-\(j\) Jordan spaces. Since \(\mathcal J^{(j)}\) is defined through normally ordered products containing \(j\) creation and \(j\) annihilation operators, it is natural to relate its image to the linear span of all such normally ordered monomials. This characterization yields convenient spanning sets and bases for the Jordan spaces, from which we determine their dimensions and prove that the order-\(n\) Jordan space coincides with the full Hermitian operator space on \(\mathcal H_{n,m}\).
}

\begin{definition}[Order-\(j\) Jordan--Schwinger map]
\label{def:Jj}
For \(1\le j\le n\) and matrices \(h_1,\dots,h_j\in \mathrm M_m(\mathbb C)\), define
\begin{equation}
\mathcal J^{(j)}(h_1,\dots,h_j)
:=
\norder{\prod_{i=1}^j \mathcal J^{(1)}(h_i)}
=
\sum_{\mathbf s,\mathbf t\in [m]^j}
\left(\prod_{i=1}^j (h_i)_{s_i t_i}\right)
\hat a_{s_1}^\dagger \cdots \hat a_{s_j}^\dagger
\hat a_{t_1} \cdots \hat a_{t_j},
\label{eq:Jj-def}
\end{equation}
where \([m]:=\{1,\dots,m\}\), \(\mathbf s=(s_1,\dots,s_j)\), and \(\mathbf t=(t_1,\dots,t_j)\). The surrounding colons denote the normal ordering. By convention, we set \(\mathcal J^{(0)}:=I\).
\end{definition}

\begin{lemma}[Hermiticity for Hermitian inputs]
\label{lem:Jj-herm}
If each \(h_i\) is Hermitian, then \(\mathcal J^{(j)}(h_1,\dots,h_j)\) is Hermitian.
\end{lemma}

\begin{proof}
Taking the adjoint of Eq.~\eqref{eq:Jj-def} gives
\begin{align*}
\mathcal J^{(j)}(h_1,\dots,h_j)^\dagger
&=
\sum_{\mathbf s,\mathbf t\in [m]^j}
\left(\prod_{i=1}^j (h_i)_{s_i t_i}\right)^*
\hat a_{t_1}^\dagger \cdots \hat a_{t_j}^\dagger
\hat a_{s_1} \cdots \hat a_{s_j}.
\end{align*}
Since each \(h_i\) is Hermitian, \((h_i)_{s_i t_i}^*=(h_i)_{t_i s_i}\). Relabeling \(\mathbf s \leftrightarrow \mathbf t\), we obtain
\[
\mathcal J^{(j)}(h_1,\dots,h_j)^\dagger
=
\sum_{\mathbf s,\mathbf t\in [m]^j}
\left(\prod_{i=1}^j (h_i)_{s_i t_i}\right)
\hat a_{s_1}^\dagger \cdots \hat a_{s_j}^\dagger
\hat a_{t_1} \cdots \hat a_{t_j}
=
\mathcal J^{(j)}(h_1,\dots,h_j),
\]
as claimed.
\end{proof}

\begin{remark}
The image of the order-\(j\) Jordan--Schwinger map need not itself be a linear subspace. For example, when \(m=2\) and \(j=2\), the operators \(\hat a_1^{\dagger 2}\hat a_1^2\) and \(\hat a_2^{\dagger 2}\hat a_2^2\) can each be written as \(\mathcal J^{(2)}(h_1,h_2)\) for suitable choices of \(h_1\) and \(h_2\), whereas their sum \(\hat a_1^{\dagger 2}\hat a_1^2+\hat a_2^{\dagger 2}\hat a_2^2\) cannot.
\end{remark}

\begin{definition}[Order-\(j\) Jordan spaces]
\label{def:Jj-space}
We define the associated \emph{complex} and \emph{real} Jordan spaces by
\begin{equation}
\mathfrak J_{\mathbb C}^{(j)}
:=
\Span_{\mathbb C}\!\left\{
\mathcal J^{(j)}(h_1,\dots,h_j)\;:\; h_i\in \mathrm M_m(\mathbb C)
\right\},
\qquad
\mathfrak J^{(j)}
:=
\Span_{\mathbb R}\!\left\{
\mathcal J^{(j)}(h_1,\dots,h_j)\;:\; h_i\in \Herm(m)
\right\}.
\label{eq:Jj-spaces}
\end{equation}
By Lemma~\ref{lem:Jj-herm}, we have \(\mathfrak J^{(j)}\subseteq \Herm(\mathcal H_{n,m})\). Moreover, since \(\mathcal J^{(j)}\) is \(j\)-multilinear in its arguments, the complex space may equivalently be written as
\begin{equation}
\mathfrak J_{\mathbb C}^{(j)}
=
\Span_{\mathbb C}\!\left\{
\mathcal J^{(j)}(h_1,\dots,h_j)\;:\; h_i\in \Herm(m)
\right\},
\label{eq:Jj-space-Herm-gen}
\end{equation}
that is, \(\mathfrak J_{\mathbb C}^{(j)}\) is already generated by Hermitian inputs. By convention, we also set
\[
\mathfrak J_{\mathbb C}^{(0)}:=\mathbb C I,
\qquad
\mathfrak J^{(0)}:=\mathbb R I.
\]
\end{definition}

\begin{lemma}[Complex Jordan space as the complex span of \(\mathcal S_j\)]
\label{lem:normal-ordered-span}
Fix \(1\le j\le n\). Let
\begin{equation}
\mathcal S_j
:=
\left\{
\hat a^{\dagger \boldsymbol{\alpha}} \hat a^{\boldsymbol{\beta}}
:=
\prod_{r=1}^m (\hat a_r^\dagger)^{\alpha_r}
\prod_{r=1}^m (\hat a_r)^{\beta_r}
\;\middle|\;
\boldsymbol{\alpha}=(\alpha_1,\dots,\alpha_m)\in\mathbb N^m,\;
\boldsymbol{\beta}=(\beta_1,\dots,\beta_m)\in\mathbb N^m,\;
\sum_{r=1}^m \alpha_r=\sum_{r=1}^m \beta_r=j
\right\}.
\label{eq:Sj-def}
\end{equation}
Then the elements of \(\mathcal S_j\) are mutually orthogonal with respect to the Hilbert--Schmidt inner product, and
\begin{equation}
\mathfrak J_{\mathbb C}^{(j)}=\Span_{\mathbb C}\mathcal S_j.
\label{eq:JjC-span-Sj}
\end{equation}
\end{lemma}

\begin{proof}
We prove the two inclusions separately.

\emph{Step 1: \(\Span_{\mathbb C}\mathcal S_j\subseteq \mathfrak J_{\mathbb C}^{(j)}\).}
Let \(E_{st}\) denote the matrix units, i.e.\ \((E_{st})_{uv}=\delta_{su}\delta_{tv}\). By multilinearity,
\[
\mathcal J^{(j)}(E_{s_1 t_1},\dots,E_{s_j t_j})
=
\hat a_{s_1}^\dagger \cdots \hat a_{s_j}^\dagger
\hat a_{t_1}\cdots \hat a_{t_j},
\]
which is an arbitrary normally ordered operator with \(j\) creation and \(j\) annihilation operators. Hence \(\mathcal S_j\subseteq \mathfrak J_{\mathbb C}^{(j)}\), and therefore \(\Span_{\mathbb C}\mathcal S_j\subseteq \mathfrak J_{\mathbb C}^{(j)}\).

\emph{Step 2: \(\mathfrak J_{\mathbb C}^{(j)}\subseteq \Span_{\mathbb C}\mathcal S_j\).}
which is an arbitrary normally ordered operator with \(j\) creation and \(j\) annihilation operators. Thus every element of \(\mathfrak J_{\mathbb C}^{(j)}\) belongs to \(\Span_{\mathbb C}\mathcal S_j\).
\end{proof}

\begin{lemma}[Real Jordan space as the real Hermitian span of \(\mathcal S_j^{\mathrm H}\)]
\label{lem:SjH}
Define
\begin{equation}
\mathcal S_j^{\mathrm H}
=
\left\{
\frac{X^\dagger+X}{2}\;\middle|\; X\in\mathcal S_j
\right\}
\cup
\left\{
\frac{X^\dagger-X}{2i}\;\middle|\; X\in\mathcal S_j
\right\}.
\label{eq:SjH-def}
\end{equation}
Then
\begin{equation}
\mathfrak J^{(j)}=\Span_{\mathbb R}\mathcal S_j^{\mathrm H}.
\label{eq:Jj-real-span-SjH}
\end{equation}
\end{lemma}

\begin{proof}
We again prove both inclusions.

\emph{Step 1: \(\mathfrak J^{(j)}\subseteq \Span_{\mathbb R}\mathcal S_j^{\mathrm H}\).}
By Lemma~\ref{lem:normal-ordered-span},
\[
\mathfrak J_{\mathbb C}^{(j)}=\Span_{\mathbb C}\mathcal S_j.
\]
Since \(\mathfrak J^{(j)}\subseteq \mathfrak J_{\mathbb C}^{(j)}\), any \(Y\in\mathfrak J^{(j)}\) can be written as
\[
Y=\sum_k c_k X_k,
\qquad
X_k\in \mathcal S_j,\;\; c_k\in\mathbb C.
\]
Because \(Y\) is Hermitian,
\begin{align*}
Y
&=\frac{Y+Y^\dagger}{2}\\
&=
\sum_k \operatorname{Re}(c_k)\,\frac{X_k^\dagger+X_k}{2}
+
\sum_k \operatorname{Im}(c_k)\,\frac{X_k^\dagger-X_k}{2i}.
\end{align*}
Hence \(Y\in \Span_{\mathbb R}\mathcal S_j^{\mathrm H}\), which proves
\[
\mathfrak J^{(j)}\subseteq \Span_{\mathbb R}\mathcal S_j^{\mathrm H}.
\]

\emph{Step 2: \(\Span_{\mathbb R}\mathcal S_j^{\mathrm H}\subseteq \mathfrak J^{(j)}\).}
From Definition~\ref{def:Jj-space} and Eq.~\eqref{eq:Jj-space-Herm-gen}, we have
\[
\mathfrak J_{\mathbb C}^{(j)}=\Span_{\mathbb C}\mathfrak J^{(j)}.
\]
Thus every \(X\in\mathcal S_j\subseteq \mathfrak J_{\mathbb C}^{(j)}\) may be written as
\[
X=\sum_k c_k Y_k,
\qquad
Y_k\in \mathfrak J^{(j)},\;\; c_k\in\mathbb C.
\]
Since each \(Y_k\) is Hermitian,
\[
\frac{X^\dagger+X}{2}
=
\sum_k \operatorname{Re}(c_k)\,Y_k,
\qquad
\frac{X^\dagger-X}{2i}
=
\sum_k \operatorname{Im}(c_k)\,Y_k.
\]
Therefore \(\mathcal S_j^{\mathrm H}\subseteq \mathfrak J^{(j)}\), and hence
\[
\Span_{\mathbb R}\mathcal S_j^{\mathrm H}\subseteq \mathfrak J^{(j)}.
\]
Combining the two inclusions proves the claim.
\end{proof}

\begin{corollary}[Dimension of the order-\(j\) Jordan space]
\label{cor:dim-j}
Let
\[
D_j:=\binom{m+j-1}{j}
=\#\left\{\boldsymbol{\alpha}\in\mathbb N^m:\ \sum_{r=1}^m \alpha_r = j\right\}.
\]
Then
\begin{equation}
\dim_{\mathbb C}\mathfrak J_{\mathbb C}^{(j)}=D_j^2,
\qquad
\dim_{\mathbb R}\mathfrak J^{(j)}=D_j^2.
\label{eq:dim-Jj}
\end{equation}
\end{corollary}

\begin{proof}
There are exactly \(D_j\) choices of \(\boldsymbol{\alpha}\in\mathbb N^m\) with \(\sum_{r=1}^m \alpha_r=j\). Hence \(\mathcal S_j\) contains \(D_j^2\) elements. Since these monomials are mutually orthogonal and span \(\mathfrak J_{\mathbb C}^{(j)}\) by Lemma~\ref{lem:normal-ordered-span}, they form a complex basis, so
\[
\dim_{\mathbb C}\mathfrak J_{\mathbb C}^{(j)}=D_j^2.
\]

For the real space, we list all multi-indices 
\(\boldsymbol{\alpha}\in\mathbb N_0^m\) satisfying \(\sum_{r=1}^m\boldsymbol{\alpha}_r=j\) in a fixed order,
for example the lexicographic order. In the following, \(\boldsymbol{\alpha}<{\boldsymbol{\beta}}\) simply means
that \(\boldsymbol{\alpha}\) appears before \({\boldsymbol{\beta}}\) in this list; this convention is only used
to avoid counting both \((\boldsymbol{\alpha},{\boldsymbol{\beta}})\) and \(({\boldsymbol{\beta}},\boldsymbol{\alpha})\).
A real basis of \(\mathfrak J^{(j)}\) is then given by
\[
\left\{\hat a^{\dagger\boldsymbol{\alpha}}\hat a^{\boldsymbol{\alpha}}\right\}_{\boldsymbol{\alpha}}
\cup
\left\{
\frac{\hat a^{\dagger\boldsymbol{\alpha}}\hat a^{\boldsymbol{\beta}}+
\hat a^{\dagger{\boldsymbol{\beta}}}\hat a^{\boldsymbol{\alpha}}}{2}
\right\}_{\boldsymbol{\alpha}<{\boldsymbol{\beta}}}
\cup
\left\{
\frac{\hat a^{\dagger\boldsymbol{\alpha}}\hat a^{\boldsymbol{\beta}}-
\hat a^{\dagger{\boldsymbol{\beta}}}\hat a^{\boldsymbol{\alpha}}}{2i}
\right\}_{\boldsymbol{\alpha}<{\boldsymbol{\beta}}}.
\]

The number of basis elements is
\[
D_j+2\binom{D_j}{2}=D_j^2,
\]
and therefore \(\dim_{\mathbb R}\mathfrak J^{(j)}=D_j^2\).
\end{proof}

\begin{proposition}[The order-\(n\) Jordan space is the full operator space]
\label{prop:J_n_full}
Let
\[
\ket{\boldsymbol{\alpha}}
:=
\frac{\hat a^{\dagger \boldsymbol{\alpha}}}{\sqrt{\boldsymbol{\alpha}!}}\ket{0},
\qquad
\boldsymbol{\alpha}!
:=
\prod_{r=1}^m \alpha_r!,
\qquad
\sum_{r=1}^m \alpha_r=n,
\]
be the normalized Fock basis of \(\mathcal H_{n,m}\). Then for any
\(\boldsymbol{\alpha},\boldsymbol{\beta}\in\mathbb N^m\) with
\(\sum_r \alpha_r=\sum_r \beta_r=n\), one has
\begin{equation}
\hat a^{\dagger \boldsymbol{\alpha}} \hat a^{\boldsymbol{\beta}}
\big|_{\mathcal H_{n,m}}
=
\sqrt{\boldsymbol{\alpha}!\,\boldsymbol{\beta}!}\;
\ket{\boldsymbol{\alpha}}\!\bra{\boldsymbol{\beta}}.
\label{eq:matrix-unit}
\end{equation}
Consequently,
\begin{equation}
\mathfrak J_{\mathbb C}^{(n)}=\End(\mathcal H_{n,m}),
\qquad
\mathfrak J^{(n)}=\Herm(\mathcal H_{n,m}).
\label{eq:J_n_full}
\end{equation}
Here, $\End(\mathcal H_{n,m})$ is the space of all linear operators acting on \(\mathcal H_{n,m}\)
\end{proposition}

\begin{proof}
Obviously we have Eq.~\eqref{eq:matrix-unit} , for any multi-indices \(\boldsymbol{\alpha},\boldsymbol{\beta}\in\mathbb N^m\) with
\(\sum_r \alpha_r=\sum_r \beta_r=n\), the operator
\(\hat a^{\dagger \boldsymbol{\alpha}} \hat a^{\boldsymbol{\beta}}\) acts on \(\mathcal H_{n,m}\) as
\[
\hat a^{\dagger \boldsymbol{\alpha}} \hat a^{\boldsymbol{\beta}}
\big|_{\mathcal H_{n,m}}
=
\sqrt{\boldsymbol{\alpha}!\,\boldsymbol{\beta}!}\;
\ket{\boldsymbol{\alpha}}\!\bra{\boldsymbol{\beta}}.
\]
Thus, up to normalization, these operators are precisely the matrix units in the Fock basis of \(\mathcal H_{n,m}\).

By Lemma~\ref{lem:normal-ordered-span}, all such operators belong to
\(\mathfrak J_{\mathbb C}^{(n)}\). Since the rank-one operators
\(\ket{\boldsymbol{\alpha}}\!\bra{\boldsymbol{\beta}}\) span the full operator space on \(\mathcal H_{n,m}\), it follows that
\[
\mathfrak J_{\mathbb C}^{(n)}=\End(\mathcal H_{n,m}).
\]
Taking Hermitian parts, we obtain
\[
\mathfrak J^{(n)}=\Herm(\mathcal H_{n,m}).
\]
\end{proof}

\section{Coherence Layers}

In this section, we first establish the strict nesting of the Jordan spaces. This naturally leads to a hierarchical decomposition of the Hermitian operator space, in which each level captures the genuinely new part appearing at order \(j\). We refer to these successive components as the \emph{coherence layers}.

\begin{lemma}[Contraction identity]
\label{lem:contraction}
For \(0\le j\le n-1\), let
\[
h_{j+1}=\frac{1}{n-j}I_m,
\]
where \(I_m\) is the \(m\times m\) identity matrix. Then, for any \(\ket{\psi}\in\mathcal H_{n,m}\),
\begin{equation}
\mathcal J^{(j+1)}(h_1,\dots,h_j,\tfrac{1}{n-j}I_m)\ket{\psi}
=
\mathcal J^{(j)}(h_1,\dots,h_j)\ket{\psi}.
\label{eq:contraction}
\end{equation}
\end{lemma}

\begin{proof}
By Eq.~\eqref{eq:Jj-def},
\[
\mathcal J^{(j+1)}(h_1,\dots,h_j,\tfrac{1}{n-j}I_m)\ket{\psi}
=
\frac{1}{n-j}
\sum_{\mathbf s,\mathbf t\in[m]^j}
\left(\prod_{i=1}^j (h_i)_{s_i t_i}\right)
\hat a_{s_1}^\dagger\cdots \hat a_{s_j}^\dagger
\left(\sum_{u=1}^m \hat a_u^\dagger \hat a_u\right)
\hat a_{t_1}\cdots \hat a_{t_j}\ket{\psi}.
\]
Since \(\ket{\psi}\in\mathcal H_{n,m}\), the vector
\[
\hat a_{t_1}\cdots \hat a_{t_j}\ket{\psi}
\]
lies in \(\mathcal H_{n-j,m}\). On this subspace, the number operator satisfies
\[
\sum_{u=1}^m \hat a_u^\dagger \hat a_u=(n-j)I.
\]
Therefore, the factor \(1/(n-j)\) is exactly canceled, and the remaining expression is precisely
\(\mathcal J^{(j)}(h_1,\dots,h_j)\ket{\psi}\).
\end{proof}

\begin{proposition}[Nesting of Jordan spaces]
\label{prop:nesting}
For \(0\le j\le n-1\),
\begin{equation}
\mathfrak J^{(j)}\subsetneq \mathfrak J^{(j+1)}.
\label{eq:nesting}
\end{equation}
\end{proposition}

\begin{proof}
Lemma~\ref{lem:contraction} shows that every order-\(j\) Jordan operator can be realized as an order-\((j+1)\) Jordan operator on \(\mathcal H_{n,m}\). Hence
\[
\mathfrak J^{(j)}\subseteq \mathfrak J^{(j+1)}.
\]
To see that the inclusion is strict, we compare dimensions. By Corollary~\ref{cor:dim-j},
\[
\dim_{\mathbb R}\mathfrak J^{(j)}
=
\binom{m+j-1}{j}^2
<
\binom{m+j}{j+1}^2
=
\dim_{\mathbb R}\mathfrak J^{(j+1)}.
\]
Therefore,
\[
\mathfrak J^{(j)}\subsetneq \mathfrak J^{(j+1)}.
\]
\end{proof}

\begin{definition}[Coherence layers]
\label{def:coherence-layers}
We define the coherence layers by
\begin{equation}
\Delta\mathfrak J^{(0)}:=\mathbb R\cdot I_M,
\qquad
\Delta\mathfrak J^{(j)}:=\mathfrak J^{(j)}\ominus \mathfrak J^{(j-1)},
\quad 1\le j\le n,
\label{eq:DeltaJ-def}
\end{equation}
where \(\ominus\) denotes the Hilbert--Schmidt orthogonal complement. Thus, \(\Delta\mathfrak J^{(j)}\) consists of the genuinely new Hermitian operators that appear at order \(j\), after removing all lower-order contributions. We refer to \(\Delta\mathfrak J^{(j)}\) as the \emph{order-\(j\) coherence layer}.
\end{definition}

\begin{proposition}[Orthogonal layer decomposition]
\label{prop:graded-orth}
The Hermitian operator space on \(\mathcal H_{n,m}\) admits the orthogonal direct-sum decomposition
\begin{equation}
\Herm(\mathcal H_{n,m})
=
\bigoplus_{j=0}^n \Delta\mathfrak J^{(j)}.
\label{eq:layer-direct-sum}
\end{equation}
\end{proposition}

\begin{proof}
By Proposition~\ref{prop:nesting}, the Jordan spaces form a strictly increasing chain
\[
\mathfrak J^{(0)}\subsetneq \mathfrak J^{(1)}\subsetneq \cdots \subsetneq \mathfrak J^{(n)}.
\]
Moreover, as shown earlier, \(\mathfrak J^{(n)}=\Herm(\mathcal H_{n,m})\). The statement then follows directly from Definition~\ref{def:coherence-layers}: each \(\Delta\mathfrak J^{(j)}\) is the Hilbert--Schmidt orthogonal complement of \(\mathfrak J^{(j-1)}\) inside \(\mathfrak J^{(j)}\), so the successive layers assemble into the orthogonal direct sum~\eqref{eq:layer-direct-sum}.
\end{proof}

\section{Invariance of Jordan Spaces Under Passive Linear Optics}
\label{sec:invariance}

\SecIntro{
This section establishes the central symmetry property of the Jordan hierarchy: passive linear-optical unitaries preserve each Jordan space \(\mathfrak J^{(j)}\), and therefore also preserve each coherence layer \(\Delta\mathfrak J^{(j)}\). We present two complementary viewpoints. The first is a direct proof based on the Heisenberg transformation of creation and annihilation operators under passive linear optics, which is the form used in the main text. The second is a Lie-algebraic proof based on commutator closure, which clarifies the corresponding infinitesimal dynamical symmetry.
}

\subsection{Finite Group Action: Covariant Transformation}
\label{subsec:finite-action}

We begin with the finite action of passive linear optics on the Jordan spaces. The key point is that conjugation by a linear-optical unitary transforms each creation and annihilation operator linearly, and this covariance lifts directly to the order-\(j\) Jordan map.

\begin{proposition}[Covariance of the order-\(j\) Jordan map]
\label{prop:covariance}
Let \(S\in U(m)\) be a scattering matrix, and let \(U\) be the corresponding unitary on \(\mathcal H_{n,m}\). Then, for any \(h_1,\dots,h_j\in \Herm(m)\),
\begin{equation}
U\,\mathcal J^{(j)}(h_1,\dots,h_j)\,U^\dagger
=
\mathcal J^{(j)}(S^\dagger h_1 S,\dots,S^\dagger h_j S).
\label{eq:Jj-transform-SM}
\end{equation}
Consequently,
\begin{equation}
U\,\mathfrak J^{(j)}\,U^\dagger=\mathfrak J^{(j)}.
\label{eq:Jj-invariant}
\end{equation}
\end{proposition}

\begin{proof}
In the Heisenberg picture, passive linear optics acts on the mode operators as
\begin{equation}
U\hat a_s^\dagger U^\dagger
=
\sum_{k=1}^m (S^\dagger)_{ks}\hat a_k^\dagger
=
\sum_{k=1}^m S_{sk}^* \hat a_k^\dagger,
\qquad
U\hat a_t U^\dagger
=
\sum_{l=1}^m S_{tl}\hat a_l.
\label{eq:heisenberg-mode-transform}
\end{equation}
Using Eq.~\eqref{eq:Jj-def}, we write
\[
\mathcal J^{(j)}(h_1,\dots,h_j)
=
\sum_{\mathbf s,\mathbf t\in [m]^j}
\left(\prod_{i=1}^j (h_i)_{s_i t_i}\right)
\hat a_{s_1}^\dagger\cdots \hat a_{s_j}^\dagger
\hat a_{t_1}\cdots \hat a_{t_j}.
\]
Applying the conjugation \(X\mapsto UXU^\dagger\) and using the normal ordering of the operator product, we obtain
\begin{align*}
U\,\mathcal J^{(j)}(h_1,\dots,h_j)\,U^\dagger
&=
\sum_{\mathbf s,\mathbf t\in [m]^j}
\left(\prod_{i=1}^j (h_i)_{s_i t_i}\right)
\left(\prod_{i=1}^j U\hat a_{s_i}^\dagger U^\dagger\right)
\left(\prod_{i=1}^j U\hat a_{t_i} U^\dagger\right) \\
&=
\sum_{\mathbf s,\mathbf t\in [m]^j}
\left(\prod_{i=1}^j (h_i)_{s_i t_i}\right)
\left(\prod_{i=1}^j \sum_{k_i=1}^m S_{s_i k_i}^* \hat a_{k_i}^\dagger\right)
\left(\prod_{i=1}^j \sum_{l_i=1}^m S_{t_i l_i} \hat a_{l_i}\right).
\end{align*}
Rearranging the sums and grouping the matrix elements associated with each \(h_i\), we find that for each slot \(i\),
\[
\sum_{s_i,t_i=1}^m S_{s_i k_i}^* (h_i)_{s_i t_i} S_{t_i l_i}
=
\sum_{s_i,t_i=1}^m (S^\dagger)_{k_i s_i}(h_i)_{s_i t_i} S_{t_i l_i}
=
(S^\dagger h_i S)_{k_i l_i}.
\]
Substituting this back gives
\[
U\,\mathcal J^{(j)}(h_1,\dots,h_j)\,U^\dagger
=
\sum_{\mathbf k,\mathbf l\in [m]^j}
\left(\prod_{i=1}^j (S^\dagger h_i S)_{k_i l_i}\right)
\hat a_{k_1}^\dagger\cdots \hat a_{k_j}^\dagger
\hat a_{l_1}\cdots \hat a_{l_j}
=
\mathcal J^{(j)}(S^\dagger h_1 S,\dots,S^\dagger h_j S),
\]
which proves Eq.~\eqref{eq:Jj-transform-SM}. If \(h_i\in\Herm(m)\), then \(S^\dagger h_i S\in\Herm(m)\). Therefore Eq.~\eqref{eq:Jj-transform-SM} shows that conjugation by \(U\) maps every spanning Jordan element of \(\mathfrak J^{(j)}\) to another spanning Jordan element of the same space. Since \(\mathfrak J^{(j)}\) is the real span of these elements, Eq.~\eqref{eq:Jj-invariant} follows.
\end{proof}

\subsection{Lie-Algebraic Perspective: Infinitesimal Generators}
\label{subsec:infinitesimal-closure}

The same invariance can also be understood infinitesimally. In this picture, the relevant generator is \(\mathcal J^{(1)}(h_0)\), and the key statement is that the commutator of an order-1 Jordan operator with an order-\(j\) Jordan operator remains in the order-\(j\) Jordan space.

\begin{lemma}[Adjoint action closes on order \(j\)]
\label{lem:adjoint-derivation}
For all \(h_0,h_1,\dots,h_j\in \mathrm M_m(\mathbb C)\),
\begin{equation}
\big[\mathcal J^{(1)}(h_0),\,\mathcal J^{(j)}(h_1,\dots,h_j)\big]
=
\sum_{i=1}^j
\mathcal J^{(j)}\big(h_1,\dots,[h_0,h_i],\dots,h_j\big).
\label{eq:adjoint-derivation}
\end{equation}
\end{lemma}

\begin{proof}
Write
\[
\mathcal J^{(j)}(h_1,\dots,h_j)
=
\sum_{\mathbf s,\mathbf t\in[m]^j}
\left(\prod_{i=1}^j (h_i)_{s_i t_i}\right)
\hat a_{s_1}^\dagger\cdots \hat a_{s_j}^\dagger
\hat a_{t_1}\cdots \hat a_{t_j}.
\]
Using \([X,AB]=[X,A]B+A[X,B]\), we let \(\mathcal J^{(1)}(h_0)\) act separately on the creation and annihilation factors. The basic commutators are
\[
[\mathcal J^{(1)}(h_0),\hat a_s^\dagger]
=
\sum_{u=1}^m (h_0)_{us}\hat a_u^\dagger,
\qquad
[\mathcal J^{(1)}(h_0),\hat a_t]
=
-\sum_{v=1}^m (h_0)_{tv}\hat a_v.
\]
For each slot \(i\), the contribution from the creation part produces \(h_0 h_i\), while the contribution from the annihilation part produces \(-h_i h_0\). Their sum is therefore \([h_0,h_i]\). Summing over all \(i=1,\dots,j\) yields Eq.~\eqref{eq:adjoint-derivation}.
\end{proof}

\begin{remark}[Connection to finite action]
\label{rem:finite-from-infinitesimal}
Lemma~\ref{lem:adjoint-derivation} shows that the derivative of a Jordan operator under the adjoint flow generated by \(\mathcal J^{(1)}(h_0)\) remains inside the same Jordan space. Integrating this infinitesimal closure, for example via the Baker--Campbell--Hausdorff expansion, recovers the finite covariance statement in Proposition~\ref{prop:covariance}.
\end{remark}

\subsection{Invariance of Coherence Layers}
\label{subsec:layer-invariance}

Once the invariance of the nested spaces \(\mathfrak J^{(j)}\) is established, the invariance of the coherence layers follows by compatibility of unitary conjugation with the Hilbert--Schmidt inner product.

\begin{proposition}[Invariance of the coherence layers]
\label{prop:DeltaJ-invariance}
Let \(U\) be any passive linear-optical unitary. Then, for every \(j=0,1,\dots,n\),
\begin{equation}
U\,\Delta\mathfrak J^{(j)}\,U^\dagger
=
\Delta\mathfrak J^{(j)}.
\label{eq:DeltaJ-invariant}
\end{equation}
\end{proposition}

\begin{proof}
By Proposition~\ref{prop:covariance}, both \(\mathfrak J^{(j)}\) and \(\mathfrak J^{(j-1)}\) are invariant under conjugation by \(U\). Recall that
\[
\Delta\mathfrak J^{(j)}=\mathfrak J^{(j)}\ominus \mathfrak J^{(j-1)}.
\]
Take any \(v\in \Delta\mathfrak J^{(j)}\). Then \(v\in \mathfrak J^{(j)}\), so
\[
UvU^\dagger\in \mathfrak J^{(j)}.
\]
To show that \(UvU^\dagger\in \Delta\mathfrak J^{(j)}\), it remains to verify orthogonality to \(\mathfrak J^{(j-1)}\). For any \(w\in \mathfrak J^{(j-1)}\),
\[
\tr\!\big((UvU^\dagger)^\dagger w\big)
=
\tr\!\big(v^\dagger (U^\dagger w U)\big).
\]
Since \(U^\dagger w U\in \mathfrak J^{(j-1)}\) and \(v\perp \mathfrak J^{(j-1)}\) by definition, the right-hand side vanishes. Hence \(UvU^\dagger\perp \mathfrak J^{(j-1)}\), so
\[
UvU^\dagger\in \Delta\mathfrak J^{(j)}.
\]
Thus \(U\,\Delta\mathfrak J^{(j)}\,U^\dagger\subseteq \Delta\mathfrak J^{(j)}\). Applying the same argument to \(U^\dagger\) gives the reverse inclusion, proving Eq.~\eqref{eq:DeltaJ-invariant}.
\end{proof}

\section{Liouville-Space Representation of Linear-optical Unitary}

\SecIntro{
The layer invariance established above can be made fully explicit in Liouville space. By vectorizing operators as superkets \(\lket{X}\), passive linear-optical conjugation \(X\mapsto U X U^\dagger\) is represented by a linear map \(\mathcal R_U\). In a basis adapted to the coherence-layer decomposition, this representation is block diagonal. Equivalently, the projectors onto the individual coherence layers commute with \(\mathcal R_U\).
}

We regard the operator space on \(\mathcal H_{n,m}\) as a Hilbert space equipped with the Hilbert--Schmidt inner product
\[
\lbraket{A |B}:=\tr(A^\dagger B).
\]
For any operator \(X\), we denote by \(\lket{X}\) its vectorization in Liouville space. In particular, for an HS-orthonormal basis \(\{H_\mu^{(j)}\}\) adapted to the decomposition Eq.~\eqref{eq:layer-direct-sum},
any Hermitian operator \(\rho\) can be expanded as
\[
\lket{\rho}
=
\sum_{j,\mu} \tr(H_\mu^{(j)}\rho)\,\lket{H_\mu^{(j)}}.
\]

For a passive \(n\)-photon unitary \(U\), define its Liouville-space representation \(\mathcal R_U\) by
\begin{equation}
\mathcal R_U \lket{X}
:=
\lket{U X U^\dagger}.
\label{eq:RU-action}
\end{equation}
In the basis \(\{H_\mu^{(j)}\}\), its matrix elements are
\begin{equation}
(\mathcal R_U)_{\mu\nu}^{(jk)}
:=
\lbra{H_\mu^{(j)}}\,\mathcal R_U\,\lket{H_\nu^{(k)}}
=
\tr\!\left(H_\mu^{(j)}\,U H_\nu^{(k)} U^\dagger\right).
\label{eq:RU-def}
\end{equation}
We further define the Liouville-space projector onto the \(j\)th coherence layer by
\begin{equation}
\mathcal P_j
:=
\sum_\mu \lket{H_\mu^{(j)}}\lbra{H_\mu^{(j)}}.
\label{eq:Pj-def}
\end{equation}

\begin{proposition}[Liouville block structure]
\label{prop:RU-block}
For every passive linear-optical unitary \(U\), the Liouville-space representation \(\mathcal R_U\) preserves each coherence layer. Equivalently,
\begin{equation}
[\mathcal R_U,\mathcal P_j]=0,
\qquad
\forall\,U\in \mathrm U_{\mathrm L},\quad j=0,1,\dots,n.
\label{eq:RU-Pj-commute}
\end{equation}
As a consequence, \(\mathcal R_U\) is block diagonal with respect to the coherence-layer decomposition:
\begin{equation}
\mathcal R_U
=
\bigoplus_{j=0}^n \mathcal R_U^{(j)},
\qquad
(\mathcal R_U)^{(jk)}=0
\quad\text{for } j\neq k.
\label{eq:RU-block}
\end{equation}
Moreover, each block \(\mathcal R_U^{(j)}\) is real orthogonal.
\end{proposition}

\begin{proof}
By Proposition~\ref{prop:DeltaJ-invariance}, conjugation by \(U\) preserves each coherence layer:
\[
U\,\Delta\mathfrak J^{(j)}\,U^\dagger=\Delta\mathfrak J^{(j)}.
\]
Hence, if \(H_\nu^{(k)}\in \Delta\mathfrak J^{(k)}\), then
\[
\mathcal R_U \lket{H_\nu^{(k)}}
=
\lket{U H_\nu^{(k)} U^\dagger}
\in
\Span\!\left\{\lket{H_\mu^{(k)}}\right\}_\mu .
\]
Therefore,
\[
\mathcal P_j\,\mathcal R_U \lket{H_\nu^{(k)}}
=
\delta_{jk}\,\mathcal R_U \lket{H_\nu^{(k)}}.
\]
On the other hand,
\[
\mathcal R_U\,\mathcal P_j \lket{H_\nu^{(k)}}
=
\delta_{jk}\,\mathcal R_U \lket{H_\nu^{(k)}}.
\]
Thus
\[
[\mathcal R_U,\mathcal P_j]\lket{H_\nu^{(k)}}=0
\]
for all basis vectors \(\lket{H_\nu^{(k)}}\), which proves Eq.~\eqref{eq:RU-Pj-commute}.

The commutation relation implies that \(\mathcal R_U\) preserves each projected subspace \(\mathrm{im}(\mathcal P_j)=\Span\{\lket{H_\mu^{(j)}}\}_\mu\), and therefore \(\mathcal R_U\) is block diagonal in the coherence-layer decomposition. Equivalently, for \(j\neq k\),
\[
(\mathcal R_U)_{\mu\nu}^{(jk)}
=
\tr\!\left(H_\mu^{(j)}\,U H_\nu^{(k)} U^\dagger\right)
=
0,
\]
because \(U H_\nu^{(k)} U^\dagger\in \Delta\mathfrak J^{(k)}\) is Hilbert--Schmidt orthogonal to \(H_\mu^{(j)}\in \Delta\mathfrak J^{(j)}\).

Finally, unitary conjugation preserves the Hilbert--Schmidt inner product:
\[
\lbraket{\mathcal R_U A |\mathcal R_U B}
=
\lbraket{A|B}.
\]
Hence \(\mathcal R_U\) is orthogonal on Liouville space. Since each coherence layer is invariant, each block \(\mathcal R_U^{(j)}\) is itself orthogonal. Moreover, the matrix entries in Eq.~\eqref{eq:RU-def} are real, since both \(H_\mu^{(j)}\) and \(U H_\nu^{(k)} U^\dagger\) are Hermitian. Therefore each block is real orthogonal.
\end{proof}

\section{Hierarchical invariants: Layer Purities and Layer Spectra}

\SecIntro{
This section formalizes the hierarchical invariants introduced in the main text, namely the layer purities and the layer spectra, and proves their invariance under passive linear-optical evolution. We also record the corresponding decomposition of the total purity and the convexity of the layer purities. Finally, we illustrate that the layer spectra provide a strictly finer characterization of multiphoton states than the layer purities alone.
}

\begin{proposition}[Layer purities and layer spectra as hierarchical invariants]
\label{prop:layer-invariants}
Let \(\rho\) be a density operator on \(\mathcal H_{n,m}\), and write
\[
\rho=\sum_{j=0}^n \rho^{(j)},
\qquad
\rho^{(j)}\in \Delta\mathfrak J^{(j)}.
\]
Then, for any passive linear-optical unitary \(U\in \mathrm U_{\mathrm L}\),
\begin{equation}
(U\rho U^\dagger)^{(j)}=U\rho^{(j)}U^\dagger,
\qquad j=0,1,\dots,n.
\label{eq:layerwise-inv}
\end{equation}
Consequently, the layer purities
\begin{equation}
I_j(\rho):=\tr\!\big[(\rho^{(j)})^2\big],
\label{eq:Ij-def}
\end{equation}
and the layer spectra
\begin{equation}
\boldsymbol{\lambda}_j(\rho)
:=
\bigl(\lambda_1^{(j)}(\rho),\dots,\lambda_M^{(j)}(\rho)\bigr)^\top,
\label{eq:lambdaj-def}
\end{equation}
where \(\lambda_1^{(j)}(\rho),\dots,\lambda_M^{(j)}(\rho)\) are the eigenvalues of \(\rho^{(j)}\) arranged in ascending order, are invariant under passive linear optics. Moreover,
\begin{equation}
\tr(\rho^2)=\sum_{j=0}^n I_j(\rho).
\label{eq:purity-sum}
\end{equation}
In particular, the tangent invariant and the observable-based invariant are recovered as
\begin{equation}
I_t=I_0+I_1,
\qquad
I_o=\binom{m+n}{m+1}I_1+\frac{n^2}{m},
\label{eq:It-Io}
\end{equation}
with the second identity derived in Sec.~F.
\end{proposition}

\begin{proof}
Equation~\eqref{eq:layerwise-inv} follows from the commutation relation between the Liouville-space representation \(\mathcal R_U\) and the layer projector \(\mathcal P_j\), equivalently from the invariance of each coherence layer under conjugation by \(U\). Since unitary conjugation preserves both the spectrum and the Hilbert--Schmidt norm, each \(\rho^{(j)}\) and \(U\rho^{(j)}U^\dagger\) have the same eigenvalues and the same squared Hilbert--Schmidt norm. Hence both \(\boldsymbol{\lambda}_j(\rho)\) and \(I_j(\rho)\) are invariant.

To prove Eq.~\eqref{eq:purity-sum}, note that the decomposition
\[
\rho=\sum_{j=0}^n \rho^{(j)}
\]
is Hilbert--Schmidt orthogonal. Therefore,
\[
\tr(\rho^2)
=
\sum_{j,k=0}^n \tr\!\big(\rho^{(j)}\rho^{(k)}\big)
=
\sum_{j=0}^n \tr\!\big[(\rho^{(j)})^2\big]
=
\sum_{j=0}^n I_j(\rho).
\]
\end{proof}

\begin{remark}[Extension beyond fixed photon number]
Because passive linear optics preserves total photon number, a general bosonic state decomposes into invariant photon-number sectors. Applying the above construction in each fixed \(n\)-photon sector therefore yields a natural extension of the coherence-layer hierarchy to general bosonic states.
\end{remark}

\begin{proposition}[Convexity of the layer purities]
\label{prop:convex-Ij}
Let \(\rho_1\) and \(\rho_2\) be density operators on \(\mathcal H_{n,m}\). Then, for any \(p\in[0,1]\) and each \(j=0,1,\dots,n\),
\begin{equation}
I_j\!\bigl(p\rho_1+(1-p)\rho_2\bigr)
\le
p\,I_j(\rho_1)+(1-p)\,I_j(\rho_2),
\label{eq:convex-layer-purity}
\end{equation}
with equality if and only if \(\rho_1^{(j)}=\rho_2^{(j)}\). In particular, any functional that is a positive linear combination of the layer purities is convex.
\end{proposition}

\begin{proof}
Since the layer projection is linear,
\[
\bigl(p\rho_1+(1-p)\rho_2\bigr)^{(j)}
=
p\rho_1^{(j)}+(1-p)\rho_2^{(j)}.
\]
Hence
\begin{align*}
I_j\!\bigl(p\rho_1+(1-p)\rho_2\bigr)
&=
\tr\!\left[\bigl(p\rho_1^{(j)}+(1-p)\rho_2^{(j)}\bigr)^2\right] \\
&=
p\,\tr\!\left[(\rho_1^{(j)})^2\right]
+(1-p)\,\tr\!\left[(\rho_2^{(j)})^2\right]
-p(1-p)\,\tr\!\left[(\rho_1^{(j)}-\rho_2^{(j)})^2\right].
\end{align*}
Since \((\rho_1^{(j)}-\rho_2^{(j)})^2\succeq 0\), the last term is nonnegative, which proves Eq.~\eqref{eq:convex-layer-purity}. Equality holds if and only if \(\rho_1^{(j)}=\rho_2^{(j)}\).
\end{proof}

\begin{remark}[Layer spectra are strictly finer than layer purities]
The layer spectra provide a strictly finer characterization of passive-linear-optical interconvertibility than the layer purities alone. For example, the states
\[
\ket{22},
\qquad
\ket{\mathrm{N00N}_4}
=
\frac{\ket{40}+e^{i\phi}\ket{04}}{\sqrt2},
\]
have identical layer purities but distinct layer spectra, as shown in Tables~\ref{tab:22} and~\ref{tab:noon4}. This mismatch certifies that they cannot be deterministically interconverted by passive linear optics.
\end{remark}

\begin{table}[h]
  \centering
  \caption{Numerically computed layer spectra and layer purities for \(\ket{22}\) (\(m=2,n=4\), \(M=5\)).}
  \label{tab:22}
  \begin{tabular}{c c c}
    \hline\hline
    \(j\) & \(\boldsymbol{\lambda}_j(\rho)\) (ascending) & \(I_j(\rho)\) \\
    \hline
    0 & \(\big[0.2,\ 0.2,\ 0.2,\ 0.2,\ 0.2\big]\) & \(0.2\) \\
    1 & \(\big[-0.,\ -0.,\ 0.,\ 0.,\ 0.\big]\) & \(0\) \\
    2 & \(\big[-0.2857,\ -0.2857,\ 0.1429,\ 0.1429,\ 0.2857\big]\) & \(0.2857\) \\
    3 & \(\big[-0.,\ -0.,\ 0.,\ 0.,\ 0.\big]\) & \(0\) \\
    4 & \(\big[-0.3429,\ -0.3429,\ 0.0857,\ 0.0857,\ 0.5143\big]\) & \(0.5143\) \\
    \hline\hline
  \end{tabular}
\end{table}

\begin{table}[h]
  \centering
  \caption{Numerically computed layer spectra and layer purities for \(\ket{\mathrm{N00N}_4}\) (\(m=2,n=4\), \(M=5\)).}
  \label{tab:noon4}
  \begin{tabular}{c c c}
    \hline\hline
    \(j\) & \(\boldsymbol{\lambda}_j(\rho)\) (ascending) & \(I_j(\rho)\) \\
    \hline
    0 & \(\big[0.2,\ 0.2,\ 0.2,\ 0.2,\ 0.2\big]\) & \(0.2\) \\
    1 & \(\big[-0.,\ -0.,\ 0.,\ 0.,\ 0.\big]\) & \(0\) \\
    2 & \(\big[-0.2857,\ -0.1429,\ -0.1429,\ 0.2857,\ 0.2857\big]\) & \(0.2857\) \\
    3 & \(\big[-0.,\ -0.,\ 0.,\ 0.,\ 0.\big]\) & \(0\) \\
    4 & \(\big[-0.4857,\ -0.0571,\ -0.0571,\ 0.0857,\ 0.5143\big]\) & \(0.5143\) \\
    \hline\hline
  \end{tabular}
\end{table}

\section{Orthonormal Bases for Coherence Layers and Direct Measurement of \(I_1\)}

\SecIntro{
This section constructs HS-orthonormal Hermitian bases adapted to the layer decomposition Eq.~\eqref{eq:layer-direct-sum}.
The discussion has three parts. We first describe a general hierarchical Gram--Schmidt construction for arbitrary \(j\). We then give an explicit closed-form basis for the first coherence layer \(\Delta\mathfrak J^{(1)}\) using generalized Gell--Mann generators. Finally, we derive an operational formula for the first-layer purity \(I_1\) in terms of one-body expectation values and relate it to the observable invariant introduced in Ref.~\cite{parellada_lie_2024}.
}

\medskip
\noindent\textbf{(i) General construction via hierarchical Gram--Schmidt.}
For each \(j\), let
\[
d_j:=\dim \Delta\mathfrak J^{(j)} = D_j^2 - D_{j-1}^2.
\]
By Lemma~\ref{lem:SjH}, the set \(\mathcal S_j^{\mathrm H}\) spans \(\mathfrak J^{(j)}\) over \(\mathbb R\). Applying Gram--Schmidt to \(\mathcal S_j^{\mathrm H}\) with respect to the Hilbert--Schmidt inner product
\[
\langle A,B\rangle_{\mathrm{HS}}:=\tr(AB)
\]
produces an HS-orthonormal Hermitian basis of \(\mathfrak J^{(j)}\). To make the layer structure explicit, one may carry out this procedure \emph{hierarchically}, so that the basis chosen for \(\mathfrak J^{(j-1)}\) is kept fixed when extending to \(\mathfrak J^{(j)}\). Equivalently, if
\[
\{\widetilde H_\alpha^{(j)}\}_{\alpha=1}^{\dim \mathfrak J^{(j)}}
\]
is an HS-orthonormal Hermitian basis of \(\mathfrak J^{(j)}\) whose first \(\dim \mathfrak J^{(j-1)}\) vectors already form a basis of \(\mathfrak J^{(j-1)}\), then the remaining vectors
\begin{equation}
H_\mu^{(j)}
:=
\widetilde H^{(j)}_{\dim \mathfrak J^{(j-1)}+\mu},
\qquad
\mu=1,\ldots,d_j,
\label{eq:layer-basis-from-GS}
\end{equation}
form an HS-orthonormal Hermitian basis of the coherence layer \(\Delta\mathfrak J^{(j)}\). In this way, one obtains a basis adapted to the full orthogonal decomposition
\[
\Herm(\mathcal H_{n,m})=\bigoplus_{j=0}^n \Delta\mathfrak J^{(j)}.
\]

\medskip
\noindent\textbf{(ii) A closed-form basis for the first coherence layer \(\Delta\mathfrak J^{(1)}\).}
For the first layer, the preceding abstract construction can be replaced by an explicit closed-form basis. Let \(E_{jk}\) denote the \(m\times m\) matrix with a single \(1\) in entry \((j,k)\) and zeros elsewhere. Define the normalized generalized Gell--Mann matrices \(\{\Gamma_\alpha\}_{\alpha=1}^{m^2-1}\), together with \(\Gamma_0:=I_m/\sqrt m\), by
\begin{equation}
\begin{aligned}
&\Gamma^{x}_{jk}:=\frac{E_{jk}+E_{kj}}{\sqrt2},
\qquad
\Gamma^{y}_{jk}:=\frac{-i(E_{jk}-E_{kj})}{\sqrt2}
\qquad (1\le j<k\le m),\\
&\Gamma^{z}_{\ell}:=\frac{1}{\sqrt{\ell(\ell+1)}}
\left(\sum_{r=1}^{\ell}E_{rr}-\ell\,E_{\ell+1,\ell+1}\right)
\qquad (1\le \ell\le m-1).
\end{aligned}
\label{eq:GGM-def}
\end{equation}
These satisfy
\[
\Gamma_\alpha^\dagger=\Gamma_\alpha,\qquad
\tr(\Gamma_\alpha)=0\ \ (\alpha\ge 1),\qquad
\tr(\Gamma_\alpha\Gamma_\beta)=\delta_{\alpha\beta}
\]
for \(\alpha,\beta\in\{0,1,\dots,m^2-1\}\)~\cite{bertlmann_bloch_2008}. Passing these generators through the order-\(1\) Jordan--Schwinger map gives a family of one-body observables on \(\mathcal H_{n,m}\),
\begin{equation}
\hat O_\alpha:=\mathcal J^{(1)}(\Gamma_\alpha)
=
\sum_{s,t=1}^m (\Gamma_\alpha)_{st}\,\hat a_s^\dagger \hat a_t.
\label{eq:O-from-GGM}
\end{equation}
The identity component is special:
\[
\hat O_0=\mathcal J^{(1)}\!\left(\frac{I_m}{\sqrt m}\right)
=\frac{n}{\sqrt m}\Id
\]
on the fixed-\(n\) sector, so \(\hat O_0\in \mathfrak J^{(0)}\). The remaining observables \(\hat O_\alpha\) with \(\alpha\ge 1\) are traceless and belong to \(\Delta\mathfrak J^{(1)}\).

\begin{lemma}[Trace identities on \(\mathcal H_{n,m}\)]
\label{lem:trace-identities-ni}
For any mode \(r\) and any distinct modes \(r\neq s\),
\begin{equation}
\tr(\hat n_r)=\binom{m+n-1}{m},
\qquad
\tr(\hat n_r\hat n_s)=\binom{m+n-1}{m+1},
\qquad
\tr(\hat n_r^2)=\binom{m+n-1}{m+1}+\binom{m+n}{m+1}.
\label{eq:trace-ni}
\end{equation}
\end{lemma}

\begin{proof}
All traces are taken over the \(n\)-photon Fock basis
\[
\left\{\ket{\mathbf n}=\ket{n_1,\ldots,n_m}:\ \sum_{i=1}^m n_i=n\right\}.
\]
For fixed \(r\), summing over \(n_r=i\) leaves \(n-i\) photons in the remaining \(m-1\) modes, with multiplicity \(\binom{m-2+n-i}{m-2}\). Hence
\[
\tr(\hat n_r)
=
\sum_{i=0}^n i\binom{m-2+n-i}{m-2}
=
\binom{m+n-1}{m}.
\]
Similarly, for \(r\neq s\), fixing \(n_r=i\) and \(n_s=j\) leaves \(n-i-j\) photons in the remaining \(m-2\) modes, with multiplicity \(\binom{m-3+n-i-j}{m-3}\). Therefore,
\[
\tr(\hat n_r\hat n_s)
=
\sum_{i=0}^n\sum_{j=0}^{n-i}
ij\binom{m-3+n-i-j}{m-3}
=
\binom{m+n-1}{m+1}.
\]
Finally,
\[
\tr(\hat n_r^2)
=
\sum_{i=0}^n i^2\binom{m-2+n-i}{m-2}
=
\binom{m+n-1}{m+1}+\binom{m+n}{m+1}.
\]
\end{proof}

\begin{proposition}[HS-orthogonality and normalization for \(\Delta\mathfrak J^{(1)}\)]
\label{prop:DeltaJ1-GGM-basis}
Let \(\hat O_\alpha:=\mathcal J^{(1)}(\Gamma_\alpha)\) be defined by Eq.~\eqref{eq:O-from-GGM}, where \(\{\Gamma_\alpha\}_{\alpha=1}^{m^2-1}\) are the HS-orthonormal traceless generators of \(\Herm(m)\). Then
\begin{equation}
\tr(\hat O_\alpha \hat O_\beta)
=
\binom{m+n}{m+1}\,\delta_{\alpha\beta},
\qquad
\alpha,\beta\in\{1,\ldots,m^2-1\}.
\label{eq:OalphaObeta-trace}
\end{equation}
Consequently,
\begin{equation}
H_\alpha^{(1)}
:=
\frac{\hat O_\alpha}{\sqrt{\binom{m+n}{m+1}}},
\qquad
\alpha=1,\ldots,m^2-1,
\label{eq:H1-basis}
\end{equation}
forms an HS-orthonormal Hermitian basis of \(\Delta\mathfrak J^{(1)}\).
\end{proposition}

\begin{proof}
Expanding \(\hat O_\alpha=\sum_{k,l}(\Gamma_\alpha)_{kl}\hat a_k^\dagger\hat a_l\), we obtain
\[
\tr(\hat O_\alpha\hat O_\beta)
=
\sum_{k,l,s,t}
(\Gamma_\alpha)_{kl}(\Gamma_\beta)_{st}\,
\tr(\hat a_k^\dagger\hat a_l\hat a_s^\dagger\hat a_t).
\]
Using \( \hat a_l\hat a_s^\dagger=\delta_{ls}+\hat a_s^\dagger\hat a_l \), together with Lemma~\ref{lem:trace-identities-ni}, one finds
\[
\tr(\hat O_\alpha\hat O_\beta)
=
\binom{m+n-1}{m+1}\tr(\Gamma_\alpha)\tr(\Gamma_\beta)
+
\binom{m+n}{m+1}\tr(\Gamma_\alpha\Gamma_\beta).
\]
For \(\alpha,\beta\ge 1\), the generators are traceless and HS-orthonormal, so
\[
\tr(\Gamma_\alpha)=0,
\qquad
\tr(\Gamma_\alpha\Gamma_\beta)=\delta_{\alpha\beta},
\]
which proves Eq.~\eqref{eq:OalphaObeta-trace}. Since \(\dim \Delta\mathfrak J^{(1)}=m^2-1\), the normalized observables in Eq.~\eqref{eq:H1-basis} form an HS-orthonormal Hermitian basis of \(\Delta\mathfrak J^{(1)}\).
\end{proof}

\medskip
\noindent\textbf{(iii) Operational meaning: direct measurement of \(I_1\).}
The explicit basis \eqref{eq:H1-basis} gives a direct measurement formula for the first-layer purity. Since
\[
\rho^{(1)}
=
\sum_{\alpha=1}^{m^2-1}
\tr(\rho H_\alpha^{(1)})\,H_\alpha^{(1)},
\]
we obtain
\begin{equation}
I_1(\rho)
=
\tr\!\big[(\rho^{(1)})^2\big]
=
\sum_{\alpha=1}^{m^2-1}\tr(\rho H_\alpha^{(1)})^2
=
\frac{1}{\binom{m+n}{m+1}}
\sum_{\alpha=1}^{m^2-1}\tr(\rho \hat O_\alpha)^2.
\label{eq:I1-observable-form}
\end{equation}
Thus \(I_1\) can be determined directly from the expectation values of a complete set of traceless one-body observables.

\medskip
\noindent\textbf{Relation to the observable invariant of Ref.~\cite{parellada_lie_2024}.}
Ref.~\cite{parellada_lie_2024} defines an observable invariant from the non-orthogonal family
\begin{equation}
\hat O_j^{\prime(1)}=\hat n_j,
\qquad
\hat O_{jk}^{\prime(2)}=\frac{\hat a_j^\dagger\hat a_k+\hat a_k^\dagger\hat a_j}{\sqrt2},
\qquad
\hat O_{jk}^{\prime(3)}=\frac{i(\hat a_j^\dagger\hat a_k-\hat a_k^\dagger\hat a_j)}{\sqrt2},
\label{eq:observables-previous}
\end{equation}
and sets
\[
I_o:=\sum_i \tr(\rho \hat O_i')^2,
\]
where the sum runs over all \(m^2\) observables. The following proposition makes its relation to \(I_1\) explicit.

\begin{proposition}[Connection between \(I_o\) and the first-layer purity \(I_1\)]
\label{prop:Io-vs-I1}
For any state \(\rho\) on \(\mathcal H_{n,m}\),
\begin{equation}
I_1(\rho)
=
\frac{I_o-\frac{n^2}{m}}{\binom{m+n}{m+1}}.
\label{eq:Io-I1}
\end{equation}
Equivalently, the tangent purity \(I_t:=I_0+I_1\) satisfies
\begin{equation}
I_t
=
\frac{1}{\binom{m+n}{m+1}}\,I_o
-
\frac{n-1}{\binom{m+n}{m}}.
\label{eq:Io-It}
\end{equation}
\end{proposition}

\begin{proof}
Let \(v\in\mathbb R^m\) collect the number expectations,
\[
v_j:=\tr(\rho \hat n_j),\qquad j=1,\ldots,m.
\]
The diagonal generators \(\Gamma_0,\Gamma_1^z,\ldots,\Gamma_{m-1}^z\) form an orthonormal basis of the diagonal subspace of \(\Herm(m)\). Applying Parseval's identity to the vector \(v\) therefore gives
\[
\sum_{j=1}^m v_j^2
=
\tr(\rho \hat O_0)^2
+
\sum_{\ell=1}^{m-1}\tr(\rho\,\mathcal J^{(1)}(\Gamma_\ell^z))^2
=
\frac{n^2}{m}
+
\sum_{\ell=1}^{m-1}\tr(\rho\,\mathcal J^{(1)}(\Gamma_\ell^z))^2.
\]
For the off-diagonal generators, Eq.~\eqref{eq:GGM-def} shows that
\[
\mathcal J^{(1)}(\Gamma_{jk}^x)=\hat O_{jk}^{\prime(2)},
\qquad
\mathcal J^{(1)}(\Gamma_{jk}^y)= -\,\hat O_{jk}^{\prime(3)},
\]
so the sign difference is irrelevant after squaring expectation values. It follows that
\[
I_o
=
\frac{n^2}{m}
+
\sum_{\alpha=1}^{m^2-1}\tr(\rho \hat O_\alpha)^2.
\]
Substituting this into Eq.~\eqref{eq:I1-observable-form} yields Eq.~\eqref{eq:Io-I1}. Finally, using \(I_0=1/M\) with \(M=\binom{m+n-1}{n}\), together with
\[
\binom{m+n}{m}=\frac{m+n}{m}M,
\]
gives Eq.~\eqref{eq:Io-It}.
\end{proof}

\section{Example: Two-Photon Two-Mode}
\label{subsec:htm22}

\SecIntro{
This section presents an explicit HS-orthonormal operator basis adapted to the decomposition
\[
\Delta\mathfrak J^{(0)} \oplus \Delta\mathfrak J^{(1)} \oplus \Delta\mathfrak J^{(2)}
\]
for the smallest nontrivial case \(n=2\), \(m=2\), and derives the corresponding Hermitian transfer matrix (HTM) blocks. The strategy is entirely explicit: we first construct a layer-adapted HS-orthonormal basis from Pauli/GGM-generated one-body operators together with an orthonormal completion, and then evaluate \(\mathcal R_U\) by tracing \(H_\mu U H_\nu U^\dagger\) using the spin-1 representation of \(U=\varphi_2(S)\).
}

The fixed-\(n\) Hilbert space \(\mathcal H_{2,2}\) has dimension \(M=3\). Throughout this section, we use the ordered Fock basis
\[
\{\ket{\mathbf n_i}\}_{i=1}^3=\{\ket{0,2},\ket{1,1},\ket{2,0}\}.
\]

\subsection{HS-Orthonormal Basis}
\label{subsec:htm22-basis}

For \(m=2\), a convenient choice of first-order observables is obtained from the normalized Pauli/GGM generators. Concretely, the order-1 Jordan images
\[
\hat O_\mu:=\mathcal J^{(1)}(\Gamma_\mu),
\qquad
\mu=0,1,2,3,
\]
take the form
\begin{equation}
\hat O_0=\frac{1}{\sqrt2}\left(\hat n_1+\hat n_2\right),\quad
\hat O_1=\frac{1}{\sqrt2}\left(\hat n_1-\hat n_2\right),\quad
\hat O_2=\frac{1}{\sqrt2}\left(\hat a_1^\dagger \hat a_2+\hat a_2^\dagger \hat a_1\right),\quad
\hat O_3=\frac{i}{\sqrt2}\left(\hat a_1^\dagger \hat a_2-\hat a_2^\dagger \hat a_1\right),
\label{eq:observables_of_two_mode}
\end{equation}
where \(\hat n_j=\hat a_j^\dagger \hat a_j\).

For \(n=2\) and \(m=2\), one has
\[
\binom{m+n}{m+1}=\binom{4}{3}=4.
\]
Accordingly,
\[
\tr(\hat O_i\hat O_j)=4\delta_{ij},
\qquad
i,j\in\{1,2,3\},
\]
while
\[
\tr(\hat O_0^2)=6.
\]
We therefore define
\begin{equation}
H_0:=\frac{1}{\sqrt6}\,\hat O_0,
\qquad
H_i:=\frac{1}{2}\,\hat O_i,
\quad
1\le i\le 3,
\label{eq:htm22-H0H3-def}
\end{equation}
so that
\[
\tr(H_\mu H_\nu)=\delta_{\mu\nu},
\qquad
\mu,\nu\in\{0,1,2,3\}.
\]

In the ordered Fock basis, these operators are
\begin{equation}
\begin{aligned}
H_0 &=
\begin{pmatrix}
 \frac{1}{\sqrt3} & 0 & 0 \\
 0 & \frac{1}{\sqrt3} & 0 \\
 0 & 0 & \frac{1}{\sqrt3}
\end{pmatrix},
\qquad
H_1=
\begin{pmatrix}
 \frac{1}{\sqrt2} & 0 & 0 \\
 0 & 0 & 0 \\
 0 & 0 & -\frac{1}{\sqrt2}
\end{pmatrix},
\\[0.6em]
H_2 &=
\begin{pmatrix}
 0 & \frac{1}{2} & 0 \\
 \frac{1}{2} & 0 & \frac{1}{2} \\
 0 & \frac{1}{2} & 0
\end{pmatrix},
\qquad
H_3=
\begin{pmatrix}
 0 & -\frac{i}{2} & 0 \\
 \frac{i}{2} & 0 & -\frac{i}{2} \\
 0 & \frac{i}{2} & 0
\end{pmatrix}.
\end{aligned}
\label{eq:htm22-tangent-basis}
\end{equation}
Here \(H_0\) spans \(\Delta\mathfrak J^{(0)}\), while \(\{H_1,H_2,H_3\}\) forms an HS-orthonormal basis of the first coherence layer \(\Delta\mathfrak J^{(1)}\).

To complete an HS-orthonormal Hermitian basis of \(\Herm(\mathcal H_{2,2})\), we choose an HS-orthonormal basis for the orthogonal complement
\[
\Delta\mathfrak J^{(2)}
=
\left(\Delta\mathfrak J^{(0)}\oplus\Delta\mathfrak J^{(1)}\right)^\perp .
\]
A convenient choice is
\begin{equation}
\begin{aligned}
H_4 &=
\begin{pmatrix}
 \frac{1}{\sqrt6} & 0 & 0 \\
 0 & -\sqrt{\frac{2}{3}} & 0 \\
 0 & 0 & \frac{1}{\sqrt6}
\end{pmatrix},
\qquad
H_5 =
\begin{pmatrix}
 0 & \frac{1}{2} & 0 \\
 \frac{1}{2} & 0 & -\frac{1}{2} \\
 0 & -\frac{1}{2} & 0
\end{pmatrix},
\qquad
H_6 =
\begin{pmatrix}
 0 & -\frac{i}{2} & 0 \\
 \frac{i}{2} & 0 & \frac{i}{2} \\
 0 & -\frac{i}{2} & 0
\end{pmatrix},
\\[0.6em]
H_7 &=
\begin{pmatrix}
 0 & 0 & -\frac{i}{\sqrt2} \\
 0 & 0 & 0 \\
 \frac{i}{\sqrt2} & 0 & 0
\end{pmatrix},
\qquad
H_8 =
\begin{pmatrix}
 0 & 0 & \frac{1}{\sqrt2} \\
 0 & 0 & 0 \\
 \frac{1}{\sqrt2} & 0 & 0
\end{pmatrix}.
\end{aligned}
\label{eq:htm22-perp-basis}
\end{equation}
Together, the nine operators \(\{H_\mu\}_{\mu=0}^8\) form an HS-orthonormal Hermitian basis adapted to the decomposition
\[
\Delta\mathfrak J^{(0)}\oplus\Delta\mathfrak J^{(1)}\oplus\Delta\mathfrak J^{(2)}.
\]

\subsection{HTM Representation}
\label{subsec:htm22-htm}

A two-mode passive linear-optical network is specified by a scattering matrix \(S\in U(2)\). Since global phases do not affect conjugation on density operators, it is convenient to parameterize the physically relevant part by Euler angles \((\alpha,\beta,\gamma)\):
\begin{equation}
S=
\begin{pmatrix}
 e^{\frac{i}{2} (\alpha + \gamma)} \cos\!\left(\frac{\beta}{2}\right) &
 e^{\frac{i}{2} (\alpha - \gamma)} \sin\!\left(\frac{\beta}{2}\right) \\
 -e^{-\frac{i}{2} (\alpha - \gamma)} \sin\!\left(\frac{\beta}{2}\right) &
 e^{-\frac{i}{2} (\alpha + \gamma)} \cos\!\left(\frac{\beta}{2}\right)
\end{pmatrix}.
\label{eq:U2-euler}
\end{equation}

The induced two-photon unitary \(U=\varphi_2(S)\in U(3)\), acting on \(\mathcal H_{2,2}\), is explicitly
\begin{equation}
U=
\begin{pmatrix}
 e^{i (\alpha + \gamma)} \cos^2\!\left(\frac{\beta}{2}\right) &
 \sqrt{2}\,e^{i \alpha} \sin\!\left(\frac{\beta}{2}\right)\cos\!\left(\frac{\beta}{2}\right) &
 e^{i (\alpha - \gamma)} \sin^2\!\left(\frac{\beta}{2}\right) \\
 -\sqrt{2}\,e^{i (\gamma - \alpha)} \sin\!\left(\frac{\beta}{2}\right)\cos\!\left(\frac{\beta}{2}\right) &
 \cos(\beta) &
 \sqrt{2}\,e^{-i (\alpha - \gamma)} \sin\!\left(\frac{\beta}{2}\right)\cos\!\left(\frac{\beta}{2}\right) \\
 e^{-i (\alpha - \gamma)} \sin^2\!\left(\frac{\beta}{2}\right) &
 -\sqrt{2}\,e^{-i \alpha} \sin\!\left(\frac{\beta}{2}\right)\cos\!\left(\frac{\beta}{2}\right) &
 e^{-i (\alpha + \gamma)} \cos^2\!\left(\frac{\beta}{2}\right)
\end{pmatrix}.
\label{eq:spin1-U}
\end{equation}

By definition, the HTM in the basis \(\{H_\mu\}\) is
\[
(\mathcal R_U)_{\mu\nu}:=\tr(H_\mu\,U H_\nu U^\dagger),
\]
so that the coefficient vector
\[
x_\mu=\tr(H_\mu\rho)
\]
transforms as \(x'=\mathcal R_U x\).

Since \(H_0\propto I\), one immediately has
\[
(\mathcal R_U)_{00}=1,
\qquad
(\mathcal R_U)_{0\nu}=(\mathcal R_U)_{\mu 0}=0
\quad
(\mu,\nu\ge 1).
\]
Moreover, by layer invariance, \(\mathcal R_U\) is block diagonal with respect to
\[
\Delta\mathfrak J^{(0)}\oplus\Delta\mathfrak J^{(1)}\oplus\Delta\mathfrak J^{(2)}.
\]

A direct evaluation of \((\mathcal R_U)_{\mu\nu}\) using Eqs.~\eqref{eq:spin1-U} and \eqref{eq:htm22-tangent-basis} yields
\begin{equation}
\mathcal R_U^{(1)}=
\begin{pmatrix}
\cos\beta & \sin\beta\cos\gamma & \sin\beta\sin\gamma \\
-\cos\alpha\sin\beta & \cos\alpha\cos\beta\cos\gamma-\sin\alpha\sin\gamma &
\cos\alpha\cos\beta\sin\gamma+\sin\alpha\cos\gamma \\
\sin\alpha\sin\beta & -\sin\alpha\cos\beta\cos\gamma-\cos\alpha\sin\gamma &
\cos\alpha\cos\gamma-\sin\alpha\cos\beta\sin\gamma
\end{pmatrix},
\label{eq:htm22-tangent-block}
\end{equation}
which is an \(\mathrm{SO}(3)\) rotation matrix.

\section{Jordan-Spectral Distance}
\label{sec:distance-Jordan}

\SecIntro{
In this section, we define a distance between two states in terms of their layer spectra and relate it to the minimal distance along the passive linear-optical orbit.
}

In the finite-dimensional fixed-\(n\) sector \(\mathcal H_{n,m}\), this yields a sharper bound than the one obtained in Ref.~\cite{parellada_no-go_2023}. Recall that every Hermitian operator admits the layer decomposition
\[
\rho=\sum_{j=0}^n \rho^{(j)},
\qquad
\rho^{(j)}\in \Delta\mathfrak J^{(j)}.
\]
Likewise, for another state \(\sigma\), we write
\[
\sigma=\sum_{j=0}^n \sigma^{(j)},
\qquad
\sigma^{(j)}\in \Delta\mathfrak J^{(j)}.
\]

For each \(j\), let \(\boldsymbol{\lambda}_j(\rho)\) denote the ordered eigenvalue vector of the layer component \(\rho^{(j)}\). We then define the squared Jordan-spectral distance by
\begin{equation}
D_{\mathrm{spec}}^2(\rho, \sigma)
:=
\frac{1}{4}\sum_{j=0}^n
\bigl\|
\boldsymbol{\lambda}_j(\rho)-\boldsymbol{\lambda}_j(\sigma)
\bigr\|_2^2.
\label{eq:def_Dspec}
\end{equation}

\begin{proposition}[Distance between Jordan-spectral invariants]
\label{prop:Jordan-distance-bound}
Let \(\rho\) and \(\sigma\) be density operators on \(\mathcal H_{n,m}\), and let \(U\) range over all passive linear-optical unitaries on \(\mathcal H_{n,m}\). Then
\begin{equation}
D_{\mathrm{spec}}(\rho, \sigma)
\;\le\;
\min_{U\in \mathrm{U_L}}
\frac{1}{2}\bigl\|U\rho U^\dagger-\sigma\bigr\|_2
\;\le\;
\min_{U\in \mathrm{U_L}}
D_{\mathrm{tr}}(U\rho U^\dagger,\sigma),
\label{eq:bound_Dspec}
\end{equation}
where
\[
D_{\mathrm{tr}}(\rho,\sigma):=\frac{1}{2}\|\rho-\sigma\|_1
\]
denotes the trace distance. 
Here, \(\|\cdot\|_2\) denotes the Euclidean norm for vectors and the Hilbert--Schmidt norm for operators, while \(\|\cdot\|_1\) denotes the trace norm.
\end{proposition}

\begin{proof}
Fix a layer \(j\). Since \(\rho^{(j)}\) and \(\sigma^{(j)}\) are Hermitian operators on the \(M\)-dimensional space \(\mathcal H_{n,m}\), the Hoffman--Wielandt inequality~\cite{hoffman_variation_1953} implies
\begin{equation}
\bigl\|
\boldsymbol{\lambda}_j(\rho)-\boldsymbol{\lambda}_j(\sigma)
\bigr\|_2^2
\;\le\;
\bigl\|\rho^{(j)}-\sigma^{(j)}\bigr\|_2^2.
\label{eq:HW-layer-vector}
\end{equation}
Summing over \(j\) and using the Hilbert--Schmidt orthogonality of the coherence layers, we obtain
\begin{equation}
\|\rho-\sigma\|_2^2
=
\sum_{j=0}^n
\bigl\|\rho^{(j)}-\sigma^{(j)}\bigr\|_2^2.
\label{eq:Pythagoras-layers}
\end{equation}
Combining Eqs.~\eqref{eq:HW-layer-vector}, \eqref{eq:Pythagoras-layers}, and \eqref{eq:def_Dspec} yields
\begin{equation}
D_{\mathrm{spec}}(\rho,\sigma)
\;\le\;
\frac{1}{2}\|\rho-\sigma\|_2.
\label{eq:spec-vs-HS}
\end{equation}

Now let \(U\in \mathrm{U_L}\). Since each coherence layer is invariant under passive linear-optical conjugation, one has
\[
(U\rho U^\dagger)^{(j)} = U\rho^{(j)}U^\dagger.
\]
Therefore the layer spectra are unchanged,
\[
\boldsymbol{\lambda}_j(U\rho U^\dagger)=\boldsymbol{\lambda}_j(\rho)
\qquad
\text{for all }j,
\]
and hence
\[
D_{\mathrm{spec}}(U\rho U^\dagger,\sigma)
=
D_{\mathrm{spec}}(\rho,\sigma).
\]
Applying Eq.~\eqref{eq:spec-vs-HS} to the pair \((U\rho U^\dagger,\sigma)\) and minimizing over \(U\in\mathrm{U_L}\) gives the first inequality in Eq.~\eqref{eq:bound_Dspec}.

The second inequality follows from the standard norm bound
\[
\|X\|_2\le \|X\|_1.
\]
\end{proof}

\section{Quantifying the Metrological Capability of Multiphoton States via Invariants}
\label{sec:QFI-LO}

\SecIntro{
Quantum metrology studies how precisely an unknown parameter can be estimated by interrogating a quantum probe.
In the standard paradigm, one (i) prepares a probe state \(\rho\) in a well-controlled Hilbert space, (ii) encodes the parameter \(\theta\) through a parameter-dependent quantum channel, commonly a unitary encoding \(\rho_\theta = e^{-i\theta \hat H}\rho\,e^{i\theta \hat H}\), and (iii) performs a measurement on \(\rho_\theta\) to produce an estimator \(\hat\theta\).
The achievable precision is ultimately constrained by the quantum Cram\'er--Rao bound, \(\mathrm{Var}(\hat\theta)\ge 1/F_Q(\rho,\hat H)\), where \(F_Q\) is the quantum Fisher information (QFI)~\cite{braunstein_statistical_1994, giovannetti_advances_2011}.

In this section we focus on a constrained, physically motivated metrological setting tailored to multiphoton linear optics.
Specifically, the probe is restricted to the fixed-\(n\) photon sector \(\mathcal H_{n,m}\), and the parameter encoding is restricted to passive linear-optical realizable unitary encoding, i.e.\ \(\rho_\theta = e^{-i\theta \hat H}\rho\,e^{i\theta \hat H}, \hat H=\mathcal J^{(1)}(h)\) with \(h\in\Herm(m)\).

Within this setting we address two questions.

\medskip
\noindent\textbf{(i) Mean metrological capability under unknown encoding.}
Given an $n$-photon pure state $\rho$, when the parameter-encoding generator is not known \emph{a priori}, what is its \emph{average} metrological capability?
Concretely, we study $\mathbb E_{\hat H}\,F_Q(\rho,\hat H)$, where $\hat H$ is drawn Haar-randomly from the constrained family of linear-optical generators.

\medskip
\noindent\textbf{(ii) State optimization under a fixed encoding.}
Given an $n$-photon pure state $\rho$ and a fixed parameter-encoding generator $\hat H$, we allow pre-processing by passive linear optics $U$ to enhance the achievable precision, i.e.,
\[
\max_{U}\,F_Q(U\rho U^\dagger,\hat H).
\]

We derive a general closed-form equality for Question~(i), and an exact equality for Question~(ii) in the special case of two photons in two modes.}

\subsection{Covariance-Matrix Approach for Calculating QFI}
\label{subsec:QFI-setup}

We consider single-parameter unitary encoding generated by a Hermitian operator $\hat H$,
\begin{equation}
\rho_\theta = e^{-i\theta \hat H}\,\rho\,e^{i\theta \hat H}.
\end{equation}
For a pure probe $\rho=\ket{\psi}\!\bra{\psi}$, the quantum Fisher information (QFI) reduces to four times the variance,
\begin{equation}
F_Q(\ket{\psi},\hat H)=4\,\Var_\psi(\hat H)
=4 \left(\braket{\hat H^2}_\psi- \braket{\hat H}_\psi^{\,2} \right).
\label{eq:QFI-pure}
\end{equation}

\medskip
\noindent\textbf{Linear-optical generator basis.}
Let $d:=m^2-1$. Following Eqs.~\eqref{eq:GGM-def} and \eqref{eq:O-from-GGM} we choose the GGM $\{\Gamma_\alpha \}_{\alpha=1}^{d}$ as a Hilbert--Schmidt orthonormal basis of traceless elements in $\Herm(m)$. The generators realizable by LON is given by order-1 Jordan-Schwinger map:
\begin{equation}
\hat O_\alpha:=\mathcal J^{(1)}(\Gamma_\alpha)
=\sum_{s,t=1}^m (\Gamma_\alpha)_{st}\,\hat a_s^\dagger \hat a_t,
\qquad \alpha=1,\dots,d.
\end{equation}
We note that the identity component $\Gamma_0:=I_m/\sqrt m$ (not included in $\alpha=1,\dots,d$) leads to
$\hat O_0=\frac{n}{\sqrt m}\Id$ on $\mathcal H_{n,m}$, hence it induces only a global phase and is irrelevant to the QFI.

To adopt the conventional spin-$j=n/2$ notation in the two-mode sector, we define the angular-momentum operators:
\begin{equation}
\hat J_\alpha := \frac{\hat O_\alpha}{\sqrt{2}},\qquad
\hat{\mathbf J} := (\hat J_1,\dots,\hat J_{d}),
\end{equation}
and parameterize a \emph{normalized} linear-optical Hamiltonian by a unit vector
$\mathbf n\in\mathbb R^{d}$,
\begin{equation}
\hat H(\mathbf n)= \mathbf n\cdot \hat{\mathbf J}
= \frac{1}{\sqrt{2}}\,\mathbf n\cdot \hat{\mathbf O},
\qquad \norm{\mathbf n}=1.
\label{eq:H-expand}
\end{equation}

\medskip
\noindent\textbf{Covariance-matrix reduction.}
Define the covariance of two observables $\hat{A}, \hat{B}$ on $\ket{\psi}$ as
\begin{equation}
\Cov_\psi(\hat{A},\hat B):=\frac{1}{2}\braket{\hat A\hat B+\hat B\hat A}_\psi-\braket{\hat A}_\psi\braket{\hat B}_\psi.
\end{equation}
We then introduce the real symmetric covariance matrix
\begin{equation}
\mathsf C_{\alpha\beta}(\psi)
:= \Cov_\psi(\hat J_\alpha,\hat J_\beta),
\qquad \alpha,\beta=1,\dots,d,
\label{eq:cov-matrix-def}
\end{equation}
following the angular-momentum formulation of Ref.~\cite{sun_fisher_2010}.
Using $\hat H(\mathbf n)=\sum_\alpha n_\alpha \hat J_\alpha$ and bilinearity of $\Cov_\psi(\cdot,\cdot)$, we obtain
\begin{equation}
F_Q(\psi,\hat H(\mathbf n))
=4\,\Var_\psi(\mathbf n\cdot \hat{\mathbf J})
=4\,\Cov_\psi(\mathbf n\cdot \hat{\mathbf J},\,\mathbf n\cdot \hat{\mathbf J})
=4\,\mathbf n^{\mathsf T}\mathsf C(\psi)\,\mathbf n.
\label{eq:Cov_rep}
\end{equation}
In the following, Question~(i) concerns the \emph{mean} metrological capability under an unknown (random) encoding, which is governed by the average eigenvalue (equivalently, $\tr\,\mathsf C$), whereas Question~(ii) concerns the best achievable precision under a fixed encoding direction, governed by the maximal eigenvalue of $\mathsf C$.

\medskip
\noindent\textbf{Generalized spin-length.}
The generalized spin (Bloch) vector is $\braket{\hat{\mathbf J}}_\psi$, with squared length
\begin{equation} \label{eq:general_spin_length}
\norm{\braket{\hat{\mathbf J}}_\psi}^{2}
:=\sum_{\alpha=1}^{d}\braket{\hat J_\alpha}_\psi^{2}
=\frac{1}{2}\sum_{\alpha=1}^{d}\braket{\hat O_\alpha}_\psi^{2}
=\frac{1}{2}\binom{m+n}{m+1}\,I_1(\psi),
\end{equation}
where in the last equality we used Eq.~\eqref{eq:I1-observable-form} for the first-layer purity $I_1$.

\subsection{Averaged QFI for General N-Photon Pure States}
\label{subsec:QFI-haaravg}

We now address Question~(i): the \emph{mean} metrological capability of a fixed pure probe
$\ket{\psi}\in\mathcal H_{n,m}$ when the linear-optical encoding generator is drawn uniformly
under an HS constraint in the traceless mode-space algebra.

We sample $\mathbf n \in\mathbb R^{d}$ uniformly from the unit sphere
$S^{d-1}$ and set $\hat H(\mathbf n)=\mathbf n\cdot\hat{\mathbf J}$ as in Eq.~\eqref{eq:H-expand}.
Define the averaged QFI
\begin{equation}
\overline{F}_Q(\ket{\psi})
:=
\mathbb E_{\mathbf n\sim \mathrm{unif}(S^{d-1})}
\Big[F_Q\big(\ket{\psi},\hat H(\mathbf n)\big)\Big].
\label{eq:Fbar-def}
\end{equation}

\begin{lemma}[HS completeness and quadratic Casimir on $\mathcal H_{n,m}$]
\label{lem:Gamma-completeness-casimir}
Let $\Gamma_0:=I_m/\sqrt m$ and $\{\Gamma_\alpha\}_{\alpha=1}^{d}$ be traceless Hermitian generators
satisfying $\tr(\Gamma_\mu\Gamma_\nu)=\delta_{\mu\nu}$.
Then the HS completeness identity reads
\begin{equation}
\sum_{\mu=0}^{d}(\Gamma_\mu)_{ij}(\Gamma_\mu)_{kl}
=\delta_{il}\delta_{jk},
\qquad
\Longrightarrow\qquad
\sum_{\alpha=1}^{d}(\Gamma_\alpha)_{ij}(\Gamma_\alpha)_{kl}
=\delta_{il}\delta_{jk}-\frac{1}{m}\delta_{ij}\delta_{kl}.
\label{eq:Gamma-Fierz}
\end{equation}
Moreover, for $\hat O_\alpha=\mathcal J^{(1)}(\Gamma_\alpha)$ and $\hat J_\alpha=\hat O_\alpha/\sqrt2$,
\begin{equation}
\sum_{\alpha=1}^{d}\hat O_\alpha^{\,2}
=
(m-1)\hat N+\Bigl(1-\frac{1}{m}\Bigr)\hat N^2,
\qquad
\sum_{\alpha=1}^{d}\hat J_\alpha^{\,2}
=
\frac{c_{n,m}}{2}\,\Id\ \ \text{on }\ \mathcal H_{n,m},
\label{eq:Casimir-J}
\end{equation}
where $\hat N=\sum_{s=1}^m \hat a_s^\dagger\hat a_s$ and
\begin{equation}
c_{n,m}=\frac{n(m-1)(m+n)}{m}.
\label{eq:cnm}
\end{equation}
\end{lemma}

\begin{proof}

\medskip
\noindent\textbf{(i) Completeness.}
Since $\{\Gamma_\mu\}_{\mu=0}^{d}$ is an HS-orthonormal basis of $\Herm(m)$, any $X\in\Herm(m)$ expands as
$X=\sum_{\mu}\tr(\Gamma_\mu X)\Gamma_\mu$.
Applying this to the matrix unit $E_{lk}$ with $(E_{lk})_{ij}=\delta_{il}\delta_{jk}$ gives
$\tr(\Gamma_\mu E_{lk})=(\Gamma_\mu)_{kl}$ and hence
$E_{lk}=\sum_\mu (\Gamma_\mu)_{kl}\Gamma_\mu$.
Taking the $(i,j)$ entry yields the first identity in Eq.~\eqref{eq:Gamma-Fierz};
subtracting the $\mu=0$ term, $(\Gamma_0)_{ij}(\Gamma_0)_{kl}=\frac{1}{m}\delta_{ij}\delta_{kl}$,
gives the second.

\medskip
\noindent\textbf{(ii) Casimir.}
Using $\hat O_\alpha=\sum_{i,j}(\Gamma_\alpha)_{ij}\hat a_i^\dagger\hat a_j$ and Eq.~\eqref{eq:Gamma-Fierz},
\begin{align}
\sum_{\alpha=1}^{d}\hat O_\alpha^{\,2}
&=\sum_{\alpha}\sum_{i,j,k,l}(\Gamma_\alpha)_{ij}(\Gamma_\alpha)_{kl}\,
\hat a_i^\dagger\hat a_j\hat a_k^\dagger\hat a_l \nonumber\\
&=\sum_{i,j}\hat a_i^\dagger\hat a_j\hat a_j^\dagger\hat a_i
-\frac{1}{m}\sum_{i,k}\hat a_i^\dagger\hat a_i\hat a_k^\dagger\hat a_k
=\sum_{i,j}\hat a_i^\dagger\hat a_j\hat a_j^\dagger\hat a_i-\frac{1}{m}\hat N^2.
\label{eq:Casimir-step1}
\end{align}
By the CCR, $\hat a_j\hat a_j^\dagger=\Id+\hat a_j^\dagger\hat a_j$, so
\begin{align}
\sum_{i,j}\hat a_i^\dagger\hat a_j\hat a_j^\dagger\hat a_i
&=\sum_{i,j}\hat a_i^\dagger(\Id+\hat a_j^\dagger\hat a_j)\hat a_i
=m\hat N+\sum_{i,j}\hat a_i^\dagger\hat a_j^\dagger\hat a_j\hat a_i.
\end{align}
Finally, $\sum_{i,j}\hat a_i^\dagger\hat a_j^\dagger\hat a_j\hat a_i=\hat N^2-\hat N$ (e.g.\ in the Fock basis),
so Eq.~\eqref{eq:Casimir-step1} becomes
\[
\sum_{\alpha=1}^{d}\hat O_\alpha^{\,2}
=m\hat N+(\hat N^2-\hat N)-\frac{1}{m}\hat N^2
=(m-1)\hat N+\Bigl(1-\frac{1}{m}\Bigr)\hat N^2.
\]
Restricting to $\mathcal H_{n,m}$ where $\hat N=n\,\Id$ yields
$\sum_{\alpha}\hat O_\alpha^{\,2}=c_{n,m}\Id$ with $c_{n,m}$ in Eq.~\eqref{eq:cnm},
and dividing by $2$ gives Eq.~\eqref{eq:Casimir-J}.
\end{proof}

\begin{theorem}[Exact Haar-distributed averaged QFI for pure states]
\label{thm:Haar-avg-QFI}
For any pure $\ket{\psi}\in\mathcal H_{n,m}$,
\begin{equation}
\overline{F}_Q(\ket{\psi})
=
\frac{4}{m^2-1}\,\tr\!\big(\mathsf C(\psi)\big)
=
\frac{2}{m^2-1}\Bigl(c_{n,m}-\binom{m+n}{m+1}I_1(\psi)\Bigr),
\label{eq:Haar-avg-QFI-I1}
\end{equation}
where $c_{n,m}$ is given in Eq.~\eqref{eq:cnm} and $I_1(\psi)$ is the first-layer purity
defined in Eq.~\eqref{eq:Ij-def}.
\end{theorem}

\begin{proof}
From Eq.~\eqref{eq:Cov_rep} in Sec.~\ref{subsec:QFI-setup},
\begin{equation}
F_Q(\psi,\hat H(\mathbf n))=4\,\mathbf n^{\mathsf T}\mathsf C(\psi)\,\mathbf n.
\end{equation}
Rotational invariance of $\mathbf n\sim\mathrm{unif}(S^{d-1})$ implies
$\mathbb E[n_\alpha n_\beta]=\delta_{\alpha\beta}/d$ with $d=m^2-1$, hence
\begin{equation}
\overline{F}_Q(\ket{\psi})
=
4\sum_{\alpha,\beta}\mathsf C_{\alpha\beta}(\psi)\,\mathbb E[n_\alpha n_\beta]
=\frac{4}{d}\tr\!\big(\mathsf C(\psi)\big).
\label{eq:Fbar-trC}
\end{equation}
Next,
\[
\tr\!\big(\mathsf C(\psi)\big)
=\sum_{\alpha=1}^{d}\Var_\psi(\hat J_\alpha)
=\sum_{\alpha}\bra{\psi}\hat J_\alpha^{\,2}\ket{\psi}-\sum_{\alpha}\bra{\psi}\hat J_\alpha\ket{\psi}^{2}.
\]
By Lemma~\ref{lem:Gamma-completeness-casimir},
$\sum_{\alpha}\hat J_\alpha^{\,2}=\frac{c_{n,m}}{2}\Id$ on $\mathcal H_{n,m}$, so
$\sum_{\alpha}\bra{\psi}\hat J_\alpha^{\,2}\ket{\psi}=\frac{c_{n,m}}{2}$.
Moreover, the generalized spin-length in Sec.~\ref{subsec:QFI-setup} gives
\[
\sum_{\alpha}\bra{\psi}\hat J_\alpha\ket{\psi}^{2}
=\norm{\braket{\hat{\mathbf J}}_\psi}^2
=\frac{1}{2}\binom{m+n}{m+1}I_1(\psi).
\]
Therefore
$\tr(\mathsf C(\psi))=\frac12\Bigl(c_{n,m}-\binom{m+n}{m+1}I_1(\psi)\Bigr)$.
Substituting this into Eq.~\eqref{eq:Fbar-trC} yields Eq.~\eqref{eq:Haar-avg-QFI-I1}.
\end{proof}

\subsection{Optimal QFI for Two-Photon Two-Mode Pure States}
\label{subsec:QFI-22-optimal}

We now address Question~(ii) for $(n,m)=(2,2)$.
In this case $\mathcal H_{2,2}$ is three-dimensional and carries the spin-$1$ irrep of $\mathrm{SU}(2)$ induced by the Schwinger map.

\medskip
\noindent\textbf{Spin generators from the $\Gamma$ basis.}
For $m=2$ we take $\Gamma_{x},\Gamma_{y},\Gamma_{z}$ to be the HS-normalized Pauli basis
\begin{equation}
\Gamma_x=\frac{1}{\sqrt2}\begin{pmatrix}0&1\\[0.2em]1&0\end{pmatrix},\qquad
\Gamma_y=\frac{1}{\sqrt2}\begin{pmatrix}0&-i\\[0.2em]i&0\end{pmatrix},\qquad
\Gamma_z=\frac{1}{\sqrt2}\begin{pmatrix}1&0\\[0.2em]0&-1\end{pmatrix},
\label{eq:Pauli-Gamma-22}
\end{equation}
so that $\tr(\Gamma_\mu\Gamma_\nu)=\delta_{\mu\nu}$ for $\mu,\nu\in\{x,y,z\}$.
Their Jordan--Schwinger images are
\begin{equation}
\hat O_x=\frac{1}{\sqrt2}\big(\hat a_1^\dagger\hat a_2+\hat a_2^\dagger\hat a_1\big),\quad
\hat O_y=\frac{1}{\sqrt2}\,i\big(\hat a_1^\dagger\hat a_2-\hat a_2^\dagger\hat a_1\big),\quad
\hat O_z=\frac{1}{\sqrt2}\big(\hat n_1-\hat n_2\big),
\label{eq:O-Pauli-22}
\end{equation}
and with $\hat J_\mu=\hat O_\mu/\sqrt2$ we recover the standard Schwinger spin operators
\begin{equation}
\hat J_x=\frac{1}{2}\big(\hat a_1^\dagger \hat a_2+\hat a_2^\dagger \hat a_1\big),\quad
\hat J_y=\frac{1}{2i}\big(\hat a_1^\dagger \hat a_2-\hat a_2^\dagger \hat a_1\big),\quad
\hat J_z=\frac{1}{2}\big(\hat n_1-\hat n_2\big).
\label{eq:Schwinger-J-22}
\end{equation}
On $\mathcal H_{2,2}$ we have the spin-$1$ Casimir $\hat{\mathbf J}^2=\hat J_x^2+\hat J_y^2+\hat J_z^2=2\,\Id$
(equivalently, $\sum_{\mu}\hat O_\mu^2=4\,\Id$, consistent with Lemma~\ref{lem:Gamma-completeness-casimir} at $(n,m)=(2,2)$).

\begin{lemma}[Passive preprocessing $\Leftrightarrow$ generator direction]
\label{lem:state-generator-equivalence-22}
Fix any pure probe $\ket{\psi}\in\mathcal H_{2,2}$ and any admissible generator $\hat H(\mathbf n_0)$.
Then
\begin{equation}
\max_{U\in \mathrm{LON}} F_Q\!\big(U\ket{\psi},\hat H(\mathbf n_0)\big)
=
\max_{\|\mathbf n\|_2=1}F_Q\!\big(\ket{\psi},\hat H(\mathbf n)\big).
\label{eq:maxU-maxn-22}
\end{equation}
\end{lemma}

\begin{proof}
For pure states, $F_Q(\ket{\psi},\hat H)=4\,\Var_\psi(\hat H)$ [Eq.~\eqref{eq:QFI-pure}], and
$\Var_{U\psi}(\hat H)=\Var_{\psi}(U^\dagger \hat H U)$, hence
$F_Q(U\ket{\psi},\hat H)=F_Q(\ket{\psi},U^\dagger \hat H U)$.

It remains to characterize $\{U^\dagger \hat H(\mathbf n_0) U\}$.
In our Hermitian transfer matrix picture, $\mathbf{n}$ undergoes a rotation given by Eq.~\eqref{eq:htm22-tangent-block} as $\mathbf{n} = \mathcal{R}_U^{(1)} \mathbf{n}_0$, $\mathcal{R}_U^{(1)}$ can realize any $\mathrm{SO}(3)$ rotation, thus $\mathbf{n}$ sweeps the entire unit sphere. 
Since the map $\mathrm{SU}(2)\to\mathrm{SO}(3)$ is surjective, $\{\mathcal{R}_U^{(1)}\mathbf n_0\}$ sweeps the entire unit sphere. Taking the maximum over $U$ thus equals the maximum over $\mathbf n$.
\end{proof}

\begin{theorem}[Exact optimal QFI for $(n,m)=(2,2)$ pure states]
\label{thm:QFI-22-pure}
For any $\ket{\psi}\in\mathcal H_{2,2}$,
\begin{equation}
F_Q^\mathrm{max}(\ket{\psi})
:=
\max_{\|\mathbf n\|_2=1}F_Q\big(\ket{\psi},\hat H(\mathbf n)\big)
=
2\Bigl(1+\sqrt{1-r_J(\psi)^2}\Bigr).
\label{eq:Fmax-22}
\end{equation}
Equivalently,
\begin{equation}
F_Q^\mathrm{max}(\ket{\psi})
=
2\Bigl(1+\sqrt{1-2I_1(\psi)}\Bigr),
\qquad (n,m)=(2,2).
\label{eq:Fmax-22-I1}
\end{equation}
\end{theorem}

\begin{proof}
\medskip
\noindent\textbf{Step 1: ``maximize over directions'' = largest covariance eigenvalue.}
Let $\mathsf C(\psi)$ be the $3\times3$ covariance matrix of $\hat{\mathbf J}$,
$\mathsf C_{\mu\nu}(\psi)=\Cov_\psi(\hat J_\mu,\hat J_\nu)$ with $\mu,\nu\in\{x,y,z\}$.
From Eq.~\eqref{eq:Cov_rep},
\begin{equation}
F_Q\big(\psi,\hat H(\mathbf n)\big)=4\,\mathbf n^{\mathsf T}\mathsf C(\psi)\,\mathbf n,
\qquad \|\mathbf n\|_2=1.
\end{equation}
Hence
\begin{equation}
F_Q^\mathrm{max}(\ket{\psi})=4\,\lambda_{\max}\!\big(\mathsf C(\psi)\big).
\label{eq:Fmax-eig-22}
\end{equation}

\medskip
\noindent\textbf{Step 2: use $\mathrm{SU}(2)$ covariance to reach a canonical representative.}
For any passive $U$, $\hat{\mathbf J}$ transforms as a vector under the spin-$1$ irrep:
\begin{equation}
U^\dagger \hat{\mathbf J}\,U = \mathcal{R}_U^{(1)}\,\hat{\mathbf J},
\qquad \mathcal{R}_U^{(1)}\in\mathrm{SO}(3),
\end{equation}
so $\langle\hat{\mathbf J}\rangle$ rotates and $\mathsf C(\psi)$ is conjugated as
$\mathsf C(U\psi)=\mathcal{R}_U^{(1)}\mathsf C(\psi)\mathcal{R}_U^{(1) \mathsf T}$.
In particular, $\lambda_{\max}(\mathsf C)$ is invariant under such preprocessing, so we may replace
$\ket{\psi}$ by any $\mathrm{SU}(2)$-rotated representative without changing the right-hand side of
Eq.~\eqref{eq:Fmax-eig-22}.

Choose $U$ such that the mean spin is aligned with $z$:
\begin{equation}
\bra{\psi'}\hat J_x\ket{\psi'}=\bra{\psi'}\hat J_y\ket{\psi'}=0,
\qquad
\bra{\psi'}\hat J_z\ket{\psi'}=r_J(\psi),
\label{eq:mean-align-22}
\end{equation}
where $\ket{\psi'}:=U\ket{\psi}$.
In the $\hat J_z$ eigenbasis $\{\ket{1},\ket{0},\ket{-1}\}$ (identified below with $\{\ket{20},\ket{11},\ket{02}\}$),
write $\ket{\psi'}=a\ket{1}+b\ket{0}+c\ket{-1}$.
Using the ladder operator $\hat J_+=\hat J_x+i\hat J_y=\sqrt2\big(\ket{1}\!\bra{0}+\ket{0}\!\bra{-1}\big)$,
the conditions $\langle\hat J_x\rangle=\langle\hat J_y\rangle=0$ are equivalent to $\langle\hat J_+\rangle=0$, i.e.
\begin{equation}
0=\bra{\psi'}\hat J_+\ket{\psi'}
=\sqrt2\,(a^*b+b^*c).
\label{eq:Jplus-zero}
\end{equation}
If $b\neq0$, Eq.~\eqref{eq:Jplus-zero} implies $a=-c\,e^{i\vartheta}$ and therefore
$\langle\hat J_z\rangle=|a|^2-|c|^2=0$, i.e.\ $r_J(\psi)=0$.
Thus, for $r_J(\psi)\neq0$ the aligned representative necessarily has $b=0$.
When $r_J(\psi)=0$, we may still choose a rotated representative with $b=0$ (e.g.\ rotate $\ket{0}$ to $(\ket{1}+\ket{-1})/\sqrt2$);
in all cases we may take
\begin{equation}
\ket{\psi'}=\cos\eta\,\ket{1}+e^{i\chi}\sin\eta\,\ket{-1},
\qquad
\eta\in\Big[0,\frac{\pi}{2}\Big].
\label{eq:spin1-2component}
\end{equation}
Finally, the stabilizer rotation about $z$, $e^{-i\theta\hat J_z}$, changes $\chi\mapsto \chi+2\theta$ without affecting
$\langle\hat J_z\rangle$ or the spectrum of $\mathsf C$, so we set $\chi=0$.
Identifying $\ket{1}\equiv\ket{20}$ and $\ket{-1}\equiv\ket{02}$, we arrive at the canonical photonic form
\begin{equation}
\ket{\psi_*}=\cos\eta\,\ket{20}+\sin\eta\,\ket{02}.
\label{eq:psi-canonical-22}
\end{equation}
For this representative,
$r_J(\psi)=|\langle\hat J_z\rangle|=|\cos(2\eta)|$.

\medskip
\noindent\textbf{Step 3: evaluate the covariance spectrum in the canonical form.}
For $\ket{\psi_*}$ one directly computes
\begin{equation}
\Var_{\psi_*}(\hat J_x)=\frac{1+\sin(2\eta)}{2},\qquad
\Var_{\psi_*}(\hat J_y)=\frac{1-\sin(2\eta)}{2},\qquad
\Var_{\psi_*}(\hat J_z)=\sin^2(2\eta),
\label{eq:Var-spin1-22}
\end{equation}
and the covariance matrix is diagonal in the $(x,y,z)$ axes.
Hence
\begin{equation}
\lambda_{\max}\!\big(\mathsf C(\psi)\big)
=\max\!\left\{\Var_{\psi_*}(\hat J_x),\Var_{\psi_*}(\hat J_y),\Var_{\psi_*}(\hat J_z)\right\}
=\frac{1+|\sin(2\eta)|}{2}
=\frac{1+\sqrt{1-r_J(\psi)^2}}{2}.
\end{equation}
Combining with Eq.~\eqref{eq:Fmax-eig-22} yields Eq.~\eqref{eq:Fmax-22}.
Substituting $r_J(\psi)^2=\norm{\braket{\hat{\mathbf J}}_\psi}^{2} = 2I_1(\psi)$ from Eq.~\eqref{eq:general_spin_length} gives Eq.~\eqref{eq:Fmax-22-I1}.
\end{proof}

\section{Generalization to Indefinite Photon Number States}
\label{sec:generalization}

\SecIntro{In the main text, we focused on the fixed photon number sector \(\mathcal{H}_{n,m}\) for simplicity. Here, we rigorously generalize our framework to states with indefinite photon numbers (e.g., coherent states or superposition of Fock states). We show that due to the photon-number superselection rule~\cite{bartlett_reference_2007, wick_intrinsic_1952} imposed by passive linear optics, the invariants for a general state simply decompose into the direct sum of invariants for each photon-number sector.}

\subsection{Block-Diagonal Evolution via Spectral Projectors}

The Hamiltonian of a passive linear optical network (LON), \(\hat{H} = \sum_{s,t} h_{st} \hat{a}_s^\dagger \hat{a}_t\), strictly commutes with the total photon number operator \(\hat{N} = \sum_s \hat{a}_s^\dagger \hat{a}_s\):
\begin{equation}
[\hat{H}, \hat{N}] = 0.
\end{equation}
Let \(\Pi_n\) denote the projector onto the subspace \(\mathcal{H}_{n,m}\) with exactly \(n\) photons. The commutation relation implies that the unitary evolution operator \(U = e^{-i\hat{H}t}\) commutes with the spectral projectors:
\begin{equation}
[U, \Pi_n] = 0, \quad \forall n \ge 0.
\end{equation}

Consider an arbitrary density operator \(\rho\) acting on the full Fock space. We define its unnormalized projection onto the \(n\)-photon sector as:
\begin{equation}
\rho_n := \Pi_n \rho \Pi_n.
\end{equation}
Under the LON evolution \(\rho \mapsto \rho' = U \rho U^\dagger\), the projection of the evolved state is given by:
\begin{equation}
\rho'_n = \Pi_n (U \rho U^\dagger) \Pi_n.
\end{equation}
Using the commutation property \(\Pi_n U = U \Pi_n\) and the idempotence \(\Pi_n^2 = \Pi_n\), we obtain:
\begin{equation}
\rho'_n = U \Pi_n \rho \Pi_n U^\dagger = U \rho_n U^\dagger.
\end{equation}
This result establishes that each \(n\)-photon component \(\rho_n\) evolves \textit{independently} and \textit{unitarily} under the action of \(U\), restricted to the subspace \(\mathcal{H}_{n,m}\).

\subsection{Direct Sum of Invariants}

Since each component \(\rho_n\) undergoes a unitary evolution generated by a linear optical Hamiltonian, the theory developed in the main text applies individually to each \(\rho_n\).
Specifically, for a fixed \(n\), the operator \(\rho_n\) admits a unique decomposition into orthogonal coherence layers (as defined in Eq.~(3) of the main text):
\begin{equation}
\rho_n = \sum_{j=0}^n \rho_n^{(j)}, \quad \text{with } \rho_n^{(j)} \in \Delta\mathfrak{J}^{(j)}_n.
\end{equation}
According to the invariance theorem derived in the main text, the evolution \(U \rho_n U^\dagger\) preserves each layer component \(\rho_n^{(j)}\) unitarily:
\begin{equation}
(U \rho_n U^\dagger)^{(j)} = U \rho_n^{(j)} U^\dagger.
\end{equation}
Consequently, the layer-purity and layer-spectrum invariants for the general state \(\rho\) are simply the collection of the invariants for each sector:
\begin{equation}
\mathcal{I}(\rho) = \bigcup_{n=0}^{\infty} \left\{ I_j(\rho_n) := \tr[(\rho_n^{(j)})^2], \quad \mathrm{spec}(\rho_n^{(j)}) \right\}_{j=0}^n.
\end{equation}
The total purity of the state (if pure) or the sum of purities (if mixed) is recovered via \(\sum_{n,j} I_j(\rho_n)\).

\vspace{1em}
\noindent\textbf{Remark (Inter-sector Coherences).}
A general state \(\rho\) may also contain coherence terms between different photon numbers, i.e., \(\rho_{nm} = \Pi_n \rho \Pi_m\) with \(n \neq m\). While these terms evolve as \(\rho'_{nm} = U \rho_{nm} U^\dagger\), they do not contribute to the Jordan-type invariants defined here.
This is because the Jordan space \(\mathfrak{J}^{(j)}\) is spanned by operators of the form \(\hat{a}^\dagger \dots \hat{a}^\dagger \hat{a} \dots \hat{a}\) (with equal numbers of creation and annihilation operators), which are block-diagonal in the photon number basis.
Therefore, any operator \(X \in \mathfrak{J}^{(j)}\) satisfies \(\tr(X \rho_{nm}) = 0\) for \(n \neq m\).
Physically, this reflects that the internal geometry of the passive linear optical group is defined solely by photon-number-conserving generators, and thus the associated invariants are insensitive to the relative phases between different number sectors.

\section{Experiment: State Preparation with Varying Invariants}
\label{subsec:state_preparation}

\SecIntro{This section describes the experimental approach for preparing two-photon polarization-encoded states whose Jordan-layer invariants \(\{I_j\}\) are tunable. The approach is: (i) prepare a controllable input polarization superposition; (ii) use HOM interference at an NPBS and post-select the two-photon same-path events to realize \(\ket{\psi_\alpha}\); (iii) tabulate the theoretical invariant values for the prepared family; (iv) describe optical compensation to mitigate NPBS-induced phase/polarization distortions.
The experimental setup is depicted in Fig.~2 of the main text.}


\subsection{Photon Source}
Light pulses with 150~fs duration, centered at 830~nm, from an ultrafast Ti:sapphire laser (Coherent Mira-HP; 76~MHz repetition rate) are firstly frequency doubled in a \(\beta\)-type barium borate (\(\beta\)-BBO) crystal to generate a second-harmonic beam at 415~nm. The upconversion beam is then utilized to pump another \(\beta\)-BBO (type-II beam-like degenerate SPDC), producing photon pairs denoted as signal and idler.
The signal and idler photons possess distinct emergence angles and spatially separate from each other.
After passing through two narrowband (3~nm) clean-up filters, they are coupled into separate single-mode fibers and then directed into the optical circuit.

\subsection{HOM Interference with Post-Selection}
In the optical circuit, we prepare the state \(\cos2\theta\ket{H}+\sin2\theta\ket{V}\) in one path (path \(a\), transmission path) using a horizontal polarizer followed by a half-wave plate (HWP) set at an angle \(\theta\), while the other path (path \(b\), reflection path) is prepared in the state \(\ket{H}\).

Before undergoing Hong--Ou--Mandel (HOM) interference~\cite{hong_measurement_1987,branczyk_hong-ou-mandel_2017} at a non-polarizing beam splitter (NPBS), the two-photon input state is
\begin{equation}
\ket{\psi_{\rm in}}
= (\cos 2\theta\,\ket{1_H}+\sin 2\theta\,\ket{1_V})_a \ket{1_H}_b
= \big(\cos 2\theta\,\hat a_H^\dagger + \sin 2\theta\,\hat a_V^\dagger\big)\hat b_H^\dagger \ket{0}.
\label{eq:psi-in}
\end{equation}
After the NPBS, using \(\hat a^\dagger \mapsto (\hat a^\dagger+\hat b^\dagger)/\sqrt{2}\) and \(\hat b^\dagger \mapsto (\hat a^\dagger-\hat b^\dagger)/\sqrt{2}\) for each polarization mode, the output state is
\begin{equation}
\begin{aligned}
\ket{\psi_{\rm out}}
&=
\frac{1}{2}\Big[\cos 2\theta \big(\hat a_H^{\dagger 2}-\hat b_H^{\dagger 2}\big)
+\sin 2\theta \big(\hat a_V^\dagger \hat a_H^\dagger-\hat b_V^\dagger \hat b_H^\dagger\big)\Big]\ket{0}
\\
&\quad+
\frac{1}{2}\Big[\cos 2\theta \big(\hat b_H^\dagger \hat a_H^\dagger-\hat a_H^\dagger \hat b_H^\dagger\big)
+\sin 2\theta \big(\hat b_V^\dagger \hat a_H^\dagger-\hat a_V^\dagger \hat b_H^\dagger\big)\Big]\ket{0}.
\end{aligned}
\label{eq:prepare_HOM}
\end{equation}

Selecting events where both photons exit in path \(a\), we obtain
\begin{equation}
\ket{\psi_{\theta}} \propto
\big(\cos 2\theta\,\hat a_H^{\dagger 2} + \sin 2\theta\,\hat a_H^\dagger \hat a_V^\dagger\big)\ket{0}.
\label{eq:psi-theta-unnorm}
\end{equation}
After normalization,
\begin{equation}
\ket{\psi_\theta}=
\frac{1}{\sqrt{1+\cos^2 2\theta}}
\Big(\sqrt{2}\cos 2\theta\,\ket{2_H,0_V}+\sin 2\theta\,\ket{1_H,1_V}\Big).
\label{eq:post_select_state}
\end{equation}
Define
\begin{equation}
\cos\alpha = \frac{\sqrt{2}\cos 2\theta}{\sqrt{1+\cos^2 2\theta}},
\qquad
\sin\alpha = \frac{\sin 2\theta}{\sqrt{1+\cos^2 2\theta}},
\label{eq:alpha-theta}
\end{equation}
so that the prepared state can be written as
\begin{equation}
\ket{\psi_{\theta}}=\ket{\psi_\alpha}
=\cos\alpha\,\ket{2_H,0_V}+\sin\alpha\,\ket{1_H,1_V},
\qquad 0\le\theta\le \frac{\pi}{4},\quad 0\le\alpha\le \frac{\pi}{2}.
\label{eq:psi-alpha}
\end{equation}

\begin{figure}[t]
\centering
\includegraphics[width=0.5\linewidth]{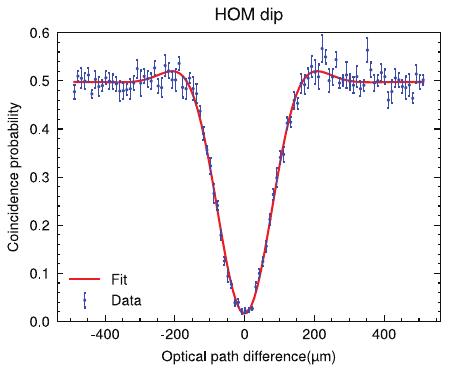}
\caption{Hong--Ou--Mandel interference dip.}
\label{fig:hom_dip}
\end{figure}

The HOM interference visibility is closely related to the tomographic fidelity of the two-photon state. To check this, we first measure the HOM interference dip by setting \(\theta=\pi/4\), which prepares \(\ket{\psi_{\rm in}}=\ket{V}_a\ket{H}_b\).
After post-selecting the case where both photons exit in path \(a\), the prepared state is \(\ket{\psi_\alpha}=\ket{1_H,1_V}\).
To observe interference in the polarization degree of freedom, we configure the evolution module as an identity operation by setting the QWP--HWP--QWP (QHQ) wave-plate group angles to Q(\(0^\circ\)), H(\(0^\circ\)), and Q(\(0^\circ\)).
The measurement module is set to Q(\(0^\circ\)) and H(\(22.5^\circ\)) to induce HOM interference in the polarization mode.

The observed HOM dip, shown in Fig.~\ref{fig:hom_dip}, is fitted using a model combining a Gaussian and a sinc function,
\begin{equation}
p_{\rm coin}(x) = a - b\, e^{-\frac{\sigma^2(x-x_0)^2}{2}}\, \operatorname{sinc}\!\big(k(x-x_0)\big),
\label{eq:hom-fit}
\end{equation}
where \(a\), \(b\), \(\sigma\), \(x_0\), and \(k\) are fitting parameters. This fitting yields an interference visibility of \(96.8\%\).

\begin{table}[!ht]
\centering
\begin{tabular}{c@{\hspace{10pt}}c@{\hspace{10pt}}c@{\hspace{10pt}}c@{\hspace{10pt}}c@{\hspace{10pt}}c@{\hspace{10pt}}c}
\hline
$\theta$ & $\alpha$ & $\ket{\psi_\alpha}$ & $I_t(=I_0+I_1)$ & $I_p(=I_2)$ & $I_{t'}(=I_1)$ & $I_o(=4I_1+2)$\\
\hline
$0^\circ$     & $0^\circ$   & $\ket{2_H,0_V}$ &
0.833 & 0.167 & 0.500 & 4\\
$7.5^\circ$   & $10.7^\circ$&
$0.983\ket{2_H,0_V}+0.186\ket{1_H,1_V}$ &
0.833 & 0.167 & 0.500 & 3.998\\
$11.25^\circ$ & $16.3^\circ$&
$0.960\ket{2_H,0_V}+0.281\ket{1_H,1_V}$ &
0.830 & 0.170 & 0.497 & 3.988\\
$15^\circ$    & $22.2^\circ$&
$0.926\ket{2_H,0_V}+0.378\ket{1_H,1_V}$ &
0.823 & 0.177 & 0.490 & 3.959\\
$22.5^\circ$  & $35.3^\circ$&
$0.816\ket{2_H,0_V}+0.577\ket{1_H,1_V}$ &
0.778 & 0.222 & 0.445 & 3.778\\
$30^\circ$    & $50.8^\circ$&
$0.632\ket{2_H,0_V}+0.775\ket{1_H,1_V}$ &
0.653 & 0.347 & 0.320 & 3.280\\
$33.75^\circ$ & $59.6^\circ$&
$0.505\ket{2_H,0_V}+0.863\ket{1_H,1_V}$ &
0.556 & 0.444 & 0.223 & 2.891\\
$37.5^\circ$  & $69.2^\circ$&
$0.354\ket{2_H,0_V}+0.935\ket{1_H,1_V}$ &
0.451 & 0.549 & 0.118 & 2.471\\
$45^\circ$    & $90^\circ$  & $\ket{1_H,1_V}$ &
0.333 & 0.667 & 0     & 2\\
\hline
\end{tabular}
\caption{
Theoretically calculated parameters of the prepared two-photon states: waveplate angle \(\theta\), angle \(\alpha\), state \(\ket{\psi_\alpha}\), tangent purity \(I_t=I_0+I_1\), perpendicular purity \(I_p = I_2\), traceless tangent purity \(I_{t'}=I_1\), and the observable invariant \(I_o\).
For \(n=2,m=2\), \(I_0=1/3\) is fixed, and \(I_o=4I_1+2\).
These states correspond to the red points shown in Fig.~1(c) of the main text.
}
\label{tab:prepared_states}
\end{table}

\subsection{Mitigation of NPBS-Induced Phase and Polarization Distortions}
In our state preparation setup, the NPBS introduces phase shifts in the transmission path and polarization distortions in the reflection path, both of which diminish the fidelity of the prepared quantum states.
To mitigate these effects, we implemented tailored compensation strategies using wave plates for each path, as shown in Fig.~2(a) by the two QHQ wave-plate groups outlined within the dashed boxes.

\medskip
\noindent\textbf{Transmission path.}
The NPBS in the transmission path induces a relative phase shift \(\varphi\) between the \(H\) and \(V\) polarization components. For an input state \(\tfrac{1}{\sqrt{2}}(\ket{H}+\ket{V})\), the transmitted state becomes \(\tfrac{1}{\sqrt{2}}(\ket{H}+e^{i\varphi}\ket{V})\).
After passing through an HWP oriented at \(22.5^\circ\), the resulting state is \(e^{i\varphi/2}\big(\cos\frac{\varphi}{2}\ket{H}-i\sin\frac{\varphi}{2}\ket{V}\big)\).
Experimentally, the \(V\)-component intensity post-interference was approximately \(5\%\) of the total intensity, corresponding to \(\varphi=25^\circ\).

To correct this phase shift, we inserted a QHQ group before the NPBS, consisting of a QWP at \(45^\circ\), an HWP at angle \(\beta\), and another QWP at \(45^\circ\). This wave-plate configuration has an effective Jones matrix
\begin{equation}
\begin{pmatrix}
e^{i2\beta} & 0 \\
0 & -e^{-i2\beta}
\end{pmatrix}.
\label{eq:QHQ-Jones}
\end{equation}
By optimizing \(\beta\), we tuned the relative phase to counteract \(\varphi\), reducing the \(V\)-component intensity to below \(0.1\%\) of the total intensity.

\medskip
\noindent\textbf{Reflection path.}
Ideally, the NPBS reflects \(H\)-polarized light exclusively. However, an unintended \(V\) component emerges, with an initial intensity of \(1.5\%\) for an input state \(\ket{H}\).
To suppress this polarization impurity, we employed a QHQ wave-plate group positioned before the NPBS. Through iterative optimization of the wave-plate angles, we reduced the \(V\)-component intensity to below \(0.1\%\).

The wave-plate sequences implemented effectively minimize phase and polarization errors induced by the NPBS.
For a more detailed analysis, quantum process tomography could be employed to fully characterize the NPBS-induced evolution.

\section{Experiment: Sampled Evolution Unitaries}
\label{subsec:state_evolution}

\SecIntro{This section describes how the passive evolutions \(U\in\mathrm U(2)\) are sampled and implemented experimentally using a QHQ wave-plate group. The strategy is: (i) use a standard parameterization of Haar-uniform \(\mathrm U(2)\); (ii) select a discrete subset of parameters; (iii) map the resulting scattering matrices to QHQ angles.}

Theoretically, the state evolution induced by any passive LON preserves all Jordan-layer invariants.
To experimentally verify this property, we implement various unitary evolutions.

In this section, we first discuss the uniform sampling of \(\mathrm{U}(2)\)~\cite{zyczkowski_random_1994} and subsequently describe the realization of the sampled \(\mathrm{U}(2)\) scattering matrices using a QHQ wave-plate configuration~\cite{simon_universal_1989}.

We parameterize \(\mathrm{U}(2)\) as follows:
\begin{equation}
\begin{aligned}
&\phi=\arcsin\sqrt{\xi}, \\
&U(\alpha,\phi,\psi,\chi):=
e^{i\alpha}
\begin{pmatrix}
e^{i\psi}\cos\phi & e^{i\chi}\sin\phi \\
-e^{-i\chi}\sin\phi & e^{-i\psi}\cos\phi
\end{pmatrix},
\end{aligned}
\label{eq:U2-sampling}
\end{equation}
where \(\alpha,\psi,\chi\in[0,2\pi]\) and \(\xi\in[0,1]\) are uniformly sampled. The resulting \(U(\alpha,\phi,\psi,\chi)\) is uniformly distributed over \(\mathrm{U}(2)\)~\cite{zyczkowski_random_1994}.

We set \(\psi,\chi=\pi/2\) or \(3\pi/2\), \(\xi=1/3\) or \(2/3\), and ignore the global phase \(\alpha\), sampling eight matrices in \(\mathrm{U}(2)\). These unitaries are then realized experimentally using the QHQ wave-plate group~\cite{simon_universal_1989}. The parameters \(\psi,\chi,\xi\) and corresponding QHQ angles \(Q(\theta_1),H(\theta_2),Q(\theta_3)\) are listed in Table~\ref{tab:QHQ_config}.

\begin{table}[ht]
\centering
\begin{tabular}{c@{\hspace{10pt}}|@{\hspace{10pt}}c@{\hspace{10pt}}c@{\hspace{10pt}}c@{\hspace{10pt}}|@{\hspace{10pt}}c@{\hspace{10pt}}c@{\hspace{10pt}}c@{\hspace{10pt}}}
\hline
Number & $\psi$ & $\chi$ & $\xi$ & $\theta_1$ & $\theta_2$ & $\theta_3$ \\
\hline
$U_1$ & $\pi/2$   & $\pi/2$   & $1/3$ & 24.1$^\circ$  & 17.6$^\circ$  & 101.2$^\circ$ \\
$U_2$ & $\pi/2$   & $\pi/2$   & $2/3$ & 149.5$^\circ$ & 27.4$^\circ$  & 175.2$^\circ$ \\
$U_3$ & $\pi/2$   & $3\pi/2$  & $1/3$ & 78.8$^\circ$  & 162.4$^\circ$ & 155.9$^\circ$ \\
$U_4$ & $\pi/2$   & $3\pi/2$  & $2/3$ & 4.8$^\circ$   & 152.6$^\circ$ & 30.5$^\circ$ \\
$U_5$ & $3\pi/2$  & $\pi/2$   & $1/3$ & 27.4$^\circ$  & 72.4$^\circ$  & 27.4$^\circ$ \\
$U_6$ & $3\pi/2$  & $\pi/2$   & $2/3$ & 81.9$^\circ$  & 62.6$^\circ$  & 133.3$^\circ$ \\
$U_7$ & $3\pi/2$  & $3\pi/2$  & $1/3$ & 152.6$^\circ$ & 107.6$^\circ$ & 152.6$^\circ$ \\
$U_8$ & $3\pi/2$  & $3\pi/2$  & $2/3$ & 98.1$^\circ$  & 117.4$^\circ$ & 46.7$^\circ$ \\
\hline
\end{tabular}
\caption{The sampled unitary parameters \(\psi,\chi,\xi\) and the corresponding QHQ angles \(Q(\theta_1),H(\theta_2),Q(\theta_3)\).}
\label{tab:QHQ_config}
\end{table}

\section{Experiment: Two-photon State Tomography}
\label{subsec:tomography}

\SecIntro{This section details the tomography procedure for a two-photon, two-mode polarization-encoded state. The strategy is: (i) implement a set of rotated photon-counting POVMs using a QH wave-plate group followed by a BD-based mode separation; (ii) realize pseudo-PNR detection with APD coincidences; (iii) verify informational completeness via rank of the measurement matrix; (iv) reconstruct \(\rho\) by least-squares fitting to measured probabilities.}

In this section, we detail the tomography procedure for a two-photon, two-mode polarization-encoded quantum state.
The reconstruction requires a set of independent POVMs to fully characterize the density matrix.

The measurement module, illustrated in Fig.~2(c) of the main text, consists of a polarization evolution and detection module.
First, the multiphoton state is evolved by a QWP--HWP (QH) wave-plate group. The evolved state is then projected onto the photon-number basis using a beam displacer (BD), which spatially separates the horizontal (\(H\)) and vertical (\(V\)) polarization components. This results in three projection basis: \(\ket{0_H,2_V}\), \(\ket{1_H,1_V}\), and \(\ket{2_H,0_V}\).

Following this projection, two fiber beam splitters and four avalanche photodiodes (APDs, Excelitas Technologies, efficiency 55\%, labeled 1--4 as in Fig.~2(c)) implement two pseudo photon-number-resolving detectors (PPNRD). Coincidence events are assigned as follows: \(\{1,2\}\) for \(\ket{2_H,0_V}\), \(\{1,3\}\), \(\{1,4\}\), \(\{2,3\}\), and \(\{2,4\}\) for \(\ket{1_H,1_V}\), and \(\{3,4\}\) for \(\ket{0_H,2_V}\).

For each QH angle configuration, the corresponding POVMs \(\{\hat E_i\}\) take the form
\begin{equation}
\hat E_1 = V_{\rm QH}^\dagger \hat \Pi_{2_H,0_V} V_{\rm QH},\quad
\hat E_2 = V_{\rm QH}^\dagger \hat \Pi_{1_H,1_V} V_{\rm QH},\quad
\hat E_3 = V_{\rm QH}^\dagger \hat \Pi_{0_H,2_V} V_{\rm QH},
\label{eq:POVMs}
\end{equation}
where \(V_{\rm QH}\) represents the multiphoton unitary induced by the QH wave-plate group, and \(\hat \Pi_{n_H,m_V}=\ket{n_H,m_V}\bra{n_H,m_V}\) are the Fock-state projectors.

By varying the QH angle configuration, we construct more independent POVMs.
It has been established that reconstructing an \(N\)-photon, \(M\)-mode state using photon-counting detection combined with an \(M\)-mode linear-optical network requires at least~\cite{banchi_multiphoton_2018}
\begin{equation}
R_{N,M} = \binom{N+M}{N} - \binom{N+M-2}{M}
\label{eq:tomography-lowerbound}
\end{equation}
distinct measurement configurations. For \(N=2\), \(M=2\), this lower bound is five independent settings.

We adopt six QH settings based on Ref.~\cite{adamson_multiparticle_2007}, with the QWP and HWP angles chosen as \((0^\circ,0^\circ)\), \((0^\circ,11.25^\circ)\), \((0^\circ,22.5^\circ)\), \((22.5^\circ,0^\circ)\), \((22.5^\circ,22.5^\circ)\), and \((45^\circ,22.5^\circ)\).

To verify POVM independence, we vectorize each \(E_i\) (matrix representation in the Fock basis) and stack them as row vectors to form a matrix \(\mathcal{M}\). The calculated rank confirms that nine independent POVMs are obtained.
The use of an overcomplete measurement set enhances fidelity and mitigates systematic errors.

Finally, the density matrix \(\rho\) is obtained by solving a least-squares optimization problem minimizing
\begin{equation}
f(\rho) = \sum_i \Big[\tr(\rho\,E_i) - p_i\Big]^2,
\label{eq:LS-objective}
\end{equation}
where \(p_i\) is the experimentally measured probability corresponding to the POVM \(E_i\).

\bibliographystyle{apsrev4-2}
\bibliography{supp_ref}